\documentclass[12pt,onecolumn]{report}
\usepackage{graphics}
\usepackage{graphicx}
\usepackage{epstopdf}
\usepackage{epsfig}
\usepackage{latexsym}
\usepackage{amsfonts}
\usepackage{here}
\usepackage{rawfonts}
\usepackage[latin1]{inputenc}
\usepackage[T1]{fontenc}
\usepackage{calc}
\usepackage{url}
\usepackage{enumerate}
\usepackage{color}
\usepackage[tbtags]{amsmath}
\usepackage{amssymb}
\usepackage{upref}
\usepackage{epic,eepic}
\usepackage{times}
\usepackage{rotating}
\usepackage{cite}

 \usepackage{a4wide}

\newcommand{\Tr}{\ensuremath{\operatorname{Tr}}}

\renewcommand{\Re}{\ensuremath{\operatorname{Re}}}
\renewcommand{\Im}{\ensuremath{\operatorname{Im}}}

\newcommand{\bi}{\begin{itemize}}
\newcommand{\ei}{\end{itemize}}

\newcommand{\be}{\begin{enumerate}}
\newcommand{\ee}{\end{enumerate}}

\newcommand{\ed}{\end{description}}
\newcommand{\bc}{\begin{center}}
\newcommand{\ec}{\end{center}}
\newcommand{\bt}{\begin{tabbing}}
\newcommand{\et}{\end{tabbing}}
\newcommand{\bfig}{\begin{figure}}
\newcommand{\efig}{\end{figure}}
\newcommand{\beq}{\begin{equation}}
\newcommand{\beqarr}{\begin{eqnarray}}
\newcommand{\beqarrn}{\begin{eqnarray*}}
\newcommand{\eeq}{\end{equation}}
\newcommand{\eeqarr}{\end{eqnarray}}
\newcommand{\eeqarrn}{\end{eqnarray*}}
\newcommand{\bflr}{\begin{flushright}\vspace{-0.2in}}
\newcommand{\eflr}{\end{flushright}}
\newcommand{\bsub}{\begin{subequations}}
\newcommand{\esub}{\end{subequations}}
\newcommand{\barr}{\begin{array}}
\newcommand{\earr}{\end{array}}

\begin{document}
\begin{titlepage}
\thispagestyle{empty}

\centering {\Large \bf STUDY AND IMPLEMENTATION OF UNITARY GATES
IN QUANTUM COMPUTATION USING SCHRODINGER DYNAMICS
}\\
\centering{Thesis submitted to the}\\
\centering{University of Delhi}\\
\centering{for the award of the degree of}\\
\centering{\LARGE{Doctor of Philosophy}}\\
\centering{in}\\
\centering{\LARGE{Electronics and Communication}} \vspace{.75 cm} \vspace{.5 cm}\\
\centering{by}\\
\centering {\Large \bf KUMAR GAUTAM}\\
\vspace {0.5cm}
\begin{centering}
DEPARTMENT OF ELECTRONICS AND COMMUNICATION ENGINEERING\\
FACULTY OF TECHNOLOGY\\ UNIVERSITY OF DELHI, NEW DELHI, INDIA\\
2016
\end{centering}
\end{titlepage}

\begin{titlepage}
\thispagestyle{empty}

\centering {\Large \bf STUDY AND IMPLEMENTATION OF UNITARY GATES
IN QUANTUM COMPUTATION USING SCHRODINGER DYNAMICS
}\\
\centering{Thesis submitted to the}\\
\centering{University of Delhi}\\
\centering{for the award of the degree of}\\
\centering{\LARGE{Doctor of Philosophy}}\\
\centering{by}\\
\centering {\Large \bf KUMAR GAUTAM}\\
\centering{Under the joint Guidance of}\\
\centering{\textbf{Prof. Harish Parthasarathy}}\\
\centering{and}\\
\centering{\textbf{Dr. Tarun Kumar Rawat}}\\
\begin{centering}
DEPARTMENT OF ELECTRONICS AND COMMUNICATION ENGINEERING\\
FACULTY OF TECHNOLOGY\\ UNIVERSITY OF DELHI, NEW DELHI, INDIA\\
2016
\end{centering}
%

\end{titlepage}


\newpage
\vspace{10cm}
\bigskip
\begin{titlepage}
\chapter*{\centering \emph{Dedicated to\\ My Mother and My Father\\along with \\ QRACE\\\big(Quantum Research And Centre of Excellence\big)}}
\end{titlepage}
\pagenumbering{roman} \vspace{-3.5 in}
\begin{centering}
{\huge Certificate\vspace{0.5cm}\\}
\end{centering}
This is to certify that the thesis entitled {\bf ``Study and
Implementation of Unitary Gates in Quantum Computation Using
Schrodinger Dynamics''} being submitted by {\bf Mr. KUMAR GAUTAM}
to the Department Of Electronics And Communication Engineering,
University of Delhi, for the award of the degree of {\bf Doctor of
Philosophy} is the record of the bona-fide research work carried
out by him under my supervision. In my opinion, the thesis has
reached the standards fulfilling the requirements of the
regulations relating to the degree.

The results contained in this thesis have not been submitted
either in part or in full to any other university or institute for
the award of any degree or diploma.
\\
\\
\noindent{\small(\textbf{Prof. Harish
Parthasarathy})\hspace{130pt}
(\textbf{Dr. Tarun Kumar Rawat})\\
Professor\hspace{225pt} Assistant Professor \\
Division of Electronics and Communication \hspace{70pt} Division of Electronics and Communication \\
Engineering \hspace{210pt} Engineering\\
Netaji Subhas Institute of Technology, Delhi \hspace{71pt} Netaji Subhas Institute of Technology, Delhi \\
Dwarka, New Delhi 110078, India\hspace{115pt} Dwarka, New Delhi
110078, India} \vspace{1cm}\begin{flushleft} {\bf Prof. Harish
Parthasarathy} \\ University Head\\
Department of ECE,\\ Faculty of Technology, University of Delhi,
India\end{flushleft}


\newpage
\vspace{115 cm}
\bigskip
\bigskip
\begin{titlepage}
\chapter*{\centering \small{There should be no boundaries to human
endeavor.\\We are all different.\\However bad life may seem, there
is always something you can do, and succeed at.\\While there's
life, there is hope.\\ by\\Stephen Hawking}}
\end{titlepage}


\newpage

\chapter*{\centering Acknowledgements}

I would like to express my special appreciation and thanks to my
advisor Professor Harish Parthasarathy and Dr. Tarun Kumar Rawat,
you have been a tremendous mentor for me. I would like to thank
you for encouraging my research and for allowing me to grow as a
research scientist. Your advice on both research as well as on my
career have been priceless.

I would like to thank Prof. Raj Senani for providing conducive
environment for research in the institute. He has been a great
source of encouragement. I am indebted to him for his valuable
suggestions and comments on my research work.

I would also like to thank my committee members, Professor
Maneesha Gupta, Professor Parul Garg for serving as my committee
members even at hardship. All of you have been there to support me
for my Ph.D. thesis. I would never forget the help extended to me
by Navneet sharma and Naman Garg during my research period at
NSIT.

I also acknowledge Dr. D Upadhya, Mr. Kuwar Singh, Mr. D V Gadare,
Prof. S. P. Singh, Dr. Sujata Sen Garg, and Dr. Joytsan Singh for
their support and encouragement through all these years.

I would like to acknowledge the help of Prof. K.R. Parthasarathy
during my research. He has been a great source of knowledge and
always ready to listen to my problems and give his suggestions.

I am also grateful to Mr. Garv Chauhan, Sonu, Vivek and Varun
Upadhyaya for their helpful suggestions and support.

Also, I would like to acknowledge Mr. Manoj and Mr. Suresh of
Communication Laboratory and Mr. Vireshwar Sharma of
Microprocessor Laboratory, Mr. Satish Kumar of DCS Laboratory, Mr
Sanjeev, Ms. Shashi Rawat, Mr. Sukhbir Singh, Mr. Vinod Kumar, Mr.
Ramdas, staff, ECE Department at NSIT.

I am grateful to Mr. Satish Kumar Tiwari, Mr. Saurabh Bharadwaj,
Mr. Vikas, Mr. Rochak, Dr. Parbhat Sharma, Meenakshi Agarwal, Rich
barsaniya, Mr. Manjeet Kumar, Ms. Richa Yadav and Ms. Urvashi
Singh of NSIT for their help.

I would like to thank all my friends and my loveable students for
helping me get through the difficult time and for their emotional
support they provided. I feel fortunate to have the support of
them with me and for making my stay at Netaji Subhas Institute of
Technology, New Delhi a memorable one. My heartfelt gratitude are
also due to faculty and staff of Netaji Subhas Institute of
Technology, New Delhi.

A special thanks to my family. Words cannot express how grateful I
am to my mother and my father for all of the sacrifices that you
have made for me. Your prayer for me was what sustained me thus
far. They have been very patient and loving when I needed it most.

\begin{flushright}
\textbf{KUMAR GAUTAM}.\\
\end{flushright}

\newpage
\chapter*{\centering Abstract}
In this work, we explore the idea of realizing quantum gates
normally used in quantum computation using physical systems like
atoms and oscillators perturbed by electric and magnetic fields.
The basic idea around which the subject of this thesis revolves is
that if a time independent Hamiltonian $H_0$ is perturbed by a
time varying Hamiltonian of the form $f(t)V$ where $f(t)$ is a
scalar function of time and $V$ is a Hermitian operator that does
not commute with $H_0$, then a very large class of unitary
operators can be realized via the Schrodinger evolution
corresponding to the time varying Hamiltonian $H_0+f(t)V$. $H_0$
by itself generates only a one dimensional class of unitary gates
while $H_0+f(t)V,\quad t\geq 0$ can generate an infinite
dimensional manifold of unitary gates. This is a consequence of
the Baker-Campbell-Hausdorff formula in Lie groups and Lie
algebras. Broadly speaking we treat two problems in this thesis
based upon the above idea. First, we take a Harmonic oscillator
and perturb it with a time independent anharmonic term. The total
Hamiltonian is then $H_1=\frac{q^2+p^2}{2}+\epsilon q^3$. We then
calculate $U_g=e^{-\iota T H_1}$ and consider this to be the
desired gate to be realized. We then perturb the harmonic
Hamiltonian with a linear time dependent term so that the overall
Hamiltonian becomes $H(t)=\frac{q^2+p^2}{2}+\epsilon f(t)q$ and
calculate the unitary evolution corresponding to $H(t)$ at time
$T$. Using the time ordering operator $T$, this gate can be
expressed as
$$
U(T)=U(T,\epsilon,f)=T\{e^{-\iota\int_0^TH(t)dt}\}
$$
$U(T)$ is calculated upto $O(\epsilon^2)$ using time dependent
perturbation theory and $f(t)$ is chosen so that $U(T,\epsilon,f)$
is as close as possible in the Frobenius norm to $U_g$ with a
power constraint on $f(t)$. This optimization problem is solved by
arriving at a linear integral equation for $f(t)$. This problem is
equivalent to perturbing a charged Harmonic oscillator with a time
varying electric field and using the electric field as our control
process to generate a gate as close as possible to the given gate.
The anharmonic gate $U_g$ is then replaced by a host of commonly
used gates in quantum computation like controlled unitary gates,
quantum Fourier transform gate etc and the control electric field
is then selected appropriately. We then apply the same formalism
to Hamiltonians consisting of an atom described by a Pauli spin
variable plus a quantum electromagnetic field Hamiltonian
described by creation and annihilation operators and an
interaction term between atom and field that is modulated by a
scalar control function. This is particularly important since
recently ion trap systems have been modeled in this way and
quantum gates realized using this scheme. In the course of
designing quantum gates using physical systems like atoms and
oscillators perturbed by electric and magnetic fields, we have
also addressed the controllability issue, that is, under what
conditions does there exist a scalar real valued function of time
$f(t), 0\leq t\leq T$ such that if $|\psi_\iota\rangle$ is any
initial wave function and $|\psi_f\rangle$ is any final wave
function, then $U(T,f)|\psi_i\rangle=|\psi_f\rangle$. We have
obtained a partial solution to this problem by replacing the
unitary evolution kernel $U(T,f)$ by its Dyson series truncated
version. In all our design procedures, the gates that actually
appear are infinite dimensional, more precisely, they are of the
form $e^{-\iota T H}$ where $H$ is an unbounded Hermitian operator
acting on an infinite dimensional Hilbert space. We have
approximated the infinite dimensional problem by a finite
dimensional one based on truncation. The primary novel feature of
this thesis, is the design of quantum gates when the system
consists of an atom/oscillator described by either position and
momentum operators or creation and annihilation operators or spin
matrices and a quantum electromagnetic field described by a
sequence of creation and annihilation operators and there is an
interaction between the atom and the electromagnetic field that is
modulated by a controllable function of time, like for example a
spin interacting with a controllable quantum magnetic field.

\newpage
  \tableofcontents
  \newpage
\listoffigures \addcontentsline{toc}{chapter}{\numberline{}List of
Figures}
 \newpage
\chapter*{Abbreviations}
\addcontentsline{toc}{chapter}{\numberline{}Abbreviations}
\begin{tabbing}
A \= B \kill

FFT \> \hspace{2cm} Fast Fourier Transform\\
DFT \> \hspace{2cm} Discrete Fourier Transform\\
QFT \> \hspace{2cm} Quantum Fourier Transform\\
NSR \> \hspace{2cm} Noise to Signal Ratio\\
NSER \> \hspace{2cm}Noise to Signal Energy Ratio\\
3-D  \> \hspace{2cm}Three- Dimensional\\

\end{tabbing}
\newpage
\chapter*{List of Symbols}
\addcontentsline{toc}{chapter}{\numberline{}List of Symbols}
\begin{tabbing}
A \= B \kill

$E$ \> \hspace{2cm} Energy eigenvalue \\
$a$ \> \hspace{2cm} Annihilation operator \\
$p$ \> \hspace{2cm} Momentum Operator \\
$q$ \> \hspace{2cm} Charge on the harmonics oscillator \\
$x$ \> \hspace{2cm} Position operator\\
$a^\dag$ \> \hspace{2cm} Creation operator\\
$H_0$ \> \hspace{2cm} Unperturbed Hamiltonian\\
$m$ \> \hspace{2cm} Mass of atom \\
$\sigma$ \> \hspace{2cm} Pauli spin matrix\\
$B_0$ \> \hspace{2cm} Unperturbed magnetic field\\
$\lambda$ \> \hspace{2cm} Lagrange Multipliers \\
$w$  \> \hspace{2cm}Bohr frequency\\

\end{tabbing}
\newpage
\pagenumbering{arabic}
\chapter{INTRODUCTION}
\section{The Need for a Quantum Theory}
A famous quote from Richard Feynman goes, "I think it is safe to
say that no one understands quantum mechanics". In this thesis
we'll pursue about quantum mechanics as a basic idea which more
deeply emphasize the conceptual structure of nature and is also
easily understood. The fundamental physical laws on the
microscopic scale (Einstein's equations for general relativity)
are expressed as partial differential equations. The state of a
system of gravitating bodies and electromagnetic fields is
determined by a set of fields satisfying these equations, and
observable quantities are functionals of these fields [1]. Thus on
the one hand, the mathematics is just that of the usual calculus:
differential equations and their real-valued solutions. On the
other hand to describe nature on a microscopic scale, the quantum
theory was developed by Rutherford and Bohr as a response to the
failure of classical mechanics and classical electromagnetics in
explaining the stability of matter because of radiation of energy
by accelerating charged particles [2]. A major part in the
creation of quantum mechanics was played by Max Planck who using
the experimental results on the spectrum of black body radiation,
postulated that the energy of a photon comes in integer multiples
of $h\nu$, where $h$ is a universal constant and $\nu$ is the
frequency of radiation. Planck's analysis was made rigorous by S.
N. Bose and A. Einstein who respectively explained Planck's black
body radiation spectrum by maximizing the entropy of
indistinguishable particles (today called Bosons) and the specific
heat of solids at low temperatures. All this work was carried out
before 1925.\par After 1925 the major players in quantum mechanics
were Heisenberg, Schr\"odinger, Dirac and Max Born. Heisenberg
proposed a new type of mechanics called matrix mechanics to
explain the spectral lines of atoms. His suggestion was that
observables like position, momentum, angular momentum and energy
should be represented not by real numbers but by matrices with the
row and column indices corresponding respectively to the initial
and final states of the atom during the radiation process [3].
Heisenberg also gave an intuitive explanation of the uncertainty
principle stating that both position and momentum of a particle
cannot be measured simultaneously with infinite accuracy. This
principle was proved much later rigorously by Hermann Weyl using
Dirac's operator theoretic formalism of quantum mechanics.
According to classical physics, the position and velocity of a
particle can be calculated simultaneously to an arbitrary
precision. But in quantum mechanics, the accuracy with which we
can measure the momentum and position simultaneously, is dictated
by Heisenberg's uncertainty principle, that is, $\Delta x \Delta
p_x\geq \frac{\hbar}{2}$. Further, De-Broglie's hypothesis states
that every moving particle has a wave function associated with it.
This wave function however spreads throughout the space and cannot
be localized. Everything came together in 1926, when E.
Schr\"odinger proposed his famous wave equation. Heisenberg
developed the theory of quantum mechanics using infinite matrices
to represent observables, and he was the first person who applied
it to the Hydrogen atom. The same year, Dirac, Born, Heisenberg
and Jordan are obtained a complete formulation of quantum
mechanics that could be applied to any quantum system.
Schr\"odinger gave the wave mechanics approach to quantum theory
using which he was able to arrive at the energy spectrum of the
Hydrogen atom by solving an eigenvalue problem while Heisenberg
gave a matrix mechanics approach to the quantum mechanics which
although being conceptually clear, was not suitable for practical
calculations [4, 5]. It was finally Dirac who unified both the
Schr\"odinger and Heisenberg pictures by showing that both
pictures lead to the same value of the average of an observable in
a state. Dirac also presented a new method to arrive at the Lie
bracket or commutator of  Heisenberg mechanics just by exploiting
the properties of the Poisson bracket of classical mechanics.
Finally, Dirac derived a relativistic wave equation of the
electron by factoring the relativistic Einstein energy-momentum
relation with linear factors using $4\times 4$ anti-commuting
matrices. Dirac's wave equation is at the heart of modern quantum
field theory as developed later by Feynman, Schwinger and
Tomonaga. Dirac in his principles of quantum mechanics, stated a
formula which said that transition amplitudes like $\langle
q_n|S_nS_{n-1}\cdots S_0|q_0\rangle$ could be calculated like
$\sum_{q_1\cdots q_{n-1}}\langle q_n|S_n|q_{n-1}\rangle \langle
q_{n-1}|S_{n-1}|q_{n-2}\rangle\cdots\langle q_1|S_n|q_0\rangle$
and that this could be generalized to infinite products by summing
over continuous paths instead of discrete paths. Feynman took the
clue from this statement of Dirac and in his Ph.D thesis,
developed a Lagrangian path integral approach to non-relativistic
quantum mechanics. This was a major breakthrough since the
Hamiltonian approach to quantum mechanics like the Schr\"odinger
and Heisenberg mechanics are not Lorentz covariant because time
occupies a special privilege as compared to the spatial variables.
Feynman on the one hand had developed the entire quantum theory of
fields involving the computation of scattering amplitudes of
electrons, positrons and photons during interactions using his
path integral approach. This approach requires both Bosonic path
integrals for the electromagnetic fields and Fermionics/Berezin
path integrals for the electron-positrons/Dirac fields. Feynman
was able to calculate very accurately, the probabilities for such
scattering processes which were verified in particle accelerators.
Feynman's approach can also be applied to quantum gate design
given a Lagrangian $L$ dependent on a control input $u(t)$ (like a
classical electromagnetic field). The idea is to choose $u(.)$ so
that the matrix
$$\left(\bigg(%
\begin{array}{c}
  \langle f|e^{\frac{\iota}{k}\int_0^TL(t, u(t))dt}|i\rangle \\
\end{array}%
\right)\bigg);\quad 1\leq n, f\leq N$$ is as close as possible to
a given unitary gate $U_g$. Schwinger and Tomonaga give an
independent method for calculating the scattering matrix based on
operator theoretic expansion of the Dirac operator field and the
quantum electromagnetic field. It was Freeman Dyson who unified
both the approaches of Feynman, Schwinger and Tomonaga using the
Dyson series expansion of the scattering matrix in the interaction
picture. We shall in our work be following the Dyson series
approximation for quantum gate design both for particle quantum
mechanics and field theoretic quantum mechanics [6, 7, 8].
\section{States, Observables and Schrodinger, Heisenberg and Dirac's
Interaction Pictures of Quantum Dynamics} In classical physics,
the state of a system is given by a point in a "phase space",
which one can think of equivalently as the space of solutions of
an equation of motion, or as (parametrizing solutions by initial
value data) the space of coordinates and momenta. Observable
quantities are just functions on this space (that is, functions of
the coordinates and momenta) [9, 10]. There is one distinguished
observable, the energy or Hamiltonian, and it determines how
states evolve in time through Hamilton's equations. The basic
structure of quantum mechanics is quite different, with the
formalism built on the following simple axioms [11, 12].
\subsection{States} \emph{The state of a quantum mechanical system is given by a nonzero
vector in a complex vector space $\mathcal{H}$ with Hermitian
inner product $\langle .,.\rangle$}. $\mathcal{H}$ may be finite
or infinite dimensional, with further restrictions required in the
infinite-dimensional case (e.g. we may want to require
$\mathcal{H}$ to be a Hilbert space) [13, 14]. The states of the
quantum system are represented as vectors in Hilbert space and
operations associated with position and momentum act like matrices
operating on these vectors. Dirac introduced the inner product
between quantum states which is described through the bra-ket
vector notation. A bra denoted as $\langle |$, is a row vector. A
ket denoted as $|\rangle$, is a column vector [15, 16].
\begin{itemize}
\item  The state space is always linear: A linear combination of
states is also a state, after appropriate normalization. \item The
state space is a \emph{complex} vector space: These linear
combinations can and do crucially involve complex numbers, in an
inescapable way. In the classical case only real numbers appear,
with complex numbers used only as an inessential calculational
tool [17, 18, 19].
\end{itemize}
\subsection{Observables}
In quantum mechanics, in order to extract quantum information from
a quantum system, we need to observe or measure the system. An
observable is a property of a physical system that can be measured
in respect of position, velocity and momentum. An observable is
associated with a Hermitian operator. The measured value of an
observable is an eigenvalue of its operator. That is, given a
Hermitian operator $X$, if the pure state $|\psi\rangle$ an
eigenvector (or eigenket) of $X$ with eigenvalue $\lambda$, then
if the system is in the state $|\psi\rangle$, the measured value
of $X$ is $\lambda$. This is different from a mixed state which is
a liner superposition of pure states represented as
$|\psi\rangle\langle \psi|$. A mixed quantum state corresponds to
a probabilistic mixture of pure states; however, different
distributions of pure states can generate equivalent (that is,
physically indistinguishable) mixed states. In other words, a
mixed state $\rho$ can be represented as
$\sum_\alpha|\psi_\alpha\rangle p_\alpha \langle \psi_\alpha|$ in
more than one way if orthogonality of the $|\psi_\alpha\rangle's$
is not imposed [20, 21, 22].
\subsection{Schrodinger, Heisenberg and Dirac's Interaction
Pictures of Quantum Dynamics} The total Hamiltonian of a perturbed
system has the form
$$H=H_0+V$$
In the Schrodinger picture, let $X$ be an observable and
$|\psi(t)\rangle$ be the state. Then $X$ remains constant while
$$
|\psi(t)\rangle=e^{-\iota tH}|\psi(0)\rangle
$$
Average value of $X$ at time $t$ is
$$\langle\psi(t)|X|\psi(t)\rangle=\langle\psi(0)|e^{\iota tH}.X.e^{-\iota tH}|\psi(0)\rangle$$
We have
$$\frac{d}{dt}|\psi(t)\rangle=-iH|\psi(t)\rangle,\quad dX/dt=0$$
In the Heisenberg picture, the average value of $X$ is the same as
in the Schrodinger picture but we assume that $|\psi\rangle$ is a
constant, that is, $|\psi(t)\rangle=|\psi(0)\rangle$ while the
observable $X$ changes with time to $X(t)$ [23, 24]. To maintain
the same average we therefore require that
$$\langle\psi(0)|X(t)|\psi(0)\rangle=\langle\psi(t)|X|\psi(t)\rangle=\langle\psi(0)|e^{\iota tH}X.e^{-\iota tH}|\psi(0)\rangle$$
Since this must be true for all states $|\psi(0)\rangle$, we
require that
$$
X(t)=e^{\iota tH}X.e^{-\iota tH}
$$
or equivalently,
$$
\frac{dX(t)}{dt}=\iota[H,X(t)]
$$
We require that in both the Schrodinger and Heisenberg pictures,
the averages of observables with time must be same since, it is
the average of the observables that we physically measure.

In the interaction picture, observables evolve according to $H_0$,
not according to $H$ while states evolve according
$V_0(t)=e^{\iota tH_0}V.e^{-\iota tH_0}$. The averages of
observables then also remain the same as the following calculation
shows. Let
$$
\frac{d}{dt}|\psi_0(t)\rangle=-\iota V_0(t)|\psi_0(t)\rangle,
$$
$$
\frac{dX_0(t)}{dt}=\iota[H_0,X_0(t)]
$$
Then,
$$
\frac{d}{dt}\langle\psi_0(t)|X_0(t)|\psi_0(t)\rangle=
$$
$$
(\frac{d}{dt}\langle\psi_0(t)|X_0(t)|\psi_0(t)\rangle+\langle\psi_0(t)|X_0(t)|\frac{d}{dt}|\psi_0(t)\rangle+\langle\psi_0(t)|X'_0(t)|\psi_0(t)\rangle
$$
$$
=\iota\langle\psi_0(t)|[V_0(t),X_0(t)]+[H_0,X_0(t)]|\psi_0(t)\rangle
$$
$$
=\iota\langle\psi_0(t)|exp(\iota tH_0)[H_0+V,X]e^{-\iota
tH_0}|\psi_0(t)\rangle
$$
$$
=\iota\langle\psi_0(t)|e^{\iota tH_0}[H,X]e^{-\iota
tH_0}|\psi_0(t)\rangle
$$
Now define
$$
|\psi(t)\rangle=e^{-\iota tH_0}|\psi_0(t)\rangle
$$
then, $|\psi(t)\rangle$ follows Schrodinger evolution, since
$$
\frac{d}{dt}|\psi(t)\rangle=-iH_0|\psi(t)\rangle-\iota e^{-\iota
tH_0}V_0(t)|\psi_0(t)\rangle
$$
$$
=-\iota(H_0+V)e^{-\iota tH_0}|\psi_0(t)\rangle=-\iota
H|\psi(t)\rangle
$$
Hence, the rate of change of the average
$\langle\psi_0(t)|X_0(t)|\psi_0(t)\rangle$ in the interaction
picture coincides with $\iota\langle\psi(t)|[H,X]|\psi(t)\rangle$,
that is, with that obtained in the Schrodinger or the Heisenberg
pictures [25, 26, 27].

\section{Realization of Finite Qubit Quantum Gates by Truncation of Infinite Dimension Quantum System}
The quantum mechanics of an atom, that is, particles and
oscillators is usually described in infinite dimensional Hilbert
spaces. The Hamiltonian is built out position and momentum
operators which are unbounded operators in a Hilbert space, hence
both the unperturbed Hamiltonian as well as its perturbation are
unbounded operators in an infinite dimensional Hilbert space. The
technique of handling unitary evolution semigroups generated by
such unbounded self adjoint operators has been dealt with
thoroughly by Kato [25, 28, 29]. Once a unitary operator in an
infinite dimensional Hilbert space $\mathcal{H}$ is known, we can
truncate it, that is, approximate $\mathcal{H}$ by
$\mathcal{H}=\text{span}{|e_\alpha\rangle,\quad \alpha=1, 2,\cdots
, N}$ a finite dimensional subspace of $\mathcal{H}$ where
$\langle e_\alpha|e_\beta\rangle=\delta_{\alpha\beta}$. Likewise,
we can approximate $U$ in $\mathcal{H}$ by $U_0$ in $H_0$ where
$U_0=((\langle e_\alpha|U|e_\beta\rangle))_{1\leq \alpha,
\beta\leq N}$. However, the truncated matrix $U_0$ will not
generally be unitary. We thus look for a unitary operator
$\widetilde{U}_0$ in $H_0$ that is closest to $U_0$ in some matrix
norm. One such approximation is obtained by applying the polar
decomposition to $U_0$ and extract $\widetilde{U}_0$ as its
unitary component. In this way finite dimensional unitary gates
can be designed. Another technique is based on using generators.
Let $U=e^{\iota H}$ where $H$ is infinite dimensional Hermitian.
Then, take $H_0=((\langle e_\alpha|H|e_\beta\rangle))_{1\leq
\alpha, \beta\leq N}$. $H_0$ is again Hermitian and we can
approximate $U$ by $U_0=e^{\iota H_0}$, which is unitary, and
finite dimensional [30, 31].
\section{Methods for Simulating Quantum Evolution}
Simulation of quantum systems gives a Hamiltonian $H(t)=H_0+V(t)$.
We simulate the wave function evaluation by directly discretizing
the continuous time Schr\"odinger equation
$$\iota \frac{d\psi(t)}{dt}=H(t)\psi(t)$$ as $$\psi(t+\Delta)=(I-\iota \Delta.H(t))\psi(t)$$
However, this is not a unitary evolution since $(I-\iota\Delta
H(t))$ is not a unitary. Hence we cannot guarantee that
$\|\psi(t)\|=1\quad \forall t$. However using the Cayley
transformation we can define an alternate unitary evolution
$\psi(t+\Delta)=(I+\frac{\iota \Delta}{2}H(t))^{-1}(I-\frac{\iota
\Delta}{2}H(t))\psi(t)$. This is a unitary evolution. Another way
is to use $\psi(t+\Delta)=e^{-\iota\Delta H(t)}\psi(t)$. The
accuracy can be improved by using
\begin{align}\nonumber\psi(t+\Delta)&=\psi(t)+\Delta\psi^{'}(t)+\frac{\Delta^2}{2}\psi{''}(t)
\\&\nonumber=\psi(t)-\iota\Delta H(t)\psi(t)+\frac{\Delta^2}{2}(-\iota
H(t)\psi(t))^{'}\\&\nonumber=\psi(t)-\iota\Delta
H(t)\psi(t)-\frac{\iota\Delta^2}{2}(H^{'}(t)\psi(t)-H^2(t)\psi(t))\\&\nonumber=[I-\iota\Delta
H(t)-\frac{\iota\Delta^2}{2}(H^{'}(t)-H^2(t))]\psi(t)\end{align}
Again this is not unitary but a Cayley transform like method can
be used to make it unitary. The Cayley transform can be applied in
the interaction picture by truncating the Dyson series to linear
orders in the perturbing potential. Let $$\psi(t)=e^{-\iota t
H_0}\varphi(t)$$ then $$\frac{|d\varphi(t)\rangle}{dt}=-\iota
\widetilde{V}(t)|\varphi(t)\rangle$$ where
$\widetilde{V}(t)=e^{\iota t H_0}Ve^{-\iota t H_0}$. This gives
the Dyson series, which was formulated by Freeman Dyson, it is a
perturbative series, and each term can be represented by Feynman
diagrams in the quantum field theory [25, 30, 31]. Consider
$$|\varphi(t)\rangle=W(t)|\psi(0)\rangle$$
where
$$W(t)=I+\sum_{n=1}^\infty\int_{0<t_n<\cdots<t_1<t}\widetilde{V}(t_1)\widetilde{V}(t_2)\cdots\widetilde{V}(t_n)dt_1\cdots dt_n$$
which can be simulated by a discrete sum
$$W(m\Delta)\approx I+\sum_{1\leq n<\infty\atop 0\leq m_k\leq m_{k-1}\leq\cdots\leq m_1\leq n}
(-\iota)^n\Delta^n\widetilde{V}(m_1\Delta)\cdots\widetilde{V}(m_k\Delta)$$
If only one term is retained, then
$$W(m\Delta)\approx I-\iota\Delta\sum_{m=0}^n\widetilde{V}(m\Delta)$$
and to make it unitary, we further apply the Cayley transform
resulting in
$$W(m\Delta)\approx \frac{I-\frac{\iota\Delta}{2}\widetilde{V}(\Delta)}{I+\frac{\iota\Delta}{2}\widetilde{V}(\Delta)}$$
and
$$|\varphi((m+1)\Delta)\rangle=W(\Delta)|\psi(m\Delta)\rangle$$
Thus
$$|\varphi(m\Delta)\rangle=W(\Delta)^m|\psi(0)\rangle$$
where
$$W(\Delta)^m=\bigg(\frac{I-\frac{\iota\Delta}{2}\widetilde{V}(\Delta)}{I+\frac{\iota\Delta}{2}\widetilde{V}(\Delta)}\bigg)^m$$
Let $$U(t)=e^{-\iota t H_0}W(t)$$ Then it is easily seen that
$$\iota\frac{dU(t)}{dt}=H(t)U(t),\quad t\geq 0$$
and $$U(0)=I$$ $U(t)$ describes the evolution from time $0$ to
time $t$ while $U(t, t_0)=U(t)U(t_0)^{-1}$ describes the evolution
from time $t_0$ to time $t>t_0$. $U(t, t_0)$ satisfies
\begin{align}\nonumber\iota\hbar\frac{\partial U(t, t_0)}{\partial t}&=H(t)U(t,
t_0)\\\nonumber U(t, t_0)&=I\end{align} We have $$U(t,
t_0)=e^{-\iota t H_0}W(t)W(t_0)^{-1}e^{\iota t_0 H}$$ writing
$W(t, t_0)=W(t)W(t_0)^{-1}$ gives
\begin{align}\nonumber\iota\frac{\partial W(t, t_0)}{\partial t}&=\widetilde{V}(t)W(t,
t_0)\\\nonumber U(t, t_0)&=I\end{align} and so
\begin{align}\nonumber W(t, t_0)=I+\sum_{n=1}^\infty
(-1)^n\int_{t_0<t_n<\cdots<t_1<t}\widetilde{V}(t_1)\cdots\widetilde{V}(t_n)dt_1\cdots
dt_n\end{align} Thus we obtain a Dyson series for $U(t, t_0)$ [32,
33, 34].
\section{Description of Some Commonly Used Quantum Gates}
The ability to generate the unitary matrix describing a quantum
computer is a huge challenge. In quantum computing and
specifically the quantum circuit model of computation, a quantum
gate (or quantum logic gate) is a basic quantum circuit operating
on a small number of qubits. They are the building blocks of
quantum circuits, like classical logic gates are for conventional
digital circuits. We have seen the enormous superiority that
qubits have over bits. This means nothing though, if we don't have
a way of manipulating the information in qubits. To manipulate
information in a qubit, quantum gates are used. Quantum logic
gates are represented by unitary matrices. The most common quantum
gate operates on spaces of one or two qubits, just like the common
classical logic gates operate on one or two bits. This means that
as matrices, quantum gates can be described by $2\times 2$ or
$4\times 4$ unitary matrices. Quantum gates are usually
represented as matrices [16, 22, 23]. A gate which acts on $k$
qubits is represented by a $2^k\times 2^k$ unitary matrix. The
number of qubits in the input and output of the gate have to be
equal. The action of the quantum gate is found by multiplying the
matrix representing the gate with the vector which represents the
quantum state [34]. Types of quantum gates are as follows,
\subsection{Identity Gate}
This is sometimes called the Pauli I gate. The function of the
gate is trivial as the output state is the same as the input
state. The matrix representing the identity gate is given by
$$I=\left [%
\begin{array}{cc}
  1 & 0 \\
  0 & 1 \\
\end{array}%
\right]$$
\subsection{Phase Shift Gate}
This is a family of single-qubit gates that leave the basis state
$|0\rangle$ unchanged and map $|1\rangle$ to $e^{\iota\theta}
|1\rangle$. The probability of measuring a $|0\rangle$ or
$|1\rangle$ is unchanged after applying this gate, however it
modifies the phase of the quantum state. This is equivalent to
tracing a horizontal circle (a line of latitude) on the Bloch
Sphere by $\theta$ radians. The matrix representing the phase
shift gate is given by
$$R_\theta=\left[%
\begin{array}{cc}
  1 & 0 \\
  0 & e^{\iota\theta} \\
\end{array}%
\right]$$
\subsection{Phase Gate}
The phase gate performs the following  mapping on the logical
states
$$S|0\rangle  = |1\rangle$$
$$S|1\rangle = \iota|0\rangle$$
It is defined by the matrix
$$S= \left(%
\begin{array}{cc}
  1 & 0 \\
  0 & \iota \\
\end{array}%
\right)$$
\subsection{The Inverter, X Gate}
This is sometimes called the Pauli X gate. The function of the
gate is to invert the logical state of the qubit much like
classical logic inverter. The difference is that the quantum
inverter can operate on superposition states. If the qubit is in
the $|0\rangle$ state, then the result will be $|1\rangle$. If the
qubit was in the $|1\rangle$ state, then the
result will be $|0\rangle$. It is defined by the matrix $$X =\left(%
\begin{array}{cc}
  0 & 1 \\
  1 & 0 \\
\end{array}%
\right)
$$
\subsection{The Y Gate}
The Pauli Y gate performs the following mapping on the logical
states $$Y|0\rangle = \iota |1\rangle$$
$$Y|1\rangle = -\iota |0\rangle$$
It is defined by the matrix $$Y =\left(%
\begin{array}{cc}
  0 & -\iota \\
  \iota & 0 \\
\end{array}%
\right)$$ \subsection{The Z Gate} The Pauli Z gate changes the
relative phase factor by $-1$, effectively negating a qubit's sign
for the $|1\rangle$ component of the state. It performs the
following mapping on the logical states.
$$Z|0\rangle =  |0\rangle$$
$$Z|1\rangle = - |1\rangle$$
It is defined by the matrix $$Z =\left(%
\begin{array}{cc}
  1 & 0 \\
  0 & -1 \\
\end{array}%
\right)$$
\subsection{The T Gate} This is sometimes called the $\frac{\pi}{8}$
for the reason that up to a certain global phase, the $T$ gate
behaves exactly as another gate which has $e^{\iota\frac{\pi}{8}}$
appearing in its diagonals. The $T$ gate is defined by the matrix
$$T=\left(%
\begin{array}{cc}
  1 & 0 \\
  0 & e^{\iota\frac{\pi}{8}} \\
\end{array}%
\right)$$
\subsection{Hadamard Gate} The quantum Hadamard gate  acts
on a single qubit [35]. The purpose of the Hadamard gate is to
create superposition states. The application of the Hadamard gate
transforms a state $|0\rangle$ and $|1\rangle$ into halfways
between this state and its negation. Specifically, the Hadamard
gates action on the states $|0\rangle$ and $|1\rangle$ is given by
\begin{align}H|0\rangle = \frac{|0\rangle+|1\rangle}{\sqrt{2}}\end{align}
and
\begin{align}H|1\rangle = \frac{|0\rangle-|1\rangle}{\sqrt{2}}\end{align}
A two-qubit Hadamard gate is defined by
\begin{align}U_H=H^{\otimes 2}|00\rangle = \frac{|00\rangle+|01\rangle+|10\rangle+|11\rangle}{2}\end{align}
\begin{align}U_H=H^{\otimes 2}|01\rangle = \frac{|00\rangle-|01\rangle+|10\rangle-|11\rangle}{2}\end{align}
\begin{align}U_H=H^{\otimes 2}|10\rangle = \frac{|00\rangle+|01\rangle-|10\rangle-|11\rangle}{2}\end{align}
\begin{align}U_H=H^{\otimes 2}|11\rangle = \frac{|00\rangle-|01\rangle-|10\rangle+|11\rangle}{2}\end{align}
\subsection{Controlled Unitary Gate}
Controlled unitary gates act on two or more qubits where one or
more qubits act as a control for some operation. If the control
qubit is in the state $|0\rangle$ then the target qubit is left
unchange [36, 37]. The gate being implemented is the following
controlled unitary gate
\begin{align}|x_1x_2x_3\rangle\longrightarrow |x_1\rangle
U_1^{x_1}|x_2\rangle U_2^{x_1x_2}|x_3\rangle\end{align}
where $U_1 =\left(%
\begin{array}{cc}
  \alpha_1 & \beta_1 \\
  -\overline{\beta}_1  & \overline{\alpha} \\
\end{array}%
\right)$ and $U_2 =\left(%
\begin{array}{cc}
  \alpha_2 & \beta_2 \\
  -\overline{\beta}_2  & \overline{\alpha}_2 \\
\end{array}%
\right)$. In other words $U_1$ is applied to the second qubits iff
the first qubits is $1$ and $U_2$ is applied to the third qubits
iff both the first and second qubits are $1$. Another way to
express the gate action is via the following formulas (we choose
$x_3$ either $0$ or $1$) $$|00x_3\rangle \longrightarrow
|00x_3\rangle$$
$$|01x_3\rangle \longrightarrow
|01x_3\rangle$$
$$|10x_3\rangle \longrightarrow
|1\rangle U_1|0\rangle|x_3\rangle$$
$$|11x_3\rangle \longrightarrow
|1\rangle U_1|1\rangle U_2|x_3\rangle$$ A complete table of
three-qubits of controlled gate is given by
$$|000\rangle \longrightarrow |000\rangle$$
$$|001\rangle \longrightarrow |001\rangle$$
$$|010\rangle \longrightarrow |010\rangle$$
$$|011\rangle \longrightarrow |011\rangle$$
$$|100\rangle \longrightarrow \beta_1|110\rangle+\overline{\alpha}_1|100\rangle$$
$$|101\rangle \longrightarrow \beta_1|111\rangle+\overline{\alpha}_1|101\rangle$$
$$|110\rangle \longrightarrow \alpha_1\beta_2|111\rangle+\alpha_1\overline{\alpha}_2|110\rangle-
\overline{\beta}_1\beta_2|101\rangle-\overline{\beta}_1\overline{\alpha}_2|100\rangle$$
$$|111\rangle \longrightarrow \alpha_1\alpha_2|111\rangle-\alpha_1\overline{\beta}_2|110\rangle-
\overline{\beta}_1\alpha_2|101\rangle+\overline{\beta}_1\overline{\beta}_2|100\rangle$$
In matrix form the controlled gate $U_c$ is given by
\[
  U_c=
\left[{\begin{array}{cccccccc}
  1       & 0       & 0       & 0       & 0       & 0       & 0       & 0       \\
  0       & 1       & 0       & 0       & 0       & 0       & 0       & 0       \\
  0       & 0       & 1       & 0       & 0       & 0       & 0       & 0       \\
  0       & 0       & 0       & 1       & 0       & 0       & 0       & 0       \\
  0       & 0       & 0       & 0       & \overline{\alpha}_1   & 0   & -\overline{\beta}_1\overline{\alpha}_2  & \overline{\beta}_1\overline{\beta}_2   \\
  0       & 0       & 0       & 0       & 0  & \overline{\alpha}_1  & -\overline{\beta}_1\beta_2  & - \overline{\beta}_1\alpha_2   \\
  0       & 0       & 0       & 0       & \beta_1  & 0  & \alpha_1\overline{\alpha}_2  & -\alpha_1\overline{\beta}_2   \\
  0       & 0       & 0       & 0       & 0  & \beta_1  & \alpha_1\beta_2   & \alpha_1\alpha_2   \\
\end{array} }\right]
\]
We can built the quantum Fourier transform gate by using
controlled unitary gate and Hadamard gate [38, 39, 40, 41].
\section{Separable and Non-separable Gates}
This thesis in particular shows that by truncating an infinite
dimensional quantum system to finite dimensions, we can realize
commonly used quantum gates. After illustrating how an arbitrary
unitary gate can be realized approximately using a perturbed
Hamiltonian, we discuss qualitatively some issues regarding how
separable and non-separable unitary gates can be realized using
respectively independent Hamiltonians and independent Hamiltonians
with an interaction. Specifically, the theory developed in this
thesis shows that given a desired unitary gate which is a small
perturbation of a separable unitary gate, we can realize the
separable component using a direct sum of two independent
Hamiltonians and then add a small interaction component to this
direct sum in such a way as to cause the error between the desired
unitary gate and the realized gate to be as small as possible. In
other words, we justify that the time dependent perturbation
theory of independent quantum systems is a natural way to realize
non-separable unitary gates which are small perturbations of
separable gates. Examples of separable and non-separable unitary
gates taken from standard textbooks on quantum computation are
given using respectively tensor products of unitaries like the
Hadamard gate and controlled unitary gates. In each case we
qualitatively discuss the realization using independent
Hamiltonians and independent Hamiltonians with a small interaction
(acting on both components of tensor product space) using the
Dyson series [42, 43, 44].
\section{Measures of Performance of Designed Gates: The Frobenius Norm, The Spectral Norm}
Various kinds of norm on spaces of matrices exist to evaluate the
performance of gates. There are $L^p$ indices norm, $p\geq 1$ and
the Frobenius norm to cite just a few. The $L^p$ indices norm are
$$\|A\|_p= \sup_{\atop \|x\|\leq 1}\frac{\|Ax\|_p}{\|x\|_p}$$
where $\|x\|_p=(\sum_{i=1}^{\infty}|x_i|^p)^\frac{1}{p}$,
$x=(x_i)$ if the vector space is $L^p(Z_+)$ or if it is
$L^p(\mathbb{R}_+)$, then $$\|x\|_p=\bigg(\int_0^\infty
|x(t)|^pdt\bigg)^\frac{1}{p}$$ where $\|.\|_p$ satisfies, apart
from the triangle inequality the matrix norm property or
submultiplicativity:
$$\|AB\|_p\leq \|A\|_p\|B\|_p$$
when $p=2$, $\|.\|_p$ is called the spectral norm defined by
$\|.\|_s$. This has the obvious property
$$\|A\|_s^2=\frac{\langle x, A^*Ax\rangle}{\langle x, x\rangle}=\sigma_{\max}(A)$$
where $\sigma_{\max}(X)$ is the maximum singular value of matrix
$X$. The other norm used in this theory is the Frobenius norm
$\|.\|_F$. It is given by $$\|A\|_F^2=\Tr(A^*A)=\sum_{\alpha,
\beta}|\langle e_\alpha |A|e_\beta\rangle|^2$$ where
$\{|e_\alpha\rangle\}_{\alpha=1}^\infty$ is an ONB for the Hilbert
space. Equivalently $$\|A\|_{F}^2=\sum_{j=1}^\infty\sigma_j(A)^2$$
where $\sigma_j(A),\quad j\geq 1$ are all the singular values of
$A$. Note that the singular values $X$ are the eigenvalues of
$(X^*X)^\frac{1}{2}$ [45, 46, 47, 48]. The Frobenius norm is
useful in that it gives a physically meaningful interpretation of
SNR in quantum gate design while the spectral norm is useful in
obtaining upper bound on matrices satisfying
differential/intergral equation (e.g. Dyson series) [49, 50].
\section{Dissertation Organization}
Chapter 2, is the ``heart'' of the thesis and details the
implementation of commonly used quantum gate.The basic ideas
needed for understanding the problems solved in the following
chapters are discussed. In this chapter, we have taken a harmonic
oscillator and calculated its eigenstates and energy eigenvalues.
Therefore, the definition of the harmonic oscillator is crucial
and will be discussed in this chapter. Here, we apply a small time
dependent perturbation of $O(\epsilon)$ to it and express the
evolution operator of this perturbation system using truncation.
This chapter deals with the approximate design of quantum unitary
gates using perturbed harmonic oscillator dynamics. The harmonic
oscillator dynamics is perturbed by a small time varying electric
field which leads to time dependent Schr\"odinger equation. The
corresponding unitary evolution after time $T$ is obtained by
approximately solving the time dependent Schr\"odinger equation.
The aim of this chapter is to minimize the discrepancy between a
given unitary gate and the gate obtained by evolving the
oscillator in the weak electric field over $[0, T]$. The proposed
algorithm shows that the approximate design is able to realize the
Hadamard gate and controlled unitary gate on three-qubit arrays
with high accuracy.

Chapter 3, we present the design of a given quantum unitary gate
by perturbing a three-dimensional ($3$-D) quantum harmonic
oscillator with a time-varying but spatially constant
electromagnetic field. The idea is based on expressing the
radiation perturbed Hamiltonian as the sum of the unperturbed
Hamiltonian and $O(e)$ and $O(e^2)$ perturbations and then solving
the Schr\"odinger equation to obtain the evolution operator at
time $T$ upto $O(e^2)$ and this is a linear-quadratic function of
the perturbing electromagnetic field values over the time interval
$[0,T]$. Setting the variational derivative of the error energy
with respect to the electromagnetic field values with an average
electromagnetic field energy constraint leads to the optimal
electromagnetic field solution: a linear integral equation. The
reliability of such a gate design procedure in the presence of
heat bath coupling is analyzed and finally an example illustrating
how atoms and molecule can be approximated using oscillators is
presented.

Chapter 4 deals with the design of quantum unitary gate by
matching the Hermitian generators. The last contribution of this
thesis focuses on a special case of a quantum gate design, that
is, the realization of non-separable systems, that is, controlled
unitary gates based on matching Hermitian generators. A given
complicated quantum controlled gate is approximated by perturbing
a simple quantum system with a small time varying potential. The
basic idea is to evaluate the generator $H_\varphi$ of the
perturbed system approximately using first order perturbation
theory in the interaction picture. $H_\varphi$ depends on a
modulating signal $\varphi(t):\quad 0\leq t\leq T$ which modulates
a known potential $V$. The generator $H_\varphi$ of the given gate
$U_g$ is evaluated using $H_g=\iota\log U_g$. The optimal
modulating signal $\varphi(t)$ is chosen so that $\|H_g -
H_\varphi\|$ is minimum. The simple quantum system chosen for our
simulation is a harmonic oscillator with charge perturbed by an
electric field that is constant in space but time varying and is
controlled externally. This is used to approximate the controlled
unitary gate obtained by perturbing the oscillator with an
anharmonic term proportional to $q^3$. Simulations results show
significantly small noise to signal ratio (NSR). Finally, we
discuss in this chapter, how the proposed method is particularly
suitable for designing some commonly used unitary gates. Another
example chosen to illustrate this method of gate design is the
ion-trap model.

In Chapter 5 prospect for the future work and the summary of our
achievement along with conclusions of this thesis are given.

\chapter{REALIZATION OF COMMONLY USE QUANTUM GATES USING PERTURBED HARMONIC OSCILLATOR}

\section{Introduction}
\label{intro}Quantum mechanics is a mathematical framework for the
accurate construction of physical theories. The physical theories
culminate in what is known as quantum electrodynamics which
describes with fantastic accuracy the interaction of atoms and
light [14, 15, 16, 17]. For years, researchers have been
interested in developing quantum computers, the theoretical next
generation of technology that will outperform conventional
computers. Instead of holding data in bits, the digital units used
by computers today, quantum computers store information in units
called `qubits'. One approach for computing with `qubits' relies
on the creation of two single photons that interfere with one
another in a device called a waveguide [18, 19].\par Since the
eighties, much effort has been put into the study of quantum
computers, and various proposals for the physical realization of
various gates. Various logic gates are proposed to synthesize the
multi-level quantum logic circuits. An important group of these
gates are controlled gates. This concept is vital to quantum
computing because all quantum transformations are unitary, and
therefore reversible. Thus, all quantum gates themselves must be
reversible. This further complicates the design of quantum
algorithms, since users only familiar with classical programming
encounter a steep learning curve when they must design algorithms
that work exclusively with reversible computations. The use of the
term gates when describing quantum gates should be taken
conceptually. As we will see, transformations on qubits are not
necessarily applied with gates in the conventional sense. Because
of the superposition phenomenon, qubit states are expressed not as
bits but as matrices of bits. Therefore, quantum gates actually
perform transformations on matrices. The simplest non-trivial
quantum logic gate is a controlled-NOT gate [22, 23]. The quantum
cNOT gate can be used as a basis to create more general quantum
gates. Quantum logic gates can be used to apply unitary
transformations to the state of qubits (their probability
amplitude vectors) without causing them to decohere, and even to
entangle and disentangle qubits.
\subsection{Time Dependent Perturbation Theory and Dyson Series}
An example of a quantum gate design using an unperturbed
Hamiltonian is to take the unperturbed Hamiltonian as
$$H_0=\sum_{k=1}^n\frac{\sigma_{z_k}eB_0}{2m}$$
where $\sigma_{z_k}$ is the $z$-component of the Pauli spin matrix
acting on the $k^{th}$ copy of $\mathcal{C}^2$ [25]. Thus, $H_0$
acts in $(\mathcal{C}^2)^{\otimes n}\cong \mathcal{C}^{2n}$.
Then $$e^{-\iota t H_0}=(e^{\frac{-\iota e B_0 t}{2m}\sigma_z})^{\otimes n}=\left[%
\begin{array}{cc}
 e^{-\iota \theta}  & 0 \\
  0 & e^{-\iota \theta} \\
\end{array}%
\right]^{\otimes n} $$ where $\theta = \frac{e B_0 t}{2m}$ This is
a separable gate which acts on the $k^{th}$ copy of
$\mathcal{C}^2$ by changing the phase of $|0\rangle _k$ by
$-\theta$ and $|1\rangle _k$ by $+\theta$. This Hamiltonian
corresponds to the energy of interaction of $n$ independent spin
$\frac{1}{2}$ particles with a constant magnetic field. This gate
takes $|x_1, x_2, \cdots x_n\rangle$ to a multiple of itself for
all $x_1, x_2, \cdots x_n \in \{0, 1\}$. Thus this gate cannot
generate mixtures of base states. In order to do so, we must
perturb it by something like $\sum_{\iota =1}^n
\frac{\sigma_{x_i}e B_x}{2m}$. A single unperturbed Hamiltonian
(constant in time) can generate only diagonal unitary gates, that
is, gates which all commute with each other. For generating
non-commuting non-diagonal gates, we must perturb it with time
dependent Hamiltonian. $H_0$ is the unperturbed Hamiltonian of a
quantum system. It can be a bounded or unbounded Hamiltonian
operator on a Hilbert space $\mathcal{H}$. If dimension
$\mathcal{H} < \infty$ then $H_0$ is always bounded. Then class of
gates realized using $H_0$ is the one parameter unitary family
$U_0(t)=e^{-\iota t H_0},\quad t \epsilon \mathbb{R}$. In
$|n\rangle,\quad n=0, 1, 2,\cdots$ are the eigenstates of $H_0$,
if $H_0|n\rangle=E_0|n\rangle,\quad n=0, 1, 2,\cdots$ then
$U_0(t)|n\rangle=e^{-\iota t E_n}|n\rangle$, so the matrix of
$U_0(t)$ in the truncated basis $|n\rangle,\quad n=0, 1, 2,\cdots,
N-1$ is $\emph{diag}[e^{-\iota t E_n}: \quad 0\leq n\leq N-1]$,
which is a unitary $N\times N$ diagonal matrix. If
$|e_\alpha\rangle,\quad \alpha=1, 2,\cdots$ is any other
orthonormal basis for $\mathcal{H}$, then the truncated matrix
$$\left(\bigg(%
\begin{array}{c}
  \langle e_\alpha|U_0(t)|e_\beta\rangle \\
\end{array}%
\right)\bigg)_{1\leq \alpha, \beta\leq N}$$ is not unitary in
general. So using the polar decomposition, we may extract the
unitary component of thus truncated matrix and treat this as the
designed gate. The group $\{U_0(t)|t\in \mathbb{R}\}$ is a one
dimensional Lie group. If however we consider the the unitary
evolution $U(t),\quad t\geq 0$ generated by a time dependent
Hamiltonian $H(t)=H_0+f(t)V,\quad t\geq 0$, that is,
$$\frac{dU(t)}{dt}=-\iota H(t)U(t),\quad t\geq 0$$ then if $[H_0, V]\neq
0$, the group of unitary operators generated by $\{U(t)\}_{t\geq
0}$ is infinite dimensional in general. Its Lie algebra contains
elements like $$\cdots (ad H_0)^n(ad V)^m(ad H_0)^p(V)\cdots (ad
H_0)^n(ad H_0)^m(H)$$ etc. Which may contain an infinite linearly
dependent set. This means that a much larger class of unitary
gates can generated using using perturbed Hamiltonians and hence
we look for time dependent perturbation for generating gates. By
writing $U(t)=U_0(t)W(t)$ we get
$$W'(t)=-\iota f(t)\widetilde{V}(t)W(t)$$
$$W(0)=I$$
where $\widetilde{V}(t)=e^{\iota t H_0}Ve^{-\iota t H_0}=e^{\iota
t adH_0}(V)$. The solution is given by the Dyson series
$$W(t)=I+\sum_{n=1}^\infty(-\iota)^n\int_{0<t_n<\cdots<t_1<t}f(t_1)\cdots f(t_n)\widetilde{V}
(t_1)\cdots\widetilde{V}(t_n)\cdots dt_1\cdots dt_n$$ and hence
\begin{align}\nonumber U(t)&=U_0(t)+\sum_{n=1}^\infty(-\iota)^n\int_{0<t_n<\cdots<t_1<t}f(t_1)\cdots
f(t_n)U_0(t-t_1)VU_0(t_1-t_2)\\&\nonumber \cdots
U_0(t_{n-1}-t_n)VU_0(t_n)\cdots dt_1\cdots dt_n\end{align} If
$\|.\|_s$ denotes the spectral norm of an operator then we get
absolute convergence of the above Dyson series.
\par The first problem studied in this thesis is important because
it can be applied to design various kinds of gates like the
quantum Fourier transform, phase gate, controlled unitary gates
etc. The quantum Fourier gate performs the DFT using $O(N)$
operations in contrast to the classical FFT algorithm which
requires $O(N\ln N)$ operation. Controlled unitary gates are used
in problems like quantum teleportation involving transmission of
quantum states using only classical bits based on entanglement
sharing. Such communication is faster than the speed of light and
is allowed quantum mechanics communication faster than time speed
of light is not possible in classical theories. Further, the
quantum Fourier transform can be used in phase estimation and
order finding which are important in signal processing. Quantum
gates have also been used in search algorithms (like Grovers
search algorithm). In short, quantum gates have found use in a
variety of signal processing and communication problem by
performing superiorly to classical algorithm and this thesis on
gate design using physical systems can find use in such problems
[19, 20, 27].
\subsection{Harmonic Oscillator Perturbed by Electric Field}
Quantum gates can be realized using various physical process like
the ion-trap scheme and our method is a general scheme based on
perturbing Hamiltonian by a time dependent potential which
includes the chosen specialized scheme such as the ion trap
method. Any physical process used to simulate a quantum gate (even
if it be the spin of a spin j-particle interacting with a magnetic
field) can be analyzed by the Hamiltonian plus perturbation method
discussed in our thesis. Our thesis in particular, is a step
forward in the practical design of quantum gates which can be used
immediately in the above applications. The disadvantage of spin
system is that the gates designed are of lower dimension. Harmonic
oscillator based gate can be of very large dimensions. The
harmonic oscillator Hamiltonian $H_0=\frac{p^2+q^2}{2}$ can be
used in the design of only a one parameter group of unitary gates
$e^{-\iota t\frac{p^2+q^2}{2}}\quad t\epsilon\Re$. The Lie algebra
of this group is in other words just one dimensional. However,
when we perturb $H_0$ by a time varying potential of the form
$f(t)V$, then we can realize a much larger class of unitary gates
with Lie algebra generated by $H_0$ and $V$, that is, gates of the
form $e^{-\iota t X}$, where $X$ is a linear combination of $(ad
H_0)^n(ad V)^m(H_0)$ and $(ad V)^n(ad H_0)^m(V)$. This is a
consequence of the  Baker-Campbell-Hausdorff formula used in Lie
group theory [16, 46]. The time dependence of $f(t)V$ makes this
possible, this is because the family of operators $H_0+f(t)V,\quad
t\geq 0$ need not commute if $[H_0, V]\neq 0$. This is the
advantage of using a time dependent perturbation term. The
dimensionality of the Lie group of generated gates gets greatly
increased, thereby facilitating the design of a larger class of
gates used in other applications.
\subsection{Coherent State}
In quantum mechanics, a coherent state is the specific quantum
state of the quantum harmonic oscillator which was first used by
Roy Glauber in the field of quantum optics [19]. This change of
state may include change in the shape of the wave function.
Coherent states are the eigenstates of the annihilation operator.
Using the eigenstates of a harmonic oscillator as a substratum for
realizing complex gates is natural since these sequences of
eigenstates can be generated by successively applying a creation
operator to the preceding eigenstates. Further by forming a linear
combination of these eigenstates defined by
\begin{align}|\alpha\rangle =
e^{\frac{-|\alpha|^2}{2}}\sum_{n=0}^\infty\frac{\alpha^n|n\rangle}{\sqrt{n!}}\end{align}
we can generate a large class of useful states called coherent
states, which are eigenfunctions of the annihilation operator with
arbitrary complex eigenvalue $\alpha$. The coherent state
wavefunction looks exactly like ground state, but shifted in
momentum and position. It then moves almost as a classical
particle, while keeping its shape fixed. So the advantage of
perturbation theory is mainly to increase the dimensionality of
the unitary group of gates realizable by a quantum physical system
from $1$ to $N$ where $N$ can even be infinity. Perturbation
theory is one of the most important methods for obtaining
approximate solution to the Schr\"odinger equation . Prior to
studying harmonic oscillators perturbed by an electric field, we
look at the general problem of computing the evolution of a
quantum system having a Hamiltonian operator of the form $H_0 +
\epsilon V(t)$ where $H_0$ is a known Hermitian operator and is
the Hamiltonian of a quantum system in the Hilbert space $
\mathcal {H} $ (which is finite dimensional), $\epsilon V(t)$; $ 0
\leq t\leq T$ is the perturbing potential where $\epsilon$ is a
small parameter. The energy dissipated in applying $V(t)$ over the
duration $[0, T]$ is the constraint to be fixed. The perturbing
time dependent potential operator $V(t)$ is chosen so that the
unitary evolution operator $U(T)$ at time $T$ upto $O(\epsilon^2)$
is as close as possible to a desired gate $U_d$ on the same
Hilbert space. This optimization is carried out without putting
any restriction on the operators $V(t) ;$ $ 0 \leq t\leq T$ except
that they may be Hermitian and satisfy energy constraints of the
form $ E=\int_0^T \mathrm{Tr}[A(t)V^2(t)]dt $, where $A(t)$ is a
known Hermitian operator valued function of time (note that
$V^\ast(t)=V(t)$) and $\mathrm{Tr}$ is trace of an operator. Most
of the Quantum gates like the Hadamard gate and controlled unitary
gates are usually designed using finite state Schr\"odinger
evolution equation. The novelty of our method is that we use an
infinite dimensional system like the quantum harmonic oscillator
to design finite dimensional gates by truncation. Finite state
systems can in practice be realized using the spin states of
elementary particles. To realize infinite dimensional gates, we
need to use observables like position $x$ and momentum $p$ that
act in $L^2(\mathbb{R})$.\par The significant contribution of the
second problem is to show how by using a real physical system such
as an atom or a molecule (modelled as a quantum harmonic
oscillator for small displacements of the electron from its
equilibrium position) we can, by applying an external field,
create unitary gates used in quantum computation with a high
degree of accuracy. This problem in particular shows that by
truncating an infinite dimensional quantum system to finite
dimensions, we can realize commonly used quantum gates. After
illustrating how an arbitrary unitary gate can be realized
approximately using a perturbed Hamiltonian, we discuss
qualitatively some issues regarding how separable and
non-separable unitary gates can be realized using respectively
independent Hamiltonians and independent Hamiltonians with an
interaction. Specifically, the theory developed in our thesis
shows that given a desired unitary gate which is a small
perturbation of a separable unitary gate, we can realize the
separable component using a direct sum of two independent
Hamiltonians and then add a small interaction component to this
direct sum in such a way as to cause the error between the desired
unitary gate and the realized gate to be as small as possible. In
other words, we justify that the time dependent perturbation
theory of independent quantum systems is a natural way to realize
non-separable unitary gates which are small perturbations of
separable gates. Examples of separable and non-separable unitary
gates taken from standard textbooks on quantum computation are
given using respectively tensor products of unitaries like the
Hadamard gate and controlled unitary gates. In each case we
qualitatively discuss the realization using independent
Hamiltonians and independent Hamiltonians with a small interaction
of order $\epsilon$ with the evolution operator computed upto
$O(\epsilon^2)$ using the Dyson series [35, 36, 41, 47, 48].
\section{Mathematical Studies of Unitary Gate Design Error Energy}
\label{sec:1} Let $U(t)$ be the unitary evolution corresponding to
the Hamiltonian $ H_0 + \epsilon V(t)$. We wish to determine
$V(t)$; $ 0 \leq t\leq T$ upto $O(\epsilon^2)$ such that $
\|U_d-U(T)\|^2$ is a minimum, where $U_d$ is a given unitary gate
(operator for finite dimension and infinite dimension quantum
system). Define
 \begin{align} \|X\|^2= \mathrm{Tr}[X^* X] = \sum_{\alpha,\beta} |X_{\alpha,\beta}|^2\end{align} which is the Frobenius
 norm, $ X_{\alpha,\beta}  = \langle{e_\alpha}|{e_\beta}\rangle $ for any
orthogonal basis ${e_\alpha}$ for $ \mathcal {H} $. If the
perturbing potential is produced by an electric field
$\overrightarrow{E}(t)$ acting on a charged quantum particle of
charge $q$, then
\begin{align} V(t)=-q\overrightarrow{E}(t)\cdot\overrightarrow{r}\end{align} where
$\overrightarrow{r}$ is the position vector operator. If the
electric field is generated by a circuit involving a resistance
$\text{R}$ then the power dissipated through $\text{R}$ is of the
form
$$\frac{d^2}{\text{R}} \int_0^T | \overrightarrow{E}(t)|^2dt$$ where $d$
is the distance between the conducting plates producing the field
and thus this dissipated energy can be expressed as $\int_0^T
\mathrm{Tr}\left[\text{A}(t)V^2(t)\right]dt$ for an appropriate
positive definite operator $\text{A}$. Using perturbation theory,
it is very difficult to find exact solutions to the Schr\"odinger
equation for Hamiltonian of even moderate complexity [25]. We
consider the time dependent Schr\"odinger equation for the
perturbed oscillator which is given by
\begin{align} i\frac{dU(t)}{dt} =  (H_0 + \epsilon V(t))U(t)
\end{align} Setting $ U(t) = e^{-iH_0t}W(t)$ gives
\begin{align}i\frac{dW(t)}{dt} = \epsilon \widetilde{V}(t)W(t)\end{align}
where $\widetilde{V}(t) = e^{iH_0t}V(t)e^{-iH_0 t}$. Since $H_0$
is known, determining the optimal $V(t)$ is equivalent to
determining the optimal potential ${\widetilde{V}(t)}$. By
expanding using Dyson series in eq. (2.5) upto second order, we
get
\begin{align}
W(t) = I-i\epsilon
\int_{0<t_1<T}\widetilde{V}(t_1)dt_1-\epsilon^2\int_{0<t_2<t_1<T}\widetilde{V}(t_1)\widetilde{V}(t_2)dt_2dt_1
+O(\epsilon^3)\end{align} The gate error energy is given by
\begin{align}\nonumber
 \mathbb{E}= \|U_d-U(T)\|^2 = \|U_d-e^{-iH_0 T}W(T)\|^2= \|\widetilde{U_d}-W(T)\|^2 \end{align}
\begin{align}\nonumber \mathbb{E}= \bigg\|\widetilde{U_d}+ i\epsilon
\int_{0<t_1<T}\widetilde{V}(t_1)dt_1
+\epsilon^2\int_{0<t_2<t_1<T}\widetilde{V}(t_1)\widetilde{V}(t_2)dt_2dt_1
+O(\epsilon^3)\bigg\|^2 \end{align} where,
$$\widetilde{U_d} = e^{iH_0t} U_d-I$$
Expanding this Frobenius norm, we get
\begin{align}\nonumber
\|U_d-U(T)\|^2 &= \|\widetilde{U_d}\|^2 \\&+ \nonumber\epsilon^2
\int_{{0<t_1},{t_2<T}}\mathrm{Tr}\left[\widetilde{V}(t_1)\widetilde{V}(t_2)\right]dt_1dt_2\\&
+\nonumber i\epsilon
\mathrm{Tr}\left[\widetilde{U_d^*}\int_{0<t_1<T}\widetilde{V}(t_1)dt_1\right]\\&-\nonumber
i\epsilon
\mathrm{Tr}\left[\widetilde{U_d}\int_{0<t_1<T}\widetilde{V}(t_1)dt_1\right]\\&+\nonumber\epsilon^2\mathrm{Tr}\left[\widetilde{U_d^*}
\int_{0<t_2<t_1<T}\widetilde{V}(t_1)\widetilde{V}(t_2)dt_2dt_1\right]\\&+\nonumber\epsilon^2\mathrm{Tr}\left[\widetilde{U_d}
\int_{0<t_2<t_1<T}\widetilde{V}(t_2)\widetilde{V}(t_1)dt_2dt_1\right]
\\&+ O(\epsilon^3)\end{align}
Note
that\begin{align}\widetilde{V}^*(t)=\widetilde{V}(t)\end{align}
\begin{align}\left(\widetilde{V}(t_1)\widetilde{V}(t_2)\right)^*=\widetilde{V}(t_2)\widetilde{V}(t_1)\end{align}
We calculate the variational derivative with respect to
$\widetilde{V}(t)$ of the last function taking into account energy
constraint using Lagrange's multiplier. The energy constraint
$$E = \int_0^T \mathrm{Tr}\left[\text{A}V^2(t)\right]dt$$
must be expressed in terms of $\widetilde{V}(t)$. Using
$$\widetilde{V}(t) = e^{iH_0t} V(t)e^{-iH_0t}$$ this constraint
becomes $$E = \int_0^T
\mathrm{Tr}\left[\text{A}(t)\widetilde{V}^2(t)\right]dt$$ where
$A(t)= e^{iH_0t} A e^{-iH_0t}$. The quantity to be minimized is
\begin{align}\|U_d-U(T)\|^2-\lambda\int_0^T \mathrm{Tr}\left[\text{A}(t)\widetilde{V}^2(t)\right]dt \end{align} where $\lambda$ is the Lagrange
multiplier. We set the variational derivative of the above
equation with respect to $\widetilde{V}(t)$ to zero. The
coefficient of $\delta\widetilde{V}(t_2)$ is
\begin{align}\nonumber2\epsilon^2\int_0^T {\widetilde{V}(t_1)}dt_1+i\epsilon U_d^*-i\epsilon U_d
+\epsilon^2 \widetilde{U}_d^*\int_{t_2}^T
{\widetilde{V}(t_1)}dt_1\\+\nonumber\epsilon^2\left(\int_0^{t_2}
{\widetilde{V}(t_1)}dt_1\right)\widetilde{U}_d^*+\epsilon^2\left(\int_{t_2}^T
{\widetilde{V}(t_1)}dt_1\right)U_d+\\\nonumber\epsilon^2\left(\int_0^{t_2}
{\widetilde{V}(t_1)}dt_1\right)U_d-\lambda\left(\text{A}(t_2)\widetilde{V}(t_2)+\widetilde{V}(t_2)\text{A}(t_2)\right)
= 0\end{align} Differentiate with respect to $t_2$ and replace it
by $t$,
\begin{align}\nonumber -\epsilon^2 \widetilde{U}_d^* \widetilde{V}(t)+\epsilon^2\widetilde{V}(t)U_d^* -\epsilon^2 \widetilde{V}(t)U_d+\epsilon^2
U_d\widetilde{V}(t)- \lambda
\left(\text{A}\widetilde{V}'(t)+\widetilde{V}'(t)\text{A}\right)=0\end{align}
We have assumed that $A(t)$ is constant operator in order to
simplify the calculations and replacing $\epsilon$ by 1,
$\widetilde{V}$ by $V$ and $\widetilde{U}_d$ by $U_d$.
\begin{align}\lambda A V'+\lambda  V' A = (U_d-U_d^*)V+V(U_d^*-U_d)\end{align}
\begin{align}\nonumber
 \mathbb{E}\nonumber = \|U_d-U(T)\|^2 = \|U_d-e^{-iH_0 T}W(T)\|^2 = \|\widetilde{U_d}-W(T)\|^2 \end{align}
 where,
$$\widetilde{U_d} = e^{iH_0t} U_d-I$$
Let $\lambda$ be the Lagrange multiplier and $A(t)=e^{iH_0t} A
e^{-iH_0t}$. Using Lagrange multiplier approach, the minimization
of the quantity
\begin{align}\|U_d-U(T)\|^2-\lambda\int_0^T \mathrm{Tr}\left[\text{A}(t)\widetilde{V}^2(t)\right]dt \end{align}
leads to
\begin{align}\lambda A V'+\lambda  V' A = (U_d-U_d^*)V+V(U_d^*-U_d)\end{align}
In the following section we are applying it to the quantum
harmonic oscillator.
\section{Design of Gates Using Time Dependent Perturbation Theory with Application to Harmonic Oscillator}
The harmonic oscillator is an extremely important and useful
concept in the quantum description of the physical word, and a
good way to begin to understand its properties is to determine the
energy eigenstates of its Hamiltonian [13, 14, 15, 16]. The
underlying Hilbert space is
$$ \mathcal{H} = L^2(R)$$ The dynamics of a single, one
dimensional harmonic oscillator is governed by the Hamiltonian:
$$ H_0 = -\frac{1}{2}\frac{d^2}{dx^2}+\frac{1}{2}x^2 = \frac{p^2+x^2}{2}$$
where $x$ and $p$ are respectively the position and momentum
operators and commutation of both the operator is $[x, p] =
i\hbar$. We can take $x$ as multiplication by $x$ and $ p =
-i\hbar\frac{d}{dx}$. Let $$a = \frac{x+ip}{\sqrt{2}} ;$$
$${a^\dagger} = \frac{x-ip}{\sqrt{2}}$$
We note that $x$ and $p$ are self adjoint operator. Where $a$ is
called annihilation operator and $a^\dagger$, its adjoint is the
creation operator. So $$aa^\dagger = H_0+\frac{1}{2}$$
$$a^\dagger a= H_0 - \frac{1}{2}$$
for any energy eigenvalue $\text{E}$. Thus the zero-point energy
$\frac{\hbar w}{2}$ is the lowest possible eigenvalue of $H_0$ and
is attainted for the eigenstates $|0\rangle$, where $$a |0\rangle
= 0$$
$$\left(x + \frac{d}{dx}\right)|0\rangle = 0$$ Thus $$|0\rangle = \pi ^{-\frac{1}{4}}e^{-\frac{x^2}{2}}$$
is the ground state wave function in the position representation
and has energy $\frac{\hbar w}{2}$. Stationary states
$$u_n(x) = H_n(x)e^{-\frac{x^2}{2}}; n = 0, 1, 2, ...$$
where $H_n(x)$ is a Hermite polynomial
$$\langle u_n, u_m\rangle= \int_{-\infty}^{\infty} u_n(x)u_m(x)dx = \delta_{nm}$$
$$ H_0 u_n = \left(n+\frac{1}{2}\right)u_n ;  n = 0, 1, 2, .....$$
Let $E_n = \left(n+\frac{1}{2}\right)$ $$\epsilon V(t) = -\epsilon
q E(t)x$$ This is the perturbing potential, where $E(t)$ is
electric field and ${q}$ is the charge on the harmonic oscillator.
The optimization is simpler; it involves determining given the
scalar electric field, ${E(t) ; 0\leq t\leq T}.$ The time
dependent Schr\"odinger equation for the perturbed oscillator is
given by
$$i\frac{dU(t)}{dt} =  \left(H_0 +
\epsilon V(t)\right)U(t)$$ or equivalently
\begin{align} i\frac{d}{dt}\langle m |U(t)| n\rangle = \langle m | H_0 U(t)| n\rangle + \epsilon \langle m |V(t)U(t)|
n\rangle\end{align} This time dependent Schr\"odinger equation
(2.14) leads to
\begin{align}\nonumber i\frac{dU_{mn}
(t)}{dt} = E_m U_{mn} (t) - \epsilon q E(t)\langle m |xU(t)|
n\rangle
\end{align} where, $\langle m |x U(t)| n\rangle = \sum_{r=0}^{N-1} x_{mr} U_{rn}
(t)$
\begin{align}\nonumber i\frac{dU_{mn} (t)}{dt} =
E_m U_{mn} (t) - \epsilon q E(t)\sum_{r=0}^{N-1} x_{mr} U_{r n}
(t)
\end{align}
$$x_{mn} = \langle m |x| n\rangle = \int_{-\infty}^\infty {xu_m(x)u_n(x)dx}$$
where \begin{align}\nonumber x &= \frac{a+a^\dag}{\sqrt{2}}
\\\nonumber x_{mn} &=\frac{1}{\sqrt{2}}\langle m
|(a+a^\dag)|n\rangle\\\nonumber x_{mn} &=
\frac{1}{\sqrt{2}}\left({\sqrt{n} \delta _{m,n-1}+ \sqrt{m} \delta
{n,m-1}}\right)\end{align} Let $U_{mn} (t) = e^{-iE_mt}W_{mn}(t).$
We get from the above,
$$W_{mn}^{'}(t) = i\epsilon q E(t)\sum_{r=0}^{N-1} x_{mn}e^{-iE_rt}W_{rn}(t)$$
So \begin{align} W_{mn}(T) = \delta_{mn}+i\epsilon q
\sum_{r=0}^{N-1} \int_o^T E(t_1)x_{mr}
e^{-iE_rt}W_{rn}(t_1)dt_1\end{align} Iterating eq. (2.15) twice,
we obtain
\begin{align}\nonumber W_{mn}(T) &= \delta_{mn}+ i\epsilon q \int_0^T x_{mn} E(t_1)
e^{-iE_nt_1}dt_1 \\&\nonumber-\epsilon^2 q^2 \int_{0<t_2<t_1<T}
E(t_1)x_{mr} e^{-iE_rt_1}\nonumber E(t_2)x_{rn}
e^{-iE_nt_2}dt_2dt_1 +O(\epsilon^3)\end{align}
\begin{align}\nonumber W_{mn}(T) &= \delta_{mn}+ i\epsilon q x_{mn}\int_0^T E(t_1)
e^{-iE_nt_1}dt_1 - \\&\nonumber\epsilon^2 q^2\sum_{r=0}^{N-1}
x_{mr}x_{rn}\int_{0<t_2<t_1<T} E(t_1)E(t_2) e^{-i(E_rt_1 +
E_nt_2)}dt_2dt_1 +O(\epsilon^3)\end{align} The gate error energy
is given by
\begin{align}\nonumber\mathbb{E} &= {\|U(T)- U_d\|}^2\\\nonumber
&= \sum_{m,n=0}^{N-1}{\|U_{m,n}(T) - U_d(m,n)\|}^2\\\nonumber
&=\sum_{m,n=0}^{N-1}{\|W_{m,n}(T) -
e^{iE_mT}U_d(m,n)\|}^2\\\nonumber &=
\sum_{m,n=0}^{N-1}{\|\widetilde{W}_{m,n}(T) -
\widetilde{U}_d(m,n)\|}^2\end{align} where \begin{align}\nonumber
\widetilde{W}_{mn}(T) &= i\epsilon q x_{mn}\int_0^T E(t_1)
e^{-iE_nt_1}dt_1 \\&\nonumber- \epsilon^2 q^2\sum_{r=0}^{N-1}
x_{mr}x_{rn}\int_{0<t_2<t_1<T} E(t_1)E(t_2)\nonumber e^{-i(E_rt_1
+ E_nt_2)}dt_2dt_1 \end{align} and
$$\widetilde{U}_d(m,n)= e^{-iE_mT}U_d(m,n)- \delta_{m,n}$$
Note that ${E(t)}$ is a real function. Expanding the gate error
energy upto $O(\epsilon^2)$, we get
\begin{align}\nonumber \mathbb{E} &= \sum_{m,n=0}^{N-1}{\left|\widetilde{U}_d(m,n)\right|}^2 +
\epsilon^2 q^2\sum_{m,n=0}^{N-1}x_{mn}^2{\left|\int_0^T
E(t_1)e^{-iE_nt_1}dt_1\right|}^2\\\nonumber& + 2\epsilon^2 q^2
\Re\bigg\{\sum_{m,r,n=0}^{N-1}\overline{\widetilde{U_d}(m,n)}x_{mr}x_{rn}\times\int_{0<t_2<t_1<T}
E(t_1)E(t_2)e^{-i(E_rt_1 +
E_nt_2)}dt_2dt_1\bigg\}\\&-2\epsilon\int\Re\left\{\overline{\widetilde{U_d}(m,n)}iqx_{mn}e^{-iE_nt_1}\right\}E(t_1)dt_1+O(\epsilon^3)
\end{align} where $\Re\{z\}$ denotes real part of the complex number
$z$. Define \begin{align}\nonumber
k_2(t_1,t_2)=q^2\bigg[x_{mn}^2e^{-iE_n(t_1-t_2)}\bigg]+
q^2\Re\left\{\sum_{m,r,n=0}^{N-1}\overline{\widetilde{U_d}(m,n)}x_{mr}x_{rn}e^{-i(E_rt_1
+E_nt_2)}\right\}\end{align}
\begin{align}\nonumber k_1(t_1) = -2q\Re\left\{\sum_{m,n=0}^{N-1}i\overline{\widetilde{U_d}(m,n)}x_{mn}e^{-iE_nt_1}\right\}\end{align}
Then \begin{align} \nonumber \mathbb{E} = {\|U_d\|}^2 +
\epsilon\int_0^T k_1(t_1)
E(t_1)dt_1+\epsilon^2\int_{0<t_2<t_1<T}k_2(t_1,t_2)
E(t_1)E(t_2)dt_2dt_1+O(\epsilon^3)
\end{align}Energy dissipation in resistor is given by
\begin{align} E_{diss} = \alpha\int_0^T E^2(t)dt\end{align}
$\mathbb{E}-\lambda E_{diss}$ is to be minimized with respect to
$E(t)$. The optimal solution is given by
\begin{align} \frac{\delta{(\mathbb{E}-\lambda E_{diss})}}{\delta E(t)}= 0
\end{align}Solving above equation with $\epsilon = 1$, we get
\begin{align}k_1(t)+\int_0^T k_2(t,\tau)E(\tau)d\tau - 2\lambda \alpha E(t) = 0 \end{align}for
$0<t<T$. We have obtained $E(t)$ by discretization. The Lagrange
multiplier $\lambda$ is determined from the dissipation constraint
$ \alpha\int_0^T E^2(t)dt = E_{diss}$.
\section{Simulation Result Showing Gate Designed Error Energy}
In this section we realize the Hadamard gate followed by a
controlled unitary gate. In both cases, we calculate via MATLAB
simulation the minimum gate error energy between the desired and
designed gate.
\subsection{Hadamard Gate}
For the simulation of the quantum Hadamard gate, we have chosen
$H_0 = \frac{p^2+x^2}{2}$, where $\epsilon = 0.1$, $x$ is a
multiplication operator and the time duration $\text{T}$ over
which the simulation has been carried out is [25, 35]. The
truncation level is $\{n = 0, 1, 3, . . . , 7\}$ where $|n\rangle$
denoted the $n^{th}$ base state of the unperturbed oscillator.
\begin{align}H_0|n\rangle = \bigg(n+\frac{1}{2}\bigg)|n\rangle, 0\leq n \leq 7\end{align}
The purpose of the Hadamard gate is to create superposition
states. The application of the Hadamard gate transforms a state
$|0\rangle$ and $|1\rangle$ into halfways between this state and
its negation. Specifically, the Hadamard gates action on the
states $|0\rangle$ and $|1\rangle$ is given by
\begin{align}H|0\rangle = \frac{|0\rangle+|1\rangle}{\sqrt{2}}\end{align}
and
\begin{align}H|1\rangle = \frac{|0\rangle-|1\rangle}{\sqrt{2}}\end{align}
A two-qubit Hadamard gate is defined by
\begin{align}U_H=H^{\otimes 2}|00\rangle = \frac{|00\rangle+|01\rangle+|10\rangle+|11\rangle}{2}\end{align}
\begin{align}U_H=H^{\otimes 2}|01\rangle = \frac{|00\rangle-|01\rangle+|10\rangle-|11\rangle}{2}\end{align}
\begin{align}U_H=H^{\otimes 2}|10\rangle = \frac{|00\rangle+|01\rangle-|10\rangle-|11\rangle}{2}\end{align}
\begin{align}U_H=H^{\otimes 2}|11\rangle = \frac{|00\rangle-|01\rangle-|10\rangle+|11\rangle}{2}\end{align}
A three-qubit state Hadamard is defined by its action on the base
states $|x_1x_2x_3\rangle$, $x_k=0,1$, where $k=1,2,3$. For
example,
\begin{align}U_H=H^{\otimes3}|000\rangle =
\frac{|000\rangle+|001\rangle+|010\rangle+|011\rangle+|100\rangle+|101\rangle+|110\rangle+|111\rangle}{2\sqrt{2}}\end{align}
In general
\begin{align}U_H|x_1x_2x_3\rangle=H|x_1\rangle\otimes H|x_2\rangle\otimes H|x_3\rangle\end{align} For
separable (independent) systems (for example $H^{\otimes 3}$) the
quantum gate has the form $U_1\otimes U_2$ where $U_1$ acts on the
first Hilbert space $\mathcal{H}_1$ and $U_2$ acts on the second
Hilbert space $\mathcal{H}_2$. Such a system can be realized using
a Hamiltonian of the form $H=H_1\otimes I_2+I_1\otimes H_2\equiv
H_1\oplus H_2$. In other words
\begin{align}e^{-\iota t H}=e^{-\iota t H_1}\otimes e^{-\iota t
H_2}=U_1(t)\otimes U_2(t)\end{align} Now consider an example of
separable system $U_d =U_{d_1}\otimes U_{d_1}$
where $$U_{d_1}=  \frac{1}{\sqrt{2}}\left[%
\begin{array}{cc}
  1 & 1 \\
  1 & -1 \\
\end{array}%
\right]= U_{d_2}$$ The Hadamard transformation is a $2\times 2$
quantum Fourier transform $\text{(QFT)}$. Since an arbitrary
$\text{QFT}$ can be built out of $2\times 2$ $\text{QFT's}$, it is
of interest to realize tensor products of the Hadamard
transformations.\par Consider a three-qubit Hadamard gate which is
formed by the tensor product of three two-qubit Hadamard gates.
\[
  H^{\otimes 3}=\frac{1}{\sqrt{2}}\left[%
\begin{array}{cc}
  1 & 1 \\
  1 & -1 \\
\end{array}%
\right]\ \otimes\frac{1}{\sqrt{2}}\left[%
\begin{array}{cc}
  1 & 1 \\
  1 & -1 \\
\end{array}%
\right]\ \otimes\frac{1}{\sqrt{2}}\left[%
\begin{array}{cc}
  1 & 1 \\
  1 & -1 \\
\end{array}%
\right]\
\]
\[
 H^{\otimes 3} =\frac{1}{2^\frac{3}{2}}
\left[{\begin{array}{cccccccc}
  1 & 1 &  1 & 1 & 1 & 1 & 1 & 1 \\
  1 & -1 & 1 & -1 & 1 & -1 & 1 & -1 \\
  1 & 1 & -1 & -1 & 1 & 1 & -1 & -1 \\
  1 & -1 &-1 & 1 & 1 & -1 & -1 & 1 \\
  1 & 1 &  1 & 1 & -1 & -1 & -1 & -1 \\
  1 & -1 & 1 & -1 & -1 & 1 & -1 & 1 \\
  1 & 1 & -1 & -1 & -1 & -1 & 1 & 1 \\
  1 & -1 &-1 & 1 & -1 & 1 & 1 & -1 \\
\end{array} }\right]
\]
The three-qubit Hadamard gate is taken and raised to the power
$\frac{1}{k}$ to get the unitary gate $G$ defined by
\begin{align}G = (H^{\otimes 3})^{\frac{1}{k}}\end{align}
Taking ln on both sides of eq. (2.30), we get
\begin{align}\ln G =  \frac{1}{k}\ln(H^{\otimes 3})= \iota\eta(k)\longrightarrow 0\end{align}
As $k\longrightarrow \infty$, $G \approx e^{\iota\eta(k)}\approx
(I+\iota\eta(k))$ where $\eta(k)$ is approximately the generator
of the unitary gate $G$ and $k\eta(k)$ is thus the approximate
generator of $H^{\otimes 3}$. For $k =50$, we obtain
\[
  G=
\left[{\begin{array}{cccccccc}
  0.9994 & 0.0003 & 0.0003 & 0.0003 & 0.0003 & 0.0003 & 0.0003 & 0.0003 \\
  0.0003 & 0.9987 & 0.0003 & -0.0003 & 0.0003 & -0.0003 & 0.0003 &-0.0003 \\
  0.0003 & 0.0003 & 0.9987 & -0.0003 & 0.0003 & 0.0003 & -0.0003 &-0.0003 \\
  0.0003 & -0.0003 & -0.0003 & 0.9994 & 0.0003 & -0.0003 & -0.0003 & 0.0003 \\
  0.0003 & 0.0003 & 0.0003 & 0.0003 & 0.9987 & -0.0003 & -0.0003 & 0.0003 \\
  0.0003 & -0.0003 & 0.0003 & -0.0003 & -0.0003 & 0.9994 & -0.0003 & 0.0003 \\
  0.0003 & 0.0003 & -0.0003 & -0.0003 & -0.0003 & -0.0003 & 0.9994 & 0.0003 \\
  0.0003 & -0.0003 & -0.0003 & 0.0003 & -0.0003 & 0.0003 & 0.0003 & 0.9987 \\
\end{array} }\right]
\]
Since $G=(H^{\otimes 3})^{\frac{1}{k}}=(H^{\frac{1}{k}})^{\otimes
3}$ we can realize $H^{\frac{1}{k}}$ separately using a two
dimensional quantum system and then take the three fold tensor
product of $H^{\frac{1}{k}}$ with itself.\par However if the
system is not separable like controlled unitary gate, then to
realize it, we must use a non-separable interaction potential
$V_{12}(t)$ which acts on $H_1\otimes H_2$, that is, $U_d$ is not
of the form $U_1\otimes U_2$. Then to realize $U_d$, we must use a
Hamiltonian of the form
\begin{align}H=H_1\otimes I_2+I_1\otimes H_2+\epsilon
V_{12}(t)\equiv H_1\oplus H_2+\epsilon V_{12}(t)\end{align} Then
if we calculate $e^{-\iota t H}$ upto $O(\epsilon^2)$ form eq.
(2.6), we get a non-separable evolution operator which is given by
\begin{align}\nonumber e^{-\iota t H}=& \big(U_1(t)\otimes U_2(t)\big)
\bigg(I-\iota\epsilon\int_0^T(U_1(-\tau_1)\otimes U_2(-\tau_1))V_{12}(\tau_1)(U_1(\tau_1)\otimes
U_2(\tau_1))d\tau_1\\&\nonumber+\epsilon^2\int_{0<\tau_2<\tau_1<T}(U_1(-\tau_1)\otimes
U_2(-\tau_1))V_{12}(\tau_1)(U_1(\tau_1)\otimes
U_2(\tau_1))(U_1(-\tau_2)\otimes
U_2(-\tau_2))\\&V_{12}(\tau_2)(U_1(\tau_2)\otimes
U_2(\tau_2))d\tau_2d\tau_1+O(\epsilon^3)\bigg)\end{align} For
example, taking \begin{align}H_0 =\frac{c}{2}I_2\otimes
I_1+I_2\otimes \frac{c}{2}I_1\end{align} where
$H_1=\frac{c}{2}I_1$ and $H_2=\frac{c}{2}I_2$ are Hermitian
matrices of size $2\times 2$ and $c_1c_2=c $. $I_1$ and $I_2$ are
identity matrices of size $2\times 2$ and the interaction
potential $V_{12}(t)$ is chosen as a Hermitian matrix of size
$4\times 4$ given by
$$V_{12}(t) = \left[%
\begin{array}{cccc}
    0.3922  &  0.3437  &  0.4973 &   0.3702\\
    0.3437  &  0.2769  &  0.3705 &   0.2679\\
    0.4973  &  0.3705  &  0.3171 &   0.6659\\
    0.3702  &  0.2679  &  0.6659 &   0.7655\\
\end{array}%
\right]$$ Simulation of eq. (2.33) yields
$$U(t)=e^{-itH}\approx \left[%
\begin{array}{cccc}
    1.0173 &  -0.4946 &   0.9986 &  -0.3806\\
   -0.4924 &  -0.3307 &  -0.3804 &  -0.1897\\
    0.9878 &  -0.3422 &   0.5479 &  -1.0027\\
   -0.3350 &  -0.1655 &  -0.9937 &  -1.0481\\
\end{array}%
\right]$$ It can be verified by numerical simulation that this
unitary gate is not separable. Such non-separable gates like
controlled unitary gates give a facility to simulate a larger
class of quantum gates. So we shall describe controlled unitary
gates (which are examples of non-separable gates) in the
subsection 4.2. It is worth noting the following correlation with
the main body of the second problem:\par Let $U_d$ be a
non-separable gate of the form \begin{align}U_d =(U_{d_1}\otimes
U_{d_1})(I+\epsilon X)=U_{d_1}\otimes U_{d_1}+\epsilon
(U_{d_1}\otimes U_{d_1})X\end{align} This gate is a small
perturbation of a separable gate. We realize $U_{d_1}$ using a
harmonic oscillator Hamiltonian $H_1$, then realize $U_{d_2}$
using another harmonic oscillator Hamiltonian $H_2$ and finally
choose $V_{12}$ so that
\begin{align}U(t)=e^{(-\iota t(H_1\otimes I_2+I_1\otimes H_2+\epsilon
V_{12}(t)))}\end{align} is closest to $U_d$ upto $O(\epsilon^2)$.
\par Remark: The $r$-qubit Hadamard gate is
given by \begin{align}H^{\otimes r}| X_1 X_2 \cdots X_r\rangle =
\frac{1}{2^{\frac{r}{2}}}\sum_{\substack{Y_1Y_2\cdots
Y_r\in\{0,1\}\\X_1X_2\cdots X_r\in\{0,1\}}}
(-1)^{X_1Y_1+\cdots+X_rY_r}\big|Y_1Y_2\cdots
Y_r\big\rangle\end{align}
\subsection{Controlled Unitary Gate}
Controlled unitary gates act on two or more qubits where one or
more qubits act as a control for some operation. If the control
qubit is in the state $|0\rangle$ then the target qubit is left
unchange [36, 37]. The gate being implemented is the following
controlled unitary gate
\begin{align}U_c\quad :\quad|x_1x_2x_3\rangle\longrightarrow |x_1\rangle
U_1^{x_1}|x_2\rangle U_2^{x_1x_2}|x_3\rangle\end{align}
where $U_1 =\left(%
\begin{array}{cc}
  \alpha_1 & \beta_1 \\
  -\overline{\beta}_1  & \overline{\alpha} \\
\end{array}%
\right)$ and $U_2 =\left(%
\begin{array}{cc}
  \alpha_2 & \beta_2 \\
  -\overline{\beta}_2  & \overline{\alpha}_2 \\
\end{array}%
\right)$. In other words $U_1$ is applied to the second qubits iff
the first qubits is $1$ and $U_2$ is applied to the third qubits
iff both the first and second qubits are one. Another way to
express the gate action is via the following formulas (we choose
$x_3$ either $0$ or $1$) $$|00x_3\rangle \longrightarrow
|00x_3\rangle$$
$$|01x_3\rangle \longrightarrow
|01x_3\rangle$$
$$|10x_3\rangle \longrightarrow
|1\rangle U_1|0\rangle|x_3\rangle$$
$$|11x_3\rangle \longrightarrow
|1\rangle U_1|1\rangle U_2|x_3\rangle$$ A complete table of
three-qubits of controlled gate is given by
$$|000\rangle \longrightarrow |000\rangle$$
$$|001\rangle \longrightarrow |001\rangle$$
$$|010\rangle \longrightarrow |010\rangle$$
$$|011\rangle \longrightarrow |011\rangle$$
$$|100\rangle \longrightarrow \beta_1|110\rangle+\overline{\alpha}_1|100\rangle$$
$$|101\rangle \longrightarrow \beta_1|111\rangle+\overline{\alpha}_1|101\rangle$$
$$|110\rangle \longrightarrow \alpha_1\beta_2|111\rangle+\alpha_1\overline{\alpha}_2|110\rangle-
\overline{\beta}_1\beta_2|101\rangle-\overline{\beta}_1\overline{\alpha}_2|100\rangle$$
$$|111\rangle \longrightarrow \alpha_1\alpha_2|111\rangle-\alpha_1\overline{\beta}_2|110\rangle-
\overline{\beta}_1\alpha_2|101\rangle+\overline{\beta}_1\overline{\beta}_2|100\rangle$$
In matrix form the controlled gate $U_c$ is given by
\[
  U_c=
\left[{\begin{array}{cccccccc}
  1       & 0       & 0       & 0       & 0       & 0       & 0       & 0       \\
  0       & 1       & 0       & 0       & 0       & 0       & 0       & 0       \\
  0       & 0       & 1       & 0       & 0       & 0       & 0       & 0       \\
  0       & 0       & 0       & 1       & 0       & 0       & 0       & 0       \\
  0       & 0       & 0       & 0       & \overline{\alpha}_1   & 0   & -\overline{\beta}_1\overline{\alpha}_2  & \overline{\beta}_1\overline{\beta}_2   \\
  0       & 0       & 0       & 0       & 0  & \overline{\alpha}_1  & -\overline{\beta}_1\beta_2  & - \overline{\beta}_1\alpha_2   \\
  0       & 0       & 0       & 0       & \beta_1  & 0  & \alpha_1\overline{\alpha}_2  & -\alpha_1\overline{\beta}_2   \\
  0       & 0       & 0       & 0       & 0  & \beta_1  & \alpha_1\beta_2   & \alpha_1\alpha_2   \\
\end{array} }\right]
\]
As pointed out in the previous section, $U_c$ cannot be realized
using a separable Hamiltonian $H_0=H_1\otimes I_2+I_1\otimes H_2$.
It can however be realized by adding an interaction potential with
the overall Hamiltonian which is given in eq. (2.32).
$$H=H_0+\epsilon V_{12}(t)=H_1\otimes I_2+I_1\otimes H_2+\epsilon
V_{12}(t)$$ In other words, we fix $H_1$ and $H_2$ and then expand
$e^{-\iota t H}$ upto $O(\epsilon^2)$. The error energy between
the resulting gate and $U_c$ can then be minimized with respect to
$V_{12}(t)$ and we can get the required approximation. In this
context, it is worth examining how to approximate the above
controlled unitary gate $U_c$ as
$$U_c\approx (U_1\otimes U_2)(I+\epsilon X)$$ where $U_1$ and
$U_2$ are unitary and $X$ is skew-Hermitian. Then we could use two
independent harmonic oscillator Hamiltonian, plus a small
interaction potential energy get a good approximation to $U_c$.
\par We here note that other modification of the harmonics oscillator like the $q$-deformed oscillator algebra
have been used in the literature to design qubit gates and that
the gates generator in can also be realized using time dependent
harmonic oscillator. The authors begin by noting that Schwinger's
response can be used to represent the angular momentum eigenstates
$|j, m\rangle,\quad |m|\leq j$, of a spin $j$ particle having $Z$-
component of angular momentum $m$ harmonics oscillator creation
and annihilation operators for $j$ and $m$ separately. The
formalism can be derived using the Ladder operators $J_+=J_1+\iota
J_2$, $J_-=J_1-\iota J_2$ of angular momentum with the commutator
relations $[J^2, J_+]=[J^2, J_-]=0,\quad [J_+, J_-]=J_3,\quad
[J_+, J_3]=-J_+,\quad [J_-, J_3]=J_-$. From these commutation
relations, it is easily shown that $J_+|j, m\rangle \propto | j,
m+1\rangle,\quad J_-|j, m\rangle \propto | j, m-1\rangle$. They
represent $q$-bit states using $|j, m\rangle$ which is in term
realized by acting on the $|0, 0\rangle$ state with
$\frac{a_1^{\dag j+m}a_2^{\dag
j+m}}{\sqrt{(j+m)!}\sqrt{j-m}!}$.\par This enables the authors to
represent all the familiar quantum gates by their actions on
$a_1^{\dag \alpha}a_2^{\dag \beta}|0 0\rangle$. If we take our
hamiltonian as $H_0=c_1a_1^\dag a_1+c_2a_2^\dag a_2$, then we can
find the matrix of $e^{-\iota t H_0}$ (this will be diagonal)
relative to the $|j, m\rangle$ and realized this gate. We can also
perturbed this Hamiltonian to
$H(t)=H_0+f_1(t)\frac{(a_1+a_1^\dag)}{2}+f_2(t)\frac{(a_2+a_2^\dag)}{2}$
and compute $f_1(t)$ and $f_2(t)$ so that the evolution operator
at the $T$ approximates a given gate. After that, the authors
design gate using $q$-deformed boson algebra wherein super
commutation like rules are satisfied by the creator and
annihilator operators $a_qa_q^\dag-qa_q^\dag a_q=q^{-N}$, where
$q=1$, it reduces to the familiar harmonic oscillator case. They
replace the qubit states $a_1^{\dag x}a_2^{\dag(1-x)}|0 0\rangle$,
$x=0,1$ in the Schwinger representation by their $q$-deformed
versions $a_{1q}^{\dag x} a_{2q}^{\dag(1-x)} |0 0\rangle$. They
formulate rules for constructly the $q$-deformed operators $a_q
a_q^\dag$ from the undeformed ones $a, a^\dag$ and then write down
the action of the familiar quantum gates on the $q$-deformed
states. If there existed a physical world in which apart from
Boson and fermions, there are $q$-Boson ($q=1$ gives Boson and
$q=-1$ gives Fermions), then it would be an interesting idea to
see whether a $q$-Boson Hamiltonian like $a_q^\dag a_q$ can be
perturbed to $a_qa_q^\dag f(t)\frac{(a_q+a_q^\dag)}{2}$ and gates
realized using this perturbed Hamiltonian. In this case,
Heisenberg's matrix mechanics would get replaced by
$$\frac{dX_t}{dt}=\iota (H_0 X_t-q X_tH)$$ and it is an immediately
not clear how Schr\"odinger equation in the interaction picture
can be formulated. The advantage of the $q$-Boson states is that
one may hope to generate a wider class of quantum gates and even
non-unitary gates by varying $q$.
\subsection{Results}
Fig. 1 shows a plot of noise to signal ratio $\text{(NSR)}$ versus
time $\text{T}$ from eq. (2.19) given by
$$\text{NSR(T)}= \frac{\mathbb{E}}{\|U_d\|^2}=\frac{\|U_d-U(T)\|^2}{\|U_d\|^2}$$ where $U_d$ is the
derived quantum Hadamard unitary gate $G(=(H^{\otimes
2})^\frac{1}{k}$ or $(H^{\otimes 3})^\frac{1}{k})$ and
$U_0(-T)U(T)$ is the simulated gate with the unperturbed gate
$U_0(T)$ removed by inversion. For small perturbing potentials
$\epsilon V(t)$, $U_0(-T)U(T)=e^{\iota T H_0}U(T)$ is close to the
identity operator and so is $G$ for large values of $k$. The graph
shows that $\text{NSR(T)}$ decreases rapidly with increasing time
$T$ and finally converges to a steady state minimum value. Zero
$\text{NSR(T)}$ is not attainable because quantum mechanical
harmonic oscillator is an infinite dimensional quantum system and
we are using a second order truncated Dyson series to approximate
the given gate and such a truncated series cannot be exactly
unitary. Decrease of $\text{NSR(T)}$ with time $\text{T}$ occurs
because for larger values of $\text{T}$, we have more degrees of
freedom $E(t)$, $0\leq t\leq T$ to choose from. More precisely,
the gate $W_g$ is being approximated by the operator
$$W(T)=I+\iota\epsilon\int_0^T q
E(t)\widetilde{X}(t)dt+(\iota\epsilon)^2\int_{0<t_2<t_1<T}q^2E(t_1)E(t_2)\widetilde{X}(t_1)\widetilde{X}(t_2)dt_1dt_2$$
where $X$ is the position operator and
$\widetilde{X}(t)=U_0^*(t)U(t)$, $U(t)=e^{-\iota t H_0}$ (The $q$
appearing here is the charge of the harmonic oscillator
\textbf{and not} the parameter in $q$-deformed systems). As
$\text(T)$ increases the span of the operators $\{W(T), 0\leq
t\leq T\}$ varying increases. This is because if $T_2>T_1$, then
$W(T_2)= W(T_1)$ with $E(t)=0$ for $t>T_2$. Hence for $T_2>T_1$
every operator $W(T_1)$ ia also of the form $W(T_2)$ for a special
choice of the electric field. Hence as $\text{T}$ increases, we
are bound to get a better approximation for the given unitary
gate. Thus,
$$\substack{min\\\text{E}(t)\\0\leq t \leq T_1}\|W(T_1)-W_g\|\geq
\substack{min\\\text{E}(t)\\0\leq t \leq T_2}\|W(T_2)-W_g\| $$
Hence our graphs show decreasing $\text{NSR(T)}$ with increasing
$\text{T}$.
%
\section{Conclusions and Scope for Future Work}
We have in the first problem, minimized the discrepancy between a
given unitary gate and the gate obtained by evolving a quantum
harmonic oscillator in a weak electric field. In carrying out this
optimization, an energy constraint on the field of the form has
been imposed.
$$ E_{diss}=\int_0^T  E^2(t)dt $$ The approach to unitary gate design in is based on using exponential co-ordinates
that is, it is a Lie algebra based method, where a predefined set
of quantum gates can be expressed directly in terms of the
physical control parameters via exponential co-ordinates on the
group of transformation. Here we don't use exponential
co-ordinates. We have instead used an approximate perturbation
theoretical approach to gate design. This is more appropriate
while dealing with real physical systems like atoms, molecules and
harmonics oscillators. Each implementation proposal for a quantum
computer has its own method to generate gates, relying on the
already existing techniques for state manipulation. We look at the
general problem of computing the evolution of a quantum system
having a Hamiltonian operator of the form $H_0 + \epsilon V(t) $
and putting some constraints on $V(t)$ in the form of unknown
scalar function of time which are determined by minimizing the
gate error energy, obtained from approximate quantum evolution. A
forthcoming problem will extend the method to design a quantum
gate in the presence of noise introduced in the potential upto
$O(\epsilon^ k)$.

\chapter{REALIZATION OF QUANTUM GATES BASED ON THREE DIMENSIONAL HARMONIC OSCILLATOR IN A TIME VARYING ELECTROMAGNETIC FIELD}
\section{Introduction}
\label{intro}When a $3$-D quantum harmonic oscillator with a
charge on an oscillator mass is subject to an external
electromagnetic field, then the Hamiltonian gets perturbed by an
$O(e)$ term and an $O(e^2)$ term where $e$ is the charge on the
oscillator mass. The $O(e)$ term is a function of $(t, r)$ plus a
vector field of the form $A^k\frac{\delta}{\delta x^k}$. The first
component, namely the function of $(t, r)$ is a linear function of
the magnetic vector potential and the electric scalar potential.
The $O(e^2)$ term is a quadratic function of the magnetic vector
potential. We assume that the perturbing electromagnetic field is
constant in space, but time dependent, then the magnetic vector
potential and electric scalar potential are both expressed easily
in terms of the electromagnetic field components. Using time
dependent second order perturbation theory applied to this
perturbed oscillator, we compute the unitary evolution operator
upto $O(e^2)$ [23]. This approximate unitary evolution operator at
time $T$ is expressible as a linear plus a quadratic function of
the electromagnetic field $(E(t),B(t),0 \leq t \leq T)$ and hence
we are able to compute upto $O(e^2)$, the error energy between
this approximate unitary gate and a given unitary gate. This error
energy is now a linear-quadratic function of the electromagnetic
field $(E(t),B(t),0 \leq t \leq T)$. We minimize this function
with respect to $(E(t),B(t),0 \leq t \leq T)$ subject to an
average energy constraint,
$$\frac{1}{T}\int_0^T \left[\frac{\epsilon_0|E(t)|^2}{2}+\frac{|B(t)|^2}{2\mu_0}\right]dt$$
The result of this constrained optimization is the optimal
electromagnetic field $(E(t), B(t), \quad 0 \leq t \leq T)$ with
which the oscillator is to be perturbed so that subject to energy
constraint, a good approximation to the desired unitary gate is
obtained. A method for finding an acceptable and good approximate
solution to the optimal integral equations for the electromagnetic
field is needed.

A $3$-D quantum harmonic oscillator carrying charge $e$ has
Hamiltonian
$$H_0=\frac{1}{2}\sum_{\alpha=1}^3(p_\alpha^2+\omega_\alpha^2q_\alpha^2)$$
where the $p_\alpha^{'s}$ and $q_\alpha^{'s}$ satisfy canonical
commutation relation $[q_\alpha, q_\beta]=[p_\alpha, p_\beta]=0$
and $[q_\alpha, p_\beta]=\iota\delta_{\alpha\beta}$. It is
perturbed by an electromagnetic field that is constant in space
but time varying. The corresponding $4$-vector potential is
approximately given by
$$\Phi(t, q)=-e\sum_{\alpha=1}^3E_\alpha(t)q_\alpha$$
and
$$A(t, q)=\frac{1}{2}B(t)\times q$$
We note that $\nabla \times A=B(t)$ and $\nabla\Phi-\frac{\partial
A}{\partial t}=E(t)-\frac{1}{2}B'(t)\times q$ so the approximation
is good only if $|B'(t)||q|<<|E(t)|$ for $q$ raising over the
atomic dimensions. The perturbed Hamiltonian is
$$
H(t)=\frac{1}{2}\sum_{\alpha=1}^3(p_\alpha+eA_\alpha(t,
q))^2+\omega_\alpha^2q_\alpha^2-e\Phi(t, q)
$$
This can be expressed as
$$H(t)=H_0+eV_1(t)+e^2V_2(t)$$
where $V_2(t)$ is a multiplication operator, that is,
$V_2(t)=\frac{1}{2}\sum_{\alpha=1}^3A_\alpha^2$ and
$V_1(t)=-\Phi(t, q)+\frac{1}{2}(p, A)+(A, p)$ is a first order
partial differential operator (E(t), B(t)) constitute six control
inputs. Using second order perturbation theory, the evolution
operator $U(t)$ which satisfies Schr\"{o}dinger equation shall
soon be derived [25, 27].
\subsection{Time Dependent Perturbation Theory for Atoms and Oscillators}
\par Most of the problems on quantum gate design
deal with abstract perturbation of a given Hamiltonian by an
operator. The idea of abstract perturbation theory is to begin
with a reasonably good starting Slater determinant and then by an
iterative scheme, based on the generalized Bloch equation,
introduce corrections to this Slater determinant to finally arrive
at the true wave function. If we achieve this, we say that we have
performed the perturbation expansion to all orders. The second
problem of the thesis discussed here an exact physical model
involving derivation of the actual Hamiltonian perturbation
operator in terms of the applied electromagnetic potential and the
position-momenta of the oscillator system [16, 17, 19]. In other
words instead of working with abstract matrices of the form
$H_0+\epsilon V(t)$ for the Hamiltonian, we work with a specific
physical model taking
\begin{align}\nonumber\langle m|H_0|n\rangle=\langle m|\frac{p^2+q^2}{2}| n\rangle\end{align}
as the unperturbed matrix and
\begin{align}\nonumber V(t)=\langle m|\frac{e}{2}((p, A(t, q)+A(t, q), p))-e\Phi(t, q)+e^2\frac{A^2}{2}|n\rangle\end{align}
as the perturbed matrix \bigg(which arises from the formula
$\frac{(p+eA)^2}{2}+\frac{q^2}{2}-e\Phi=\frac{p^2+q^2}{2}+\bigg(\frac{e}{2}((p,
A)+(A, p))\bigg)-e\Phi+e^2\frac{A^2}{2}$\bigg). Where, $A(t, q)$
is the magnetic vector potential and $\Phi(t, q)$ is the electric
scalar potential. The perturbing operators coming into the picture
in our thesis are derived from basic physics involving the
quantum-mechanical motion of the oscillator in an external
electromagnetic field. The second novel feature of the second
problem involves studying in addition to the classical
electromagnetic field the effect of a quantum electromagnetic
field coming from the heat bath and interacting with the
oscillator, on the gate performance.\par The significant
contribution of the third problem is to show how by using a real
physical system such as an atom or a molecule (modelled as a $3$-D
quantum harmonic oscillator for small displacements of the
electron from its equilibrium position), we can, by applying an
external electromagnetic field, create unitary gates used in
quantum computation with a high degree of accuracy. This problem
in particular shows that by truncating an infinite dimensional
quantum system to finite dimensions, we can realize a quantum gate
based on three dimensional harmonic oscillators in a time varying
electromagnetic field. After illustrating how an arbitrary unitary
gate can be realized approximately using a perturbed Hamiltonian,
we have already discuss qualitatively some issues regarding how
separable and nonseparable unitary gates can be realized using
respectively independent Hamiltonians and independent Hamiltonians
with an interaction [22, 35, 36]. A schematic diagram of the
Schr\"odinger dynamics and the optimal gate design is shown below:

\subsection{Interaction of Electromagnetic Field with Oscillator
Hamiltonian}The unperturbed Hamiltonian $H_0$ for a $3$-D
dimensional harmonic oscillator is given by,
\begin{align}H_0 &= \frac{p^2+q^2}{2}\end{align}
where, $p=(p_1,p_2,p_3)$ and $q=(q_1,q_2,q_3)$ are momentum and
position operators respectively. Let the system be perturbed by a
time-varying electromagnetic field whose electric and magnetic
field are $E(t)=(E_1(t),E_2(t),E_3(t))$ and
$B(t)=(B_1(t),B_2(t),B_3(t))$ respectively. The true electric
field is given by
\begin{align}E(t,{r})=-\nabla\Phi-\frac{\partial A}{\partial
t}=E(t)-\frac{1}{2}(B'(t)\times{r})\end{align} where,
$\Phi=-E(t){r}$ and ${A}(t,{r})= \frac{1}{2}(B(t)\times{r})$ but
usually $|B'(t)\times{r}|<<|E(t)|$, so $E(t,r)\approx E(t)$. The
perturbed Hamiltonian for the system is given by
\begin{align}H(t)= \frac{(p+eA)^2+q^2}{2}-e\Phi\end{align}
which approximates to
\begin{align}H(t)= \frac{(p+eA)^2+q^2}{2}-e\Phi= \frac{p^2+q^2}{2}+ \frac{e}{2}((p,A)+(A,p))-e\Phi+
\frac{e^2A^2}{2}\end{align} Since
\begin{align}(p,A)+(A,p)=-2iA.\nabla-i(\nabla.A)\end{align}
$$(\nabla.A)=\frac{1}{2}\nabla.(B(t)\times{r})=-\frac{1}{2}B(t).(\nabla\times{r})=0$$
we get
\begin{align}H(t)=H_0+eA.p+\frac{e^2A^2}{2}-e\Phi=H_0+e(\frac{1}{2}B(t)\times
r.p+E(t).r)+\frac{e^2}{8}(B(t)\times r)^2\end{align} Since
\begin{align}(B(t)\times r)^2=B^2(t)q^2-(B(t),q)^2\end{align} and with $ L=r\times p
\triangleq q\times p$, we obtain
\begin{align}H(t)=H_0+e\bigg(\frac{1}{2}B(t).
L+E(t).q)+\frac{e^2}{8}(B^2(t)q^2-(B(t),q)^2\bigg)\end{align}
Thus, $H(t)$ obtained in eq. (3.8) represents the perturbed
Hamiltonian of the quantum harmonic oscillator in the presence of
external electromagnetic field [44].
\section{Mathematical Modeling of Quantum Unitary Gate}
\label{sec:2}The moral equivalent in quantum computing to partial
recursive functions are unitary operators. As every classically
computable problem can be reformulated as calculating the value of
a partial recursive function, each quantum computation must have a
corresponding unitary operator [13, 14, 16]. Let $U(t)$ be the
unitary evolution operator corresponding to the Hamiltonian
$H(t)$, which is given by
\begin{align}\iota U'(t)=H(t)U(t),\quad t\geq 0\end{align}
Define \begin{align}a_k=\frac{q_k+ip_k}{\sqrt{2}}\end{align}
\begin{align}a^{\dag}_k=\frac{q_k-ip_k}{\sqrt{2}}\end{align}
Then, we can write $$[a_k,a_j^{\dag}] = \delta_{kj}$$
$$\sum_{k=1}^3 a_ka^{\dag}_k=H_0+\frac{3}{2},\quad\sum_{k=1}^3
a^{\dag}_ka_k=H_0-\frac{3}{2}$$ Now,
$$L_1=q_2p_3-q_3p_2=\frac{(a_2+a^\dag_2)}{\sqrt2}\frac{(a_3-a^\dag_3)}{\sqrt2\iota}-\frac{(a_3+a^\dag_3)}{\sqrt2}\frac{(a_2-a^\dag_2)}{\sqrt2\iota}$$
\begin{align}\nonumber \iota L_1=\frac{1}{2}{\left\{a_2a_3-a_2a_3^\dag +a_2^\dag a_3-a_2^\dag
a_3^\dag -a_3a_2+a_3a_2^\dag -a_3^\dag a_2+a_3^\dag
a_2^\dag\right\}}=a_2^\dag a_3-a_2a_3^\dag\end{align}
Therefore,$$L_1=\iota (a_2a_3^\dag-a_2^\dag a_3)$$
\begin{align}L_2=\iota (a_3a_1^\dag-a_3^\dag a_1)\end{align}
$$L_3=\iota(a_1a_2^\dag-a_1^\dag a_2)$$
Let ${\left\{|n_1,n_2,n_3\rangle , 0\leq n_1,n_2,n_3<\infty
\right\}}$ be the eigenstates of the $3$-D harmonic oscillator.
Therefore,
\begin{align}H_0|n_1,n_2,n_3\rangle=(n_1+n_2+n_3+\frac{3}{2})|n_1,n_2,n_3\rangle\end{align}
Let us convert the angular momentum and position operators to
their dynamical counterparts in the interaction picture.
$$L(t)=exp(itH_0).L.exp(-itH_0)$$
$$q(t)=exp(itH_0).q.exp(-itH_0)$$
If we denote evolution operator in interaction picture by $W(t)$,
given by
\begin{align}U(t)=exp(-itH_0)W(t)\end{align}
\begin{align}iW'(t)=V(t)W(t)\end{align}
where
\begin{align}V(t)=e\bigg(\frac{1}{2}B(t).L(t)+E(t).q(t)\bigg)+\frac{e^2}{8}(B(t)\times q(t))^2\end{align}
It denotes the interaction part of the Hamiltonian. By expanding
using Dyson series in eq. (3.15) the evolution operator is upto
second order, given by
$$W(t)=I- i\int_0^tV(t_1)dt_1-\int_{0<t_2<t_1<t}V(t_1)V(t_2)dt_1dt_2+O(e^2)$$
The transition probability amplitude between stationary states is
given by
\begin{align}\nonumber\langle n_1n_2n_3|U(t)|m_1m_2m_3\rangle &=\langle
n_1n_2n_3|exp(-itH_0)W(t)|m_1m_2m_3\rangle\\&=exp(-itE(n_1n_2n_3))\langle
n_1n_2n_3|W(t)|m_1m_2m_3\rangle \end{align} where,
$E(n_1n_2n_3)=(n_1+n_2+n_3+\frac{3}{2})$, using  eq. (3.16), we
get
\begin{align}\nonumber W(t) &=I-\nonumber ie\int_0^t\bigg(\frac{1}{2}B(t_1).L(t_1)+E(t_1).q(t_1)\bigg)dt_1
-\nonumber i\frac{e^2}{8}\int_0^t\bigg(B(t_1)\times
q(t_1)\bigg)^2dt_1\\&-\nonumber e^2
\int_{0<t_2<t_1<t}\bigg(\frac{1}{2}B(t_1).L(t_1)+E(t_1).q(t_1)\bigg)\bigg(\frac{1}{2}B(t_2).L(t_2)+E(t_2).q(t_2)\bigg)dt_1dt_2\\&
+O(e^3)\end{align}
Denoting \begin{align}\langle
n_1n_2n_3|L_1(t)|m_1m_2m_3\rangle & \triangleq \langle
n|L_1(t)|m\rangle=e^{itE(n,m)}\langle n|L_1|m\rangle
\end{align} where
$$E(n,m)=E(n)-E(m)=\sum_{k=1}^3{(n_k-m_k)}$$
The general matrix element of the angular momentum operator $L_1$,
is given by
$$\langle n|L_1|m\rangle=i\langle n|a_2a_3^\dag-a_2^\dag
a_3|m\rangle $$ with
$$ a_3|m_1m_2m_3\rangle=\sqrt m_3|m_1,m_2,m_3-1\rangle$$
$$\langle n_1n_2n_3|a_2^\dag =\sqrt n_2\langle n_1,n_2-1,n_3|$$
$$\langle n|a_2^\dag a_3|m\rangle=\sqrt
{n_2m_3}\delta[n_1-m_1]\delta[n_2-1-m_2]\delta[n_3-m_3+1]$$
$$\langle n|a_2a_3^\dag |m\rangle=\langle n|a_3^\dag
a_2|m\rangle=\sqrt
{n_3m_2}\delta[n_1-m_1]\delta[n_2-m_2+1]\delta[n_3-1-m_3]$$ For
the concise formulation of the equations, we define an operator
$\mathcal{Z}^{-1}_k$, which acts on function
$f:\mathbb{Z}^3_+\rightarrow\mathbb{R}$ by the rule
$$\mathcal{Z}^{-1}_1 f(n_1n_2n_3)=f(n_1-1,n_2,n_3)$$
$$\mathcal{Z}^{-1}_2 f(n_1n_2n_3)=f(n_1,n_2-1,n_3)$$
$$\mathcal{Z}^{-1}_3 f(n_1n_2n_3)=f(n_1,n_2,n_3-1)$$
Thus, we have
$$\langle n|L_1|m\rangle=-i(\sqrt{n_2m_3}\mathcal{Z}^{-1}_2\mathcal{Z}_3-\sqrt{n_3m_2}\mathcal{Z}_2\mathcal{Z}^{-1}_3)\delta[n-m]$$
Likewise,
$$\langle n|L_2|m\rangle=-i(\sqrt{n_3m_1}\mathcal{Z}^{-1}_3\mathcal{Z}_1-\sqrt{n_1m_3}\mathcal{Z}_3\mathcal{Z}^{-1}_1)\delta[n-m]$$
$$\langle n|L_3|m\rangle=-i(\sqrt{n_1m_2}\mathcal{Z}^{-1}_1\mathcal{Z}_2-\sqrt{n_2m_1}\mathcal{Z}_1\mathcal{Z}^{-1}_2)\delta[n-m]$$
The general expression can be written as
$$ \langle n|L_k|m \rangle= -i\sum_{r,s} \epsilon(krs)
\sqrt{n_rm_s}\mathcal{Z}^{-1}_r\mathcal{Z}_s \delta[n-m]$$ We also
need the matrix elements of $B(t).L(t)$
\begin{align} \nonumber
\langle n|B(t).L(t)|m\rangle & =B_(t)\langle
n|L_1(t)|m\rangle+B_2(t)\langle n|L_2(t)|m\rangle+B_3(t)\langle
n|L_3(t)|m\rangle\\&\nonumber=-\iota e^{\iota
tE(n-m)}\{B_1(t)(\sqrt{n_2m_3}\mathcal{Z}^{-1}_2\mathcal{Z}_3-\sqrt{n_3m_2}\mathcal{Z}_2\mathcal{Z}^{-1}_3)
\\&
+\nonumber
B_2(t)(\sqrt{n_3m_1}\mathcal{Z}^{-1}_3\mathcal{Z}_1-\sqrt{n_1m_3}\mathcal{Z}_3\mathcal{Z}^{-1}_1)
\\&+ B_3(t)(\sqrt{n_1m_2}\mathcal{Z}^{-1}_1\mathcal{Z}_2-\sqrt{n_2m_1}\mathcal{Z}_1\mathcal{Z}^{-1}_2)\}\delta[n-m] \end{align}
To get the general matrix elements for position matrix, we have
$$\langle n|q_k(t)|m\rangle = e^{itE(n,m)}\langle
n|q_k|m\rangle$$
\begin{align} \langle n|q_1|m\rangle &=\langle
n|\frac{(a_1+a_1^\dag)}{\sqrt{2}}|m\rangle=(\sqrt{\frac{m_1}{2}}
\mathcal{Z}_1+
\sqrt{\frac{n_1}{2}}\mathcal{Z}^{-1}_1)\delta[n-m]\end{align} and
likewise for $q_2,q_3$. The general expression comes out to be,
$$\langle n|q_k|m\rangle=(\sqrt{\frac{m_k}{2}}\mathcal{Z}_k+\sqrt{\frac{n_k}{2}}\mathcal{Z}^{-1}_k)\delta[n-m],k=1,2,3 $$
We also need matrix elements of
$$\langle n|q_k(t)q_l(t)|m\rangle=e^{itE(n,m)}\langle
n|q_kq_l|m\rangle $$ For $k \neq l$,
\begin{align}\nonumber\langle n|q_kq_l|m\rangle &=\langle
n|\frac{(a_k+a_k^\dag)}{\sqrt{2}}\frac{(a_l+a_l^\dag)}{\sqrt{2}}|m\rangle=\frac{1}{2}\{\langle
n|a_ka_l +a_ka_l^\dag +a_k^\dag a_l + a_k^\dag a_l^\dag|m \rangle
\}\\&\nonumber=\frac{1}{2}\{\sqrt{m_km_l}\mathcal{Z}_k\mathcal{Z}_l
+ \sqrt{m_kn_l}\mathcal{Z}_k\mathcal{Z}^{-1}_l
+\sqrt{m_ln_k}\mathcal{Z}^{-1}_k\mathcal{Z}_l
+\sqrt{n_kn_l}\mathcal{Z}^{-1}_k\mathcal{Z}^{-1}_l\}\delta[n-m]\end{align}
For $k=l$, the required matrix elements is given by
\begin{align}\nonumber\langle
n|q_k^2|m\rangle &=\frac{1}{2}\langle
n|(a_k+a_k^\dag)^2|m\rangle=\frac{1}{2}\{\langle n|a_k^2
+a_k^{\dag^2} +a_ka_k^\dag +a_k^\dag a_k|m \rangle
\}\\&\nonumber=\frac{1}{2}\{\sqrt{m_k(m_k-1)}\mathcal{Z}_k^2+\sqrt{n_k(n_k-1)}\mathcal{Z}^{-2}_k+(2m_k+1)\}\delta[n-m]\end{align}
We can rewrite the Dyson series eq. (3.18) as follows,
$$ W(t)=I+eW_1(t)+e^2W_2(t)+O(e^3)$$
where, $$
W_1(t)=-i\int_0^t(\frac{1}{2}B(t_1).L(t_1)+E(t_1).q(t_1))dt_1$$
and \begin{align}\nonumber \langle n|W_1(t)|m \rangle
&=-i\int_0^t\frac{1}{2}B(t_1).\langle n|L(t_1)|m \rangle dt_1
-i\int_0^t E(t_1).\langle n|q(t_1)|m \rangle dt_1\\ &=-\nonumber
\frac{i}{2}\int_0^t
e^{it_1E(n,m)t_1}\{B_1(t)(\sqrt{n_2m_3}\mathcal{Z}^{-1}_2\mathcal{Z}_3-\sqrt{n_3m_2}\mathcal{Z}_2\mathcal{Z}^{-1}_3)\\&\nonumber
+B_2(t)(\sqrt{n_3m_1}\mathcal{Z}^{-1}_3\mathcal{Z}_1-\sqrt{n_1m_3}\mathcal{Z}_3\mathcal{Z}^{-1}_1)\\&\nonumber
+B_3(t)(\sqrt{n_1m_2}\mathcal{Z}^{-1}_1\mathcal{Z}_2-\sqrt{n_2m_1}\mathcal{Z}_1\mathcal{Z}^{-1}_2)\}\delta[n-m]dt_1\\&\nonumber
-i\int_0^t
e^{it_1E(n,m)_1}\{E_1(t_1)(\sqrt{\frac{m_1}{2}}\mathcal{Z}_1+\sqrt{\frac{n_1}{2}}\mathcal{Z}^{-1}_1)\\&\nonumber
+E_2(t_1)(\sqrt{\frac{m_2}{2}}\mathcal{Z}_2+\sqrt{\frac{n_2}{2}}\mathcal{Z}^{-1}_2)\\&
+E_3(t_1)(\sqrt{\frac{m_3}{2}}\mathcal{Z}_3+\sqrt{\frac{n_3}{2}}\mathcal{Z}^{-1}_3)\}\delta[n-m]dt_1\end{align}
or equivalently, defining
$$\hat{B_k}(n,t)=\int_0^t B_k(t_1)e^{\iota n t_1}dt_1$$
and $$\hat{E_k}(n,t)=\int_0^t E_k(t_1)e^{\iota n t_1}dt_1$$ we get
the following expression for the $O(e)$ terms of the matrix
element of the evolution operator in the interaction picture:
\begin{align}\nonumber \langle n|W_1(t)|m \rangle &= -\frac{i}{2}\{
\hat{B_1}(n-m,t)(\sqrt{n_2m_3}\mathcal{Z}^{-1}_2\mathcal{Z}_3-\sqrt{n_3m_2}\mathcal{Z}_2\mathcal{Z}^{-1}_3)\\&\nonumber
+\hat{B_2}(n-m,t)(\sqrt{n_3m_1}\mathcal{Z}^{-1}_3\mathcal{Z}_1-\sqrt{n_1m_3}\mathcal{Z}_3\mathcal{Z}^{-1}_1)\\&\nonumber
+\hat{B_3}(n-m,t)(\sqrt{n_1m_2}\mathcal{Z}^{-1}_1\mathcal{Z}_2-\sqrt{n_2m_1}\mathcal{Z}_1\mathcal{Z}^{-1}_2)\}\delta[n-m]\\&
-i\sum_{k=1}^3
\hat{E_k}(n-m,t)(\sqrt{\frac{m_k}{2}}\mathcal{Z}_k+\sqrt{\frac{n_k}{2}}\mathcal{Z}^{-1}_k)\delta[n-m]\end{align}
Further, the $O(e^2)$ term $W_2(t)$ of the evolution operator in
the interaction picture is given by
\begin{align}\nonumber W_2(t)&=-\frac{i}{8}\int_0^t(B(t_1)\times q(t_1))^2dt_1
\\&\nonumber-\int_{0<t_2<t_1<t}\bigg(\frac{1}{2}B(t_1).L(t_1)+E(t_1).q(t_1)\bigg)\bigg(\frac{1}{2}B(t_2).L(t_2)+E(t_2).q(t_2)\bigg)dt_1dt_2\end{align}
On using eq. (3.7), we get
\begin{align}\nonumber\langle n|(B(t)\times q)^2|m \rangle &=e^{\iota t E(n,m)}\bigg\{B^2(t)\sum_{k=1}^3 \langle n|q_k^2|m \rangle
-\sum_{k=1}^3 B_k(t)B_r(t)\langle n|q_kq_r|m \rangle
\bigg\}\\&\nonumber=\bigg\{(B_2^2(t)+B_3^2(t)) \langle n|q_1^2|m
\rangle + (B_3^2(t)+B_1^2(t)) \langle n|q_2^2|m \rangle
\\&\nonumber + (B_1^2(t)+B_2^2(t)) \langle n|q_3^2|m \rangle-2\sum_{1<k<r<3} B_k(t)B_r(t)\langle n|q_kq_r|m \rangle \bigg\}e^{\iota
t E(n,m)t}
\end{align} To calculate $\langle n|W_2(t)|m \rangle $, we also need
$\langle n|L_k(t_1)L_r(t_2)|m \rangle $ and $\langle
n|L_k(t_1)q_r(t_2)|m \rangle$
$$L_k(t_1)L_r(t_2)=e^{\iota t_1H_0}L_kexp(-\iota(t_1-t_2)H_0)L_re^{-\iota t_2H_0}$$
So ,$$\langle n|L_k(t_1)L_r(t_2)|m \rangle
=e^{\iota(E(n)t_1-E(m)t_2)}\sum_s \langle n|L_k|s \rangle \langle
s|L_r|m \rangle e^{-\iota(t_1-t_2)E(s)}$$ The other quantity
required is
\begin{align}\nonumber T_1& = \langle n|\int_{0<t_1<t_2<t}
B(t_1).L(t_1)B(t_2).L(t_2)dt_1dt_2|m
\rangle\\&\nonumber=\sum_{k,r=1}^3 \int_{0<t_1<t_2<t}
B_k(t_1).B_r(t_2)\langle n|L_k(t_1)L_r(t_2)|m \rangle
dt_1dt_2\\&\nonumber=\sum_{k,r,s} \int_{0<t_1<t_2<t}
B_k(t_1).B_r(t_2)e^{-\iota(t_1-t_2)E(s)} \langle n|L_k|s \rangle
\langle s|L_r|m \rangle
e^{\iota(E(n)t_1-E(m)t_2}dt_1dt_2\end{align} We now substitute for
$\langle n|L_r|m \rangle$ in the above equation but firstly, let
us define $e_r=(\delta_{r1},\delta_{r2},\delta_{r3})$, $r=1,2,3$
,that is,
$$e_1=(100),e_2=(010),e_3=(001)$$
As can be seen clearly, $$\mathcal{Z}^{-1}_r\mathcal{Z}_s
\delta[n-m]=\delta[n-m-e_r+e_s]$$ So, on incorporating all the
above we have $T_1$ as
\begin{align}\nonumber T_1&=-\sum_{k,r,s,\alpha,\beta,\mu,\nu}\bigg(\int_{0<t_2<t_1<t}B_k(t_1).B_r(t_2)e^{-\iota(t_1-t_2)E(s)}
\epsilon(k\alpha\beta)\epsilon(r\mu\nu)\\&\nonumber\sqrt{n_\alpha
s_\beta} \sqrt{s_\mu m_\nu}\delta[n-s-e_\alpha+e_\beta]
\delta[s-m-e_\mu+e_\nu]e^{\iota(E(n)t_1-E(m)t_2)}\bigg)dt_1dt_2
\\&\nonumber=-\sum_{k,r,\alpha,\beta,\mu,\nu}\bigg(\int_{0<t_2<t_1<t}B_k(t_1).B_r(t_2)e^{-\iota(t_1-t_2)E(n-e_\alpha+e_\beta)}
\epsilon(k\alpha\beta)\epsilon(r\mu\nu)\\&\nonumber\sqrt{n_\alpha
(n_\beta-\delta_{\alpha\beta}+1)
m_\nu(m_\mu-1+\delta_{\mu\nu})}\delta[n-m-e_\alpha+e_\beta+e_\mu-e_\nu]\\&\nonumber
e^{\iota(E(n)t_1-E(m)t_2)}\bigg)dt_1dt_2\end{align} Defining
\begin{align}\nonumber K_{k,r}^{(n,m)}(t_1,t_2)&=\sum_s e^{-\iota(t_1-t_2)E(s)}\langle n|L_k|s \rangle \langle s|L_r|m \rangle
e^{\iota(E(n)t_1-E(m)t_2)}\\&\nonumber=-\sum_{\alpha\beta\mu\nu}
e^{-\iota(t_1-t_2)E(n-e_\alpha+e_\beta)}\epsilon(k\alpha\beta)\epsilon(r\mu\nu)\\&\nonumber\sqrt{n_\alpha
(n_\beta-\delta_{\alpha\beta}+1)
m_\nu(m_\mu-1+\delta_{\mu\nu})}\delta[n-m-e_\alpha+e_\beta+e_\mu-e_\nu]\\&\nonumber
e^{\iota(E(n)t_1-E(m)t_2)}\end{align} Thus
\begin{align}\nonumber\langle n|W_2(t)|m
\rangle &=-\frac{i}{8}\int_0^t\langle n| (B(t_1)\times
q(t_1))^2|m\rangle
dt_1\\&\nonumber-\frac{1}{8}\sum_{k,r}\int_{0<t_2<t_1<t}K_{k,r}^{(n,m)}(t_1,t_2)B_k(t_1)B_r(t_2)dt_1dt_2\\&\nonumber-\frac{1}{4}\int_{0<t_2<t_1<t}\langle
n|B(t_1).L(t_1)E(t_2).q(t_2)|m \rangle
dt_1dt_2\\&\nonumber-\frac{1}{4}\int_{0<t_2<t_1<t}\langle
n|E(t_1).q(t_1)B(t_2).L(t_2)|m \rangle
dt_1dt_2\\&\nonumber-\frac{1}{2}\int_{0<t_2<t_1<t}\langle
n|E(t_1).q(t_1)E(t_2).q(t_2)|m \rangle dt_1dt_2\end{align} We need
to calculate the general matrix element of$(B(t)\times q(t))^2$,
so
$$(B(t)\times
q(t))^2=\sum_{k\alpha\beta\mu\nu}\{\epsilon(k\alpha\beta)\epsilon(k\mu\nu)B_\alpha(t)B\mu(t)
q_\beta(t)q_\nu(t)\}$$
$$\langle n|(B(t)\times
q(t))^2|m \rangle
=\sum_{k\alpha\beta\mu\nu}\{\epsilon(k\alpha\beta)\epsilon(k\mu\nu)B_\alpha(t)B\mu(t)
\langle n|q_\beta(t)q_\nu(t)|m \rangle\}$$ Define
$$G_{\alpha\mu}^{(n,m)}(t)=\sum_{k\beta\nu}\{\epsilon(k\alpha\beta)\epsilon(k\mu\nu)
\langle n|q_\beta q_\nu|m \rangle\}e^{itE(n,m)}$$ Thus
\begin{align}\nonumber T_2\triangleq \int_o^t \langle n|(B(t)\times q(t))^2|m \rangle
dt_1=\sum_{k,r} \int_o^t
G_{kr}^{(n,m)}(t_1)B_k(t_1)B_r(t_1)dt_1\end{align}
\begin{align}\nonumber T_3 &\triangleq
\int_{0<t_2<t_1<t}\langle n|B(t_1).L(t_1)E(t_2).q(t_2)|m \rangle
dt_1dt_2\end{align} where
\begin{align}\nonumber\langle n|B(t_1).L(t_1)E(t_2).q(t_2)|m
\rangle=\sum_{k,r}B_k(t_1)E_r(t_2)\langle n|L_k(t_1)q_r(t_2)|m
\rangle\end{align} and
\begin{align}\nonumber\langle n|L_k(t_1)q_r(t_2)|m
\rangle &= \langle n|e^{\iota
t_1H_0}L_ke^{-\iota(t_1-t_2)H_0}q_re^{-\iota t_2H_0}|m
\rangle\\&\nonumber=e^{\iota(E(n)t_1-E(m)t_2)}\sum_s \langle
n|L_k|s \rangle \langle s|q_r|m \rangle
e^{-\iota(t_1-t_2)}E(s))\end{align} Let the above equation be
equal to $H_{k,r}^{(n,m)}(t_1,t_2)$. Thus
$$ T_3=\sum_{k,r}\int_{o<t_2<t_1<t}H_{k,r}^{(n,m)}(t_1,t_2)B_k(t_1)E_r(t_2)dt_1dt_2$$
\begin{align}\nonumber T_4 &\triangleq \int_{o<t_2<t_1<t} \langle
n|E(t_1).q(t_1)B(t_2).L(t_2)|m \rangle
dt_1dt_2\\&\nonumber=\sum_{k,r} \int_{o<t_2<t_1<t}
E_k(t_1)B_r(t_2) \langle n|q_k(t_1)L_r(t_2)|m \rangle
dt_1dt_2\end{align}
\begin{align}\nonumber\langle n|q_k(t_1)L_r(t_2)|m \rangle &=
e^{\iota(E(n)t_1-E(m)t_2)} \langle
n|q_ke^{-\iota(t_1-t_2)H_0}L_r|m \rangle
\\&\nonumber=e^{\iota(E(n)t_1-E(m)t_2)} \sum_s \langle n|q_k|s \rangle
\langle s|L_r|m \rangle e^{-\iota(t_1-t_2)E(s)}\\&\nonumber
\triangleq L_{k,r}^{(n,m)}(t_1,t_2)\end{align} So $T_4$ finally
becomes
$$T_4=\sum_{k,r} \int_o^t L_{kr}^{(n,m)}(t_1,t_2)E_k(t_1)B_r(t_2)dt_1dt_2$$
Similarly
\begin{align}\nonumber T_5 &\triangleq \int_{o<t_2<t_1<t}\langle
n|E(t_1).q(t_1)E(t_2).q(t_2)|m \rangle
dt_1dt_2\\&\nonumber=\sum_{k,r} \int_{o<t_2<t_1<t}
E_k(t_1)E_r(t_2) \langle n|q_k(t_1)q_r(t_2)|m \rangle
dt_1dt_2\end{align} we substitute the following in the above
equation
\begin{align}\nonumber\langle n|q_k(t_1)q_r(t_2)|m \rangle
&=e^{\iota(E(n)t_1-E(m)t_2)} \sum_s \langle n|q_k|s \rangle
\langle s|q_r|m \rangle e^{-\iota(t_1-t_2)E(s)}\\&\nonumber
\triangleq \mu_{k,r}^{(n,m)}(t_1,t_2)\end{align} So $T_5$ finally
becomes
$$T_5=\sum_{k,r} \int_o^t \mu_{kr}^{(n,m)}(t_1,t_2)E_k(t_1)E_r(t_2)dt_1dt_2$$
Combining all the above formulae, we get the matrix elements of
$W_2(t)$ as
\begin{align}\nonumber\langle n|W_2(t)|m
\rangle &=-\frac{i}{2}\sum_{k,r} \int_o^t
G_{kr}^{(n,m)}(t_1)B_k(t_1)B_r(t_1)dt_1\\&\nonumber-\frac{1}{8}\sum_{k,r}
\int_o^t
K_{kr}^{(n,m)}(t_1,t_2)B_k(t_1)B_r(t_2)dt_1dt_2\\&\nonumber-\frac{1}{4}\sum_{k,r}
\int_o^t
H_{kr}^{(n,m)}(t_1,t_2)B_k(t_1)E_r(t_2)dt_1dt_2\\&\nonumber-\frac{1}{4}\sum_{k,r}
\int_o^t
L_{kr}^{(n,m)}(t_1,t_2)E_k(t_1)B_r(t_2)dt_1dt_2\\&\nonumber-\frac{1}{2}\sum_{k,r}
\int_o^t
\mu_{kr}^{(n,m)}(t_1,t_2)E_k(t_1)E_r(t_2)dt_1dt_2\end{align}
Define the Kernels
$$ g_{11}(t_1,t_2;n,m,k,r)=-\frac{i}{2}
G_{kr}^{(n,m)}(t_1,t_2)(t_1)\delta(t_1-t_2)-\frac{1}{8}
K_{kr}^{(n,m)}\theta(t_1-t_2)$$
$$ g_{12}(t_1,t_2;n,m,k,r)=-\frac{1}{4}
H_{kr}^{(n,m)}(t_1,t_2)(t_1)\theta(t_1-t_2)-\frac{1}{4}
L_{kr}^{(n,m)}\theta(t_2-t_1)$$
$$g_{22}(t_1,t_2;n,m,k,r)=-\frac{1}{2}
\mu_{kr}^{(n,m)}(t_1,t_2)\theta(t_1-t_2) $$ The general matrix
element of $W_2(t)$ becomes
\begin{align}\langle n|W_2(t)|m
\rangle &=\sum_{k,r} \int_{[0,t]^2}
g_{11}(t_1,t_2;n,m,k,r)B_k(t_1)B_r(t_2)dt_1dt_2\\&\nonumber+\sum_{k,r}
\int_{[0,t]^2}
g_{12}(t_1,t_2;n,m,k,r)B_k(t_1)E_r(t_2)dt_1dt_2\\&\nonumber+\sum_{k,r}
\int_{[0,t]^2}
g_{22}(t_1,t_2;n,m,k,r)E_k(t_1)E_r(t_2)dt_1dt_2\end{align}
Likewise
\begin{align}\nonumber\langle n|W_1(t)|m \rangle =
-\frac{i}{2}\sum_{k} \int_o^t B_k(t_1) \langle n|L_k(t_1)|m
\rangle dt_1-i \sum_{k} \int_o^t E_k(t_1) \langle n|q_k(t_1)|m
\rangle dt_1\end{align} $$\langle n|L_k(t)|m \rangle=e^{\iota
E(n,m)t} \langle n|L_k|m \rangle \triangleq f(t,n,m,k)$$
$$\langle n|q_k(t)|m \rangle=e^{\iota E(n,m)t} \langle n|q_k|m
\rangle \triangleq g(t,n,m,k)$$ we now introduce creation kernels
which enable us to display explicitly the linear and quadratic
dependence of the evolution operator upto $O(e^2)$ on the electric
and magnetic fields. Thus, we define
$$h_1(t,n,m,k)=-\frac{i}{2}f(t,n,m,k)$$
$$h_2(t,n,m,k)=-i g(t,n,m,k)$$
\begin{align}\nonumber\langle n|W_1(t)|m \rangle=\sum_{k} \int_o^t
h_1(t,n,m,k) B_k(t_1) dt_1+\sum_{k} \int_o^t h_2(t,n,m,k) E_k(t_1)
dt_1\end{align} and
\begin{align}\nonumber\langle n|W_2(t)|m
\rangle &=\sum_{k,r} \int_{[0,t]^2}
g_{11}(t_1,t_2;n,m,k,r)B_k(t_1)B_r(t_2)dt_1dt_2\\&\nonumber+\sum_{k,r}
\int_{[0,t]^2}
g_{12}(t_1,t_2;n,m,k,r)B_k(t_1)E_r(t_2)dt_1dt_2\\&\nonumber+\sum_{k,r}
\int_{[0,t]^2}
g_{22}(t_1,t_2;n,m,k,r)E_k(t_1)E_r(t_2)dt_1dt_2\end{align} where,
$h_1, h_2, g_{11}, g_{12}, g_{22}$ are given. The expression for
general matrix element of $U(t)$ can be calculated by
\begin{align}exp\{itE(n)\}\langle n|U(t)|m \rangle= \delta{[n-m]}+e\langle
n|W_1(t)|m \rangle +e^2 \langle n|W_2(t)|m \rangle
+O(e^3)\end{align} Average value of an observable $X$ is given by
$$\langle X \rangle(t)=\Tr\{ \rho(t)X\}=Tr\{U^*(t)\rho(0)U(t)X\}$$
$$=Tr\{\rho(0)X(t)\}$$ where,
$$X(t)=U(t)XU^*(t)=exp(-itH_0)W(t)XW^*(t)exp(itH_0)$$ So, taking
$\rho(0)=|n\rangle \langle n|$, we get
$$\langle X \rangle(t)=\langle
n|exp(-itH_0)W(t)XW^*(t)exp(itH_0)|n \rangle$$
$$=\langle
n|W(t)XW^*(t)|n \rangle$$
$$=\langle
n|\{I+eW_1(t)+e^2W_2(t)\}X\{I+eW_1^*(t)+e^2W_2^*(t)\}|n \rangle$$
$$= 1+e(\langle n|W_1(t)|n \rangle + \langle n|W_1^*(t)|n
\rangle)$$
$$+e^2(\langle
n|W_1(t)XW_1^*(t)|n \rangle +\langle n|W_2(t)X)|n \rangle+\langle
n|XW_2^*(t)|n \rangle+O(e^3)$$
$$=1+2eRe(\langle n|W_1(t)|n \rangle)$$
$$+e^2\{(\sum_{s,r} \langle
n|W_1(t)|s \rangle \langle r|W_1^*(t)|n \rangle X[s,r])$$
$$+2Re(\sum_m \langle n|W_2(t)|m \rangle X[m,n])\} +O(e^3)$$
These equations can be used to estimate the electric and magnetic
fields from measurement of $\langle X(t) \rangle$. The above
formulae for $\langle X\rangle(t)$ are only of subsidiary interest
as our main aim is approximately a given unitary operator gate by
the obtained unitary evolution matrix.
\section{Determining the Optimal Electric and Magnetic Fields for Gate Design}
We now propose an approximate algorithm for determining the
electric and magnetic fields, $E(t),B(t),0 \leq t \leq T$, so that
the evolved unitary gate $U(T)$ best approximates a desired
unitary gate $U_d$ in Fig. 1. The designed gate is given by
\begin{align}U(T)=U_0(T)(I + eW_1(T)+e^2W_2(T))\end{align}
We had calculated $W_1(T)$ and $W_2(T)$ in section 3 from eq.
(3.18), which is given by
\begin{align}\langle n|W_1(T)|m\rangle=\sum_k\int_0^T(h_1(t,n,m,k)B_k(t_k)
+ h_2(t,n,m,k)E_k(t_k))dt_k\end{align} and
\begin{align}\nonumber\langle n|W_2(t)|m \rangle &=\sum_{k,r} \int_{[0,T]^2}
g_{11}(t_1,t_2;n,m,k,r)B_k(t_1)B_r(t_2)dt_1dt_2\\&\nonumber+\sum_{k,r}
\int_{[0,T]^2}
g_{12}(t_1,t_2;n,m,k,r)B_k(t_1)E_r(t_2)dt_1dt_2\\&\nonumber+\sum_{k,r}
\int_{[0,T]^2}
g_{22}(t_1,t_2;n,m,k,r)E_k(t_1)E_r(t_2)dt_1dt_2\end{align} Define
\begin{align}h(t,m,n,k)=
\begin{bmatrix}
h_1(t,n,m,k) \\
h_2(t,n,m,k)
\end{bmatrix}\end{align} and
\begin{align}\xi_k(t)=
\begin{bmatrix}
B_k(t) \\
E_k(t)
\end{bmatrix},1 \leq k \leq 3 \end{align}
Further, we have
$$h(t,n,m)=
\begin{bmatrix}
h(t,n,m,1) \\
h(t,n,m,2) \\
h(t,m,n,3)
\end{bmatrix}$$and
$$\xi(t)=
\begin{bmatrix}
\xi_1(t) \\
\xi_2(t) \\
\xi_3(t)
\end{bmatrix}=\begin{bmatrix}
B_1(t) \\
E_1(t) \\
B_2(t) \\
E_2(t) \\
B_3(t) \\
E_3(t) \\
\end{bmatrix}$$
This definition of $\xi(t)$ as the $6$- components electromagnetic
field vector will enable us to express the unitary evolution
operator explicitly as a second degree Volterra functional of
$\xi(.)$ using which the gate error energy optimization is readily
carried out.

Substituting eqs. (3.28) and (3.29) in eq. (3.27), we get
\begin{align}\langle n|W_1(T)|m \rangle =
\sum_k\int_0^Th(t,n,m,k)^T\xi_k(t)dt=\int_0^T
h(t,n,m)^T\xi(t)dt\end{align} Let $\{e_\alpha\}_{\alpha=1}^6 $ be
the standard ordered basis for
$\mathcal{R}^6:e_\alpha=[\delta_{\alpha1},\delta_{\alpha2},.,\delta_{\alpha6}]^T$
Thus, $B_k(t)$ and $E_k(t)$ concisely become
$$ B_k(t)=e_{2k-1}^T.\xi(t)$$
$$ E_k(t)=e_{2k}^T.\xi(t)$$
So, on making above substitution for $\langle n|W_2(t)|m\rangle$,
we get
\begin{align}\nonumber\langle n|W_2(t)|m \rangle &=\sum_{k,r} \int_{[0,T]^2}
g_{11}(t_1,t_2;n,m,k,r)(e_{2k-1}^T \otimes e_{2r-1}^T)(\xi(t_1)
\otimes \xi(t_2))dt_1dt_2\\&\nonumber+\sum_{k,r} \int_{[0,T]^2}
g_{12}(t_1,t_2;n,m,k,r)(e_{2k-1}^T \otimes e_{2r}^T)(\xi(t_1)
\otimes \xi(t_2))dt_1dt_2\\&+\sum_{k,r} \int_{[0,T]^2}
g_{22}(t_1,t_2;n,m,k,r)(e_{2k}^T \otimes e_{2r}^T)(\xi(t_1)
\otimes \xi(t_2))dt_1dt_2\end{align} Equivalently, define the
vectors
$$g_{11}(t_1,t_2,n,m)=\sum_{k,r}g_{11}(t_1,t_2,n,m,r)(e_{2k-1} \otimes
e_{2r-1})$$
$$g_{12}(t_1,t_2,n,m)=\sum_{k,r}g_{12}(t_1,t_2,n,m,r)(e_{2k-1} \otimes
e_{2r})$$
$$g_{22}(t_1,t_2,n,m)=\sum_{k,r}g_{22}(t_1,t_2,n,m,r)(e_{2k} \otimes
e_{2r})$$ So $W_2(t)$ finally becomes \begin{align}\langle
n|W_2(t)|m \rangle= \int_{[0,T]^2} g(t_1,t_2;n,m,)^T(\xi(t_1)
\otimes \xi(t_2))dt_1dt_2\end{align} where,
$g(t_1,t_2;n,m)=g_{11}(t_1,t_2;n,m)+g_{12}(t_1,t_2;n,m)+g_{22}(t_1,t_2;n,m)\in
\mathcal{R}^9$. The gate error energy function has been calculated
by
\begin{align}\nonumber||U_d-U(T)||^2&=||U_d -U_0(T)(I+eW_1(T)+e^2W_2(T))||^2+\mathcal{O}(e^3)\\&=||W_d
-eW_1(T)-e^2W_2(T)||^2+\mathcal{O}(e^3)\end{align} where,
$W_d=U_0(-T)U_d-I$. On expanding eq. (3.33) upto second order
terms, we get
$$||U_d-U(T)||^2=||W_d||^2+e^2(||W_1(T)||^2-2Re(Tr(W_d^*W_2(T))))$$
$$-2eRe(Tr(W_d^*W_1(T))) +\mathcal{O}(e^3)$$
The gate error energy function is given by
\begin{align}\nonumber\mathbb{E}(\xi(.))&\triangleq
||U_d-U(T)||^2-||W_d||^2=-2e\sum_{n,m}Re\{\langle n|W_d^*|m
\rangle \langle m|W_1(T)|n \rangle \}\\&\nonumber+e^2\left\{
\sum_{n,m} |\langle n|W_1(T)|m \rangle-2e\sum_{n,m}Re\{\langle
n|W_d^*|m \rangle \langle m|W_2(T)|n \rangle
\}\right\}+\mathcal{O}(e^3)\\&\nonumber=-2e\sum_{m,n}Re\{\bar{\langle
n|W_d|m \rangle} \int_0^T h(t,m,n)^T \xi(t)
dt\}+e^2\{\sum_{m,n}|\int_0^T h(t,m,n)^T \xi(t)
dt|^2\\&\nonumber-2\sum_{m,n}Re\{\bar{\langle n|W_d|m \rangle}
\int_{[0,T]^2} g(t_1,t_2,m,n)^T (\xi(t_1)\otimes \xi(t_2))
dt_1dt_2
\}\}+\mathcal{O}(e^3)\\&\nonumber=\int_0^T(-2e\sum_{m,n}Re\{\bar{\langle
n|W_d|m \rangle} h(t,m,n)\})^T\xi(t) dt\\&\nonumber+\int_{[0,T]^2}
(e^2Re\sum_{m,n}h(t_1,n,m)\otimes\bar{h}(t_2,n,m)0^T(\xi(t_1)\otimes
\xi(t_2)) dt_1 dt_2
\\&\nonumber-\int_{[0,T]^2}(2e^2\sum_{m,n}Re\{\bar{\langle n|W_d|m
\rangle} g(t_1,t_2,m,n)\})^T (\xi(t_1)\otimes \xi(t_2))dt_1 dt_2
\\&=\int_0^T \alpha(t)^T\xi(t)dt+\int_{[0,T]^2}\beta(t_1,t_2)^T
(\xi(t_1)\otimes \xi(t_2)) dt_1 dt_2\end{align} where,
$\alpha(t)=-2e\sum_{m,n}Re\{\bar{\langle n|W_d|m \rangle}
h(t,m,n)\}$ and
$\beta(t_1,t_2)=e^2\sum_{m,n}Re(h(t_1,n,m)\otimes\bar{h}(t_2,n,m))-2e^2\sum_{m,n}Re\{\bar{\langle
n|W_d|m \rangle} g(t_1,t_2,m,n))$. Minimizing the gate error
energy function $\mathbb{E}(\xi(.)$ subject to quadratic
constraint, given by
$$\int_{[0,T]^2}\xi(t_1)^T Q(t_1,t_2) \xi(t_2) dt_1 dt_2 = \varepsilon_0$$
or equivalently, $$\int_{[0,T]^2} q(t_1,t_2)^T(\xi(t_1)\otimes
\xi(t_2)) dt_1 dt_2 =\varepsilon_0$$ where,
$q(t_1,t_2)=Vec(Q(t_1,t_2)) \in \mathcal{R}^{36}$. Let $\lambda$
be the Lagrange multiplier. Using Lagrange multiplier approach,
the optimization problem is to minimize.
\begin{align}\nonumber\mathbb{E}(\xi(.),\lambda)&=\int_0^T \alpha(t)^T \xi(t)dt +
\int_{[0,T]^2}\beta(t_1,t_2)^2(\xi(t_1)\otimes \xi(t_2)) dt_1
dt_2\\&-\lambda\bigg(\int_{[0,T]^2}q(t_1,t_2)^T(\xi(t_1)\otimes
\xi(t_2)dt_1 dt_2 - \varepsilon_0\bigg)\end{align} The optimal
solution is given by $\frac{\delta \mathbb{E}}{\delta\xi(t)}=0$.
Solving the eq. (3.35) with $e=1$, we get
\begin{align}\nonumber &\alpha(t)+\int_0^T(I_6\otimes\xi(t_1))^T(\beta(t,t_1)-\lambda
q(t,t_1)) \\&+ \int_0^T(\xi(t_1)\otimes I_6)(\beta(t_1,t)-\lambda
q(t_1,t))dt_1=0,\quad 0\leq t \leq T
\end{align}
Note, we can have
\begin{align}\nonumber(I_6\otimes\xi^T)\eta =(\delta_{jk}\xi^T)\eta=(\delta_{jk}\xi^T)\sum_{\alpha,\beta=1}^6(\eta^T e_\alpha\otimes
e_\beta)e_\alpha\otimes e_\beta =\sum_{\alpha,\beta=1}^6(\eta^T
e_\alpha\otimes e_\beta)e_\alpha e_\beta^T \xi\end{align} and
\begin{align}\nonumber(\xi_T\otimes I_6)\eta=(\xi_T\otimes
I_6)\sum_{\alpha,\beta=1}^6(\eta^T e_\alpha\otimes
e_\beta)e_\alpha\otimes e_\beta =\sum_{\alpha,\beta=1}^6 \eta^T
(e_\alpha\otimes e_\beta)e_\beta e_\alpha^T \xi\end{align} Thus,
the condition for the minimization becomes
\begin{align}\alpha(t)+\int_0^T
R_\lambda(t,t_1)\xi(t_1)dt_1=0 ,0\leq t \leq T\end{align} where,
\begin{align}\nonumber R_\lambda(t,t_1)=\sum_{\alpha,\beta=1}^6(\beta(t,t_1)-\lambda
q(t,t_1))^T(e_\alpha\otimes e_\beta)(e_\alpha e_\beta^T+e_\beta
e_\alpha^T)=P(t,t_1)-\lambda S(t,t_1)\end{align} Here, $P(t,t_1)$
and $S(t,t_1)$ is given by
\begin{align}P(t,t_1)=\sum_{\alpha,\beta=1}^6 \beta(t,t_1)^T(e_\alpha\otimes
e_\beta)(e_\alpha e_\beta^T+e_\beta e_\alpha^T)\end{align} and
\begin{align}S(t,t_1)=\sum_{\alpha,\beta=1}^6 q(t,t_1)^T(e_\alpha\otimes
e_\beta)(e_\alpha e_\beta^T+e_\beta e_\alpha^T)\end{align} On
substitution of the minimization condition of eqs. (3.38) and
(3.39) in eq. (3.37), we get
\begin{align}\alpha(t) + \int_0^T(P(t,t_1)-\lambda S(t,t_1))\xi(t_1)
dt_1\end{align} Discretization of eq.(3.40) leads to
\begin{align} \nonumber(P- \lambda S)\xi +&\alpha=0\\&\xi= -(P- \lambda S)^{-1}\alpha\end{align}
Applying constraint on the gate error energy function $\mathbb{E}$
with respect to $\lambda$, that is,
$$\frac{\delta \mathbb{E}}{\delta \lambda}=0$$ This gives the
constraint
\begin{align}\int_{[0,T]^2}q(t_1,t_2)^T(\xi(t_1)\otimes
\xi(t_2))dt_1 dt_2 =\varepsilon_0\end{align} Eq. (3.42) on
discretization leads to
$$q^T(\xi \otimes \xi)= \varepsilon_0$$
Substituting for $\xi$ from eq. (3.41), we get
$$a^T((P- \lambda S)^{-1} \otimes (P- \lambda S)^{-1})(\alpha \otimes
\alpha)=-\varepsilon_0$$ or
$$a^T(\text{adj}(P- \lambda S) \otimes \text{adj}(P- \lambda S))(\alpha
\otimes \alpha)+(\text{det}(P- \lambda S)^{2}\varepsilon_0 =0$$
This is a polynomial equation for $\lambda$ and can be solved.
\section{Analysis of Reliability of the Designed Quantum Gate in the Presence of Heat Bath}
\subsection{Heat Bath Perturbation}
On heat bath perturbation [51], the oscillator interacting with a
classical electromagnetic field $E(t), B(t)$ has a Hamiltonian,
given by
\begin{align}\nonumber H(t,q,p)= \frac{1}{2}(p+eA(t,q))^2+\frac{1}{2}q^2-e
\Phi(t,q)\end{align} In addition to this coupling with the heat
bath involves modelling the bath as a quantum electromagnetic
field with creation and annihilation operator $a^+_{Bk}$ and
$a_{Bk}$, $k=1,2,3 \cdots N$ respectively. The interaction
Hamiltonian has the form
\begin{align}\nonumber H_{int}= \alpha (p,a_B+a_B^+)= \alpha \sum\limits_k p_k (a_{Bk}+a^+_{Bk})\end{align}
The heat bath itself has Hamiltonian $H_B= \beta \sum
\limits_{k=1}^3 w_k a_{Bk}^+ a_{Bk}$. The overall state of the
oscillator and bath can be expressed as linear combination of $|
n_1 n_2 n_3 m_1 m_2 m_3\rangle$, where if $a_k= \frac{q_k+
ip_k}{\sqrt{2}},$ then
$$a_k^+ a_k |n_1 n_2 n_3 m_1 m_2 m_3\rangle= n_k|n_1 \cdots
m_3\rangle$$
$$a^+_{Bk} a_{Bk}|n_1 n_2 n_3 m_1 m_2 m_3\rangle= m_k w_k|n_1 \cdots m_3\rangle$$
where, $n_k, m_k= 0,1,2, \cdots $. The overall interaction
evolution operator is now (the interaction between oscillator and
classical electromagnetic field)
\begin{align}V_T(t)= V(t) + \alpha \sum\limits_{k=1}^3
p_k(t)(a_{Bk}e^{i w_k t}+ a_{Bk}^+ e^{-i w_k t})\end{align} where,
$V(t)$ is given by eq. (3.16). The charge in the gate generator
caused by this interaction of the oscillator with the quantum
electromagnetic field is then upto linear orders in $(a_{Bk},
a_{Bk}^+)$
\begin{align}\nonumber
 & \int_0^T  w \langle n_1^{'} n_2^{'} n_3^{'} m_1^{'} m_2^{'} m_3^{'}|
\alpha \sum\limits_{k=1}^3 p_k(t)(a_{Bk}e^{iw_k t}+ a_{Bk}^+
e^{-iw_kt})|. n_1 n_2 n_3 m_1 m_2 m_3\rangle dt\\ \nonumber & =
\int_0^T w \alpha \sum_{k=1}^3 \langle
n_1^{'}n_2^{'}n_3^{'}|p_k(t)|n_1 n_2 n_3\rangle\langle m_1^{'}
m_2^{'} m_3^{'}| a_{Bk} e^{iw_k t}+ a^+_{Bk}e^{-i w_k t}| m_1 m_2
m_3\rangle\end{align} After calculating thus, we average over the
bath state described by a matrix element $\rho_B
\big[m_1^{'}m_2^{'}m_3^{'}|m_1 m_2 m_3\big]$ to get the change in
the generator due to bath oscillator interaction as
\begin{align}\nonumber
& \alpha \sum\limits_{k=1}^3\int_0 ^T \langle
n_1{'}n_2{'}n_3{'}|p_k(t)|n_1 n_2 n_3\rangle \sum\limits_{m,m^{'}}
\rho_B[m^{'}|m]\langle m^{'}|a_{Bk}e^{iw_k t}+ a^+_{Bk}e^{-i w_k
t}|m\rangle dt
\\ \nonumber &= \alpha \sum\limits_{k=1}^3
\int_0^T\langle n^{'}|p_k(t)|n\rangle T_r\{\rho_B(a_{Bk}e^{- w_k
t}+ a^+_{Bk} e^{-i w_k t})\} dt
\end{align}
Essentially this amounts to taking the partial trace of the
oscillator bath generator over the bath variable. This should
small in magnitude for reliable performance.
\subsection{The Approximation of Atoms and Molecules in Equilibrium by Oscillator Models in the
Small Oscillation Approximation} Consider an electron in an atom
or a molecule with Hamiltonian $H_0= \frac{p^2}{2m}+V(q)$. If the
electron is localized around an equilibrium  $q_0$, then $\nabla
V(q_0)=0$ and for small perturbations $\delta q$ around this
position, the Hamiltonian is approximately given by
\begin{align}\nonumber H_0=\frac{p^2}{2m}+ V(q_0)+ \frac{1}{2}\delta q^T \nabla V(q_0)+
\frac{1}{2}\delta q^T(\nabla \nabla^T V(q_0))\delta q\end{align}
and since $\nabla V(q_0)=0$, we get by neglecting the constant
$V(q_0)$, $H_0= \frac{p^2}{2m}+ \frac{1}{2} q^T K q$, where,
$\delta q$ is renamed as $q$ and $K= \nabla \nabla^T V(q_0)=
\bigg(\frac{\partial ^2 V(q_0)}{\partial q_{\alpha}\partial
q_{\beta}}\bigg)_{1\leq \alpha, \beta \leq 3}$. The Hessian matrix
of V, namely K is positive definite since V attains its minimum at
$q_0$. By making a canonical orthogonal transformation to $(q,p)$,
we get
\begin{align}\nonumber H_0 = \frac {p^2}{2m}+ \frac{1}{2}\sum\limits_{j=1^3}
\lambda_j q_j^2= \frac{1}{2m} \sum_{j=1}^3 (p^2 + \lambda_j
q_j^2)\end{align} which is a $3$-D oscillator. In other words, the
small oscillation about equilibrium approximation results in a
harmonic oscillator approximation to the Hamiltonian of the
unperturbed system. Example
$V(q)=\frac{a}{|q_0|^{\alpha}}-\frac{b}{|q_0|^{\beta}}$, describes
a molecular potential for $\alpha
> \beta$. The equilibrium $q_0$ satisfies, $\nabla V (q_0)=0$ or
\begin{align}\nonumber\frac{-\alpha a}{|q_0|^{\alpha +1}}+ \frac{\beta
b}{|q_0|^{\beta+1}}=0\end{align} or
\begin{align}\nonumber|q_0|^{\alpha-\beta}=\frac{\alpha a}{\beta
b}\end{align}or
\begin{align}\nonumber|q_0|=\bigg(\frac{\alpha a}{\beta b}\bigg)^{\frac{1}{(\alpha-\beta)}}\end{align}
The fact that oscillators are approximations to atoms and
molecules combined with the fact that experiments related to
exposing atoms and molecules to radiation fields are easily
performed in the physical laboratory and accelerators makes the
approach of our quantum gate design a very much physically
realizable possibility compared with other schemes such as ion
trap models.
\section{Conclusions and Scope for Future Work}
We have perturbed a $3$-D oscillator with a spatially constant,
but time varying electromagnetic field and by applying
time-dependent perturbation theory upto $O(e^2)$ calculated the
unitary evolution operator after time $T$ upto $O(e^2)$. This
operator depends linearly and quadratically on the six functions
$(E_1(t), E_2(t), E_3(t), B_1(t), B_2(t), B_3(t))$, namely the
perturbing electromagnetic field components. By applying a
quadratic energy constraint on the electromagnetic field we have
upto $O(e^2)$ minimized the Frobenius error norm square between
this evolution operator and a given finite dimensional unitary
operator. The evolution operator has been calculated relative to
the eigenbasis of the unperturbed oscillator Hamiltonian and has
been truncated to finite dimension. The result is a set of linear
integral equations for the optimal electromagnetic field with a
Lagrange multiplier, that is, determined numerically. As a more
sophisticated example, we finally discuss gate design using a
quantum electromagnetic field acting on a quantum harmonic
oscillator. The electromagnetic field is modelled by three
creation and annihilation operators modulated by a scalar function
of time and the gate error energy is minimized with respect to
this function.
\chapter{REALIZATION OF THE THREE-QUBIT QUANTUM CONTROLLED GATE BASED ON MATCHING HERMITIAN GENERATORS}
\section{Introduction}
\label{intro}The final contribution of this thesis is to designing
controlled gates (that is, non-separable system) using the quantum
mechanics of an electron bound to nucleus in electromagnetic
fields. But the major problems are how to control a quantum system
from its initial state to a target state. Such a control problem
can ultimately be changed to the problem of generating a series of
specific unitary evolution operators for a given target states of
a quantum system. So for realizing the gate, first we obtain the
stationary state energy levels and then the evolution of an
initial wave packet of the unperturbed quantum system. The desired
unitary operators of a given target state can be obtained by using
appropriate decompositions. These states could take the form of
the spin of an electron along a given direction is an example of a
qubit [18, 19, 40]. The study of quantum computation could lead to
a better understanding of the principles common to all quantum
systems.\par The study of the perturbed anharmonic oscillator has
a great importance in quantum mechanics. Many researchers have
done important investigations on this problem. The use of the term
gates when describing quantum gates should be taken conceptually.
As we will see, transformations on qubits are not necessarily
applied with gates in the conventional sense. Because of the
superposition phenomenon, qubit states are expressed not as bits
but as vectors of bits. Therefore, quantum gates actually perform
transformations on complex vectors.
\par In quantum mechanics, a coherent state is the specific
quantum state of the quantum harmonic oscillator which was first
used by Roy Glauber in the field of quantum optics. This change of
state may include change in the shape of the wave function.
Coherent states are the eigenstates of the annihilation operator.
Using the eigenstates of a harmonic oscillator as a substratum for
realizing complex gates is natural since these sequences of
eigenstates can be generated by successively applying a creation
operator to the preceding eigenstates. Here, the time dependent
evolution of a quantum system, introduced by Dirac, known as the
interaction representation is discussed. In this representation,
both operators and states move in time. The interaction
representation is particularly useful in problems involving time
dependent potentials acting on a system. It also provides a route
to the whole apparatus of quantum field theory. In the interaction
representation operators move with $H_0$, the unperturbed
Hamiltonian. A harmonic oscillator acted on by an external time
dependent force is interesting for two reasons. First, it is a
model for actual physical phenomena such as the quantum radiation
from a known current. Second, it provides an excellent case where
high order calculations can be carried out analytically. Prior to
studying harmonic oscillators perturbed by an electric field, we
look at the general problem of computing the evolution of a
quantum system having a Hamiltonian operator of the form $H_0 +
\epsilon V(t)$ where $H_0$ is known and is the Hamiltonian of a
quantum system in the Hilbert space $ \mathcal {H} $ (which may be
finite dimensional, in which case $H_0$ is a Hermitian operator),
$\epsilon V(t)$; $ 0 \leq t\leq T$ is the perturbing potential
where $\epsilon$ is a small parameter [25, 44].\par Due to such
perturbations, the quantum system considered is simulated and as a
consequence, changes its states. The importance of perturbing a
quantum system with Hamiltonian $H_0$ by a potential $f(t)V$ lies
in the fact that using $H_0$ alone, we can generate a one
dimensional Lie group $e^{-\iota t H_0}$, $t \in\\mathbb{Re}$ of
unitary gates while with $H_0+f(t)V$, we can generate the Lie
group having an $N$ dimensional Lie algebra spanned by $H_0$, $V$,
and all commutators of $H_0$ with $V$, which can even be an
infinite dimensional Lie algebra. So the advantage of perturbation
theory is mainly to increase the dimensionality of the unitary
group of gates realizable by a quantum physical system from $1$ to
$N$, where $N$ can even be infinity. Perturbation theory is one of
the most important methods for obtaining approximate solutions to
the Schr\"odinger equation [46].
\subsection{Novelty of Matching Generator}
The first novelty of our method is that we use an infinite
dimensional system like the quantum harmonic oscillator to design
finite dimensional gates by truncation. Finite state systems can
in practice be realized using the spin states of elementary
particles. To realize infinite dimensional gates, we need to use
observables like position $q$ and momentum $p$ that act in
$L^2(\mathbb{R})$. The second novel feature of the work is that
optimization needs to be carried out only over the discrete
frequency samples of the control input Fourier transform. The
third novelty of this work include designing gates using an ion
modelled as a spin system interacting with a classical and a
quantum electromagnetic field. The final novel feature deals with
the design of separable and weakly non-separable gates using a sum
of independent Hamiltonian without and with an interaction acting
on a tensor product space.
\par The significant contribution of the last problem is to show how by
using a real physical system such as an atom or a molecule
(modelled as a quantum harmonic oscillator for small displacements
of the electron from its equilibrium position) we can, by applying
an electric field, create unitary gates used in quantum
computation with a high degree of accuracy. This problem in
particular shows that by truncating an infinite dimensional
quantum system to finite dimensions, we can design quantum gates
using matching Hermitian generators for a non-separable system.
After illustrating how an arbitrary unitary gate can be realized
approximately using a perturbed Hamiltonian, we discuss
qualitatively some issues regarding how non-separable systems can
be realized using respectively independent Hamiltonians and
independent Hamiltonians with an interaction. Specifically, the
theory developed in our thesis shows that given a unitary gate
which is a small perturbation of a separable unitary gate, we can
realize the separable component using a direct sum of two
independent Hamiltonians and then add a small interaction
component to this direct sum in such a way as to cause the error
between the desired unitary gate and the realized gate to be as
small as possible. In other words, we justify that the time
dependent perturbation theory of independent quantum systems is a
natural way to realize non-separable unitary gates which are small
perturbations of separable gates. Examples of separable and
non-separable systems taken from standard textbooks on quantum
computation are given using respectively tensor products of
unitaries and the controlled unitary gates. In this case we
qualitatively discuss the realization using independent
Hamiltonians and independent Hamiltonians with a small interaction
of order $\epsilon$ with the evolution operator computed upto
$O(\epsilon^2)$ using the Dyson series. Both examples (namely,
first perturbing of harmonics with time dependent electric field
and second, the ion trap model involving perturbation of
spin-magnetic energy with a single mode quantum electromagnetic
field) combined in this problem illustrate the general philosophy
adopted-namely simulate the generator of a quantum gate by
perturbing a time independent system Hamiltonian with a small time
dependent interaction Hamiltonian of the system with an external
force [28, 29, 35, 36].
\section{Mathematical Studies of Matching Generator}
In this section, we shall describe how quantum gates can be
designed based on operators derived completely form the
interaction picture Hamiltonian $\widetilde{V}(t)$. Specifically,
we calculate the generator in the interaction picture where the
interaction Hamiltonian is of the form $\epsilon\varphi(t) V$ with
$\varphi(t)$ be a control function. We calculate the Dyson series
approximate evolution operator upto $O(\epsilon)$ and then
repeatedly apply this approximate interaction evolution operator
to realize the given unitary gate. The optimization for the
perturbed harmonic oscillator case is carried out conveniently
with respect to the Fourier samples of $\varphi$. The generator of
a quantum unitary gate $U_g$ of size $(N+1)\times (N+1)$ is a
Hermitian matrix $H_g\in C^{(N+1)\times(N+1)}$ which satisfies
\begin{eqnarray}U_g=e^{-\iota H_g}. \label{eq1}\end{eqnarray} Taking $\ln$ of both sides in eq. (4.1), we get
\begin{eqnarray}H_g=\iota\ln(U_g). \label{eq2}\end{eqnarray}Specifying $U_g$ is therefore
 equivalent to specifying $H_g$. For example, if $U_g$ is the DFT
 matrix \begin{eqnarray} U_g = \frac{1}{\sqrt{N}}\bigg\|e^ {\iota\frac{2\pi k n}{N}}\bigg\|_{0\leq k, n\leq
 N-1}. \label{eq3}\end{eqnarray} then one can compute $H_g$ by applying the spectral theorem
 to $U_g$. Specifically, if \begin{eqnarray} U_g = \sum _{\alpha = 1}^r e^{\iota
 \lambda_\alpha}P_\alpha. \label{eq4}\end{eqnarray} where $\{ P_\alpha\}_{\alpha = 1}^r$
 form a resolution of identity, that is, $P_\alpha ^2 = P_\alpha =
 P_\alpha^\star$, $ P_\alpha P_\beta = 0$ for $\alpha \neq \beta$
 and $\sum_{\alpha = 1}^r P_\alpha = I_N$, then we can take
\begin{eqnarray}\nonumber H_g = \sum_{\alpha = 1}^r \lambda_\alpha P_\alpha\end{eqnarray}
To design $U_g$ (or equivalently $H_g$), consider an infinite
dimensional quantum system with Hamiltonian
\begin{eqnarray}H(t)=H_0+\epsilon\varphi(t) V(t),\quad t\geq0. \label{eq5}\end{eqnarray} where $H_0$ is
unperturbed Hamiltonian matrix to evolve with a small perturbation
parameter $\epsilon$, a modulating signal $\varphi (t)$ and an
interaction potential (which is also a Hermitian matrix) $V(t)$,
that is, $\epsilon \varphi (t) V(t)$. If this system evolves for a
duration $T$, then upto $O(\epsilon)$, the corresponding unitary
evolution operator $U(t)$ satisfies the following Schrodinger
equation [11-14].
\begin{eqnarray} U^{'}(t) &= - \iota \big(H_0 + \epsilon \varphi (t)
 V\big)U(t);\quad t\geq 0\label{eq6}\end{eqnarray} \\where \begin{eqnarray} U(0)= I. \label{eq7}\end{eqnarray} The unitary evolution
operator is approximately given by \begin{eqnarray}U(T)= e^{-\iota
T H_0}W(T)\end{eqnarray} where $W(T)$ is the interaction picture
evolution operator and expanding using Dyson series upto first
order, we get
\begin{eqnarray} W(T) = I - \iota \epsilon \int_0^T \varphi
(t)\widetilde{V}(t) dt + O(\epsilon^2). \label{eq7}\end{eqnarray}
  and \begin{eqnarray}\widetilde{V}(t) = e^{\iota t H_0} V(t)
 e^{-\iota t H_0}. \label{eq8}\end{eqnarray}
In the standard terminology of quantum mechanics, $W(T)$ is the
evolution operator in the interaction picture. We can obtain
$W(T)$ from $U(T)$ by removing the effect of $H_0$, that is,
\begin{eqnarray}W(T) = e^{\iota TH_0}U(T) = U_0(-T)U(T)\end{eqnarray}
Equivalently the `interaction component' of $U(T)$ is
\begin{eqnarray}U_{\varphi,\epsilon}=e^{\iota T H_0}U(T)=W(T)\end{eqnarray} If we apply
$U_{\varphi,\epsilon}$ $m$ times then the realized unitary gate is
given by \begin{eqnarray}U_{\varphi,\epsilon}^m =
\bigg(I-\iota\epsilon \int_0^T
\varphi(t)\widetilde{V}(t)dt+O(\epsilon^2)\bigg)^m\end{eqnarray}
With $\epsilon=\frac{1}{m}$, we get
\begin{eqnarray}U_\varphi=\lim_{\atop m\rightarrow\infty}
U_{\varphi,\frac{1}{m}}^m = e^{-\iota
\int_0^T\varphi(t)\widetilde{V}(t)dt}\end{eqnarray} which is an
exact unitary gate. The generator of $U_{\varphi}$ is thus
\begin{eqnarray}H_\varphi = \iota ln U_\varphi=\int_0^T
\varphi(t)\widetilde{V}(t)dt\end{eqnarray} In other words,
$U_\varphi$ can be realized by applying
$U_{\varphi,\epsilon}=W(T)=I-\frac{\iota}{m}\int_0^T
\varphi(t)\widetilde{V}(t)dt$, $m=\frac{1}{\epsilon}$ times ($m$
is large) while the given unitary gate $U_g$ can be realized by
applying $I-\frac{\iota}{m} H_g$, $m$ times. Thus, approximating
$U_g$ by $U_\varphi$ is equivalent to approximating
$I-\frac{\iota}{m}\int_0^T \varphi(t)\widetilde{V}(t)dt$ by
$I-\frac{\iota}{m}H_g$, or equivalently by approximating $H_g$ by
$\int_0^T\varphi(t)\widetilde{V}(t)dt$.\par Therefore, the task is
to design $\{\varphi(t)\}_{0\leq t \leq T}$ so that $\|
H_g-\int_0^T\varphi(t)\widetilde{V}(t)\|^2$ is a minimum subject
to an energy constraint on $\{\varphi(t)\}_{0\leq t \leq T}$. Once
such a $\varphi(t)$ has been designed, our perturbed quantum
system with Hamiltonian
\begin{eqnarray}H(t)=H_0+\frac{1}{m}\varphi(t)V\end{eqnarray} is evolved for a
duration $T$ producing the approximate unitary operator
\begin{eqnarray}U(T)= e^{-\iota
TH_0}\bigg(I-\frac{\iota}{m}\int_0^T\varphi(t)\widetilde{V}(t)dt\bigg)\end{eqnarray}
that is
\begin{eqnarray}e^{\iota TH_0}U(T)=U_{\varphi,\frac{1}{m}}=
I-\frac{\iota}{m}\int_0^T
\varphi(t)\widetilde{V}(t)dt\end{eqnarray} Thus
\begin{eqnarray}U_{\varphi,\frac{1}{m}}^m = \bigg(I-\frac{\iota}{m}\int_0^T
\varphi(t)\widetilde{V}(t)dt\bigg)^m = e^{-\iota\int_0^T
\varphi(t)\widetilde{V}(t)dt} = e^{-\iota H_g}=U_g\end{eqnarray}
approximately for large $m$. In other words, the required gate
$U_g$ is designed by evolving the system with Hamiltonian $H(t)$
given in eq. (4.16) for a duration $T$, resulting in $U(T)$ then
applying $e^{\iota T H_0}$ to this gate resulting in $e^{\iota T
H_0}U(T)$ and finally applying this gate $m$ times resulting in
$(e^{\iota T H_o}U(T))^m=U_g$. The set of unitary gates obtained
by evolving for a duration $T$ a system with Hamiltonian
$H_0+\varphi(t)V(t), 0\leq t \leq T$ may not exhaust all possible
unitary gates. In fact it will exhaust only those gates that have
the form $e^{\iota X}$ where $X$ belongs to the Lie algebra
generated by $H_0$ and $V$. However the first order perturbed
system defined by an approximation to the wave function evolution
equation
\begin{eqnarray}\psi^{'}(t)=-\iota(H_0+\epsilon\varphi(t)V(t))\psi(t)\end{eqnarray} is given by
\begin{eqnarray}\psi(t)= \psi_0(t)+\epsilon\psi_1(t)+O(\epsilon^2)\end{eqnarray} where
\begin{eqnarray}\psi_1^{'}(t)+\iota H_0\psi_1(t)=-\iota\varphi(t)V(t)\psi_0(t),\quad t \geq 0\end{eqnarray} and
\begin{eqnarray}\psi_0(t)=e^{-\iota t H_0}\psi(0),\quad \psi_1(0)=0\end{eqnarray} that
is, given any $\psi(t)=\psi_0(T)+\epsilon \psi_{1f}$, we can
choose a modulating signal $\{\varphi(t)\}_{0\leq t \leq T}$ such
that $\psi_1(T)=\psi_{1f}$ provided that
\begin{eqnarray}\psi_{1f}=\bigg(-\iota\int_0^T e^{-\iota(T- \tau)H_0} V e^{-\iota\tau
H_0}\varphi(\tau)d\tau\bigg)\psi(0). \label{eq9}\end{eqnarray}
Here, we are approximating $H_0$ and $V$ by finite dimensional
truncations. The Cayley Hamilton theorem applied to the $(N+1)
\times (N+1)$ matrix $H_0$ shows that eq. (4.24) is feasible for
arbitrary $\psi_{1f}$ provided that $$ \texttt{Rank Col} [H_0^r V
H_0^s \psi(0): 0 \leq r, s\leq N]= N+1$$we can assume that this
condition holds. $[H_0^r V H_0^s: 0 \leq r, s\leq N]$ is $(N+1)
\times (N+1)$ matrix. Suppose $H_0$ and $V$ are randomly chosen
Hermitian matrices with each entry above the diagonal being a
uniformly distributed random variable in $[a, b]\times [c, d]$,
that is,
$$(H_0)_{\alpha\beta}=x+\iota y;{ x \in [a, b]\quad\text{is uniform} \atop  y \in [c, d]\quad \text{is uniform}}$$
Suppose further the diagonal entries of $H_0$ are uniform real
number is $[\alpha, \beta]$ and like wise for $V$. Then, it is
easy to see that the matrix
$$[H_0^r V H_0^s: 0 \leq r, s\leq N]\in \mathcal{C}^{(N+1)\times ((N+1)N^2)}$$
has with probability $1$ full row rank $=N+1$. More generally,
this result is also true if the entries of $H_0$ and $V$ have
continuous probability densities (that is, non atomic probability
distribution). This means that upto first order in $\epsilon$, our
quantum system is almost surely controllable, that is, almost
surely, any unitary gate can be realized with an error of
$O(\epsilon^2)$. Therefore, to approximate $U_g$ by $W(T)^m$, we
can approximate the generator $H_g$ of $U_g$ by the approximate
generator $H_g$ of $W(T)^m$, that is, we choose the modulating
signal $\{ \varphi(t)\}_{0\leq t\leq T}$ so that
\begin{eqnarray}\bigg\|H_g - \int_0^T
\varphi(t)\widetilde{V}(t)dt\bigg\|_F^2 = \sum _{n,m = 1}^\infty
\bigg|\langle u_m, H_g u_n\rangle - \int_0^T \varphi(t)\langle
u_m, \widetilde{V}(t)u_n\rangle
dt\bigg|^2\label{eq25}\end{eqnarray} is a minimum, where $\{| u_n
\rangle\}_{n=1}^\infty$ is an orthogonal eigenbasis for $H_0$ with
eigenvalues $\{| E_n
 \rangle\}_{n=1}^\infty$. Let $$H_g[m,n] = \big\langle u_m, H_g u_n\big\rangle$$
$$V[m,n] = \langle u_m, V u_n\rangle$$
Since $H_g^\star = H_g$ and $V^\star = V$, we have
 $\overline{H_g[m, n]} = H_g[n, m]$, and  $\overline{V[m, n]} = V[n,
m]$. It gives
\begin{eqnarray}\nonumber\langle u_m,\widetilde{V}(t)u_n\rangle = \bigg\langle u_m, e^{\iota t H_0}V
e^{-\iota t H_0} u_n\bigg\rangle &= \bigg\langle e^{-\iota t H_0}
u_m, V e^{-\iota t H_0} u_n\bigg\rangle \\&= e^{\iota t E(m,
n)}V[m, n]\end{eqnarray} where $E(m, n) = E_m - E_n$. Therefore
\begin{eqnarray}\int_0^T \varphi(t)\langle u_m, \widetilde{V}(t)u_n\rangle dt =
\bigg(\int_0^T \varphi(t)e^{\iota t E(m,n)}dt\bigg)V[m, n] =
\widehat{\varphi}_T(E(m, n))\end{eqnarray} where
$\widehat{\varphi}_T(\omega) = \int_0^T \varphi(t)e^{\iota t
E(m,n)}dt$. Hence, $\{ \varphi(t)\}_{0\leq t\leq T} $ is to be
chosen so that \begin{eqnarray}\sum _{n, m}\bigg| H_g[m, n]-
\widehat{\varphi} _T(E(m, n))V[m ,n]\bigg|^2\end{eqnarray} is a
minimum. If $H_0 = \frac{1}{2m}p^2 + \frac{1}{2}m \omega_0^2 q^2 $
is the Hamiltonian of a harmonic oscillator, then $E_n =
(n+\frac{1}{2})\omega_0$, $n = 0,1,2,\cdots \infty $ and $E_m -E_n
= (m-n)\omega_0$. In this case, the quantity to be minimized is
\begin{eqnarray}\nonumber\sum _{n, m\geq 0}\bigg| H_g[m, n]-
\widehat{\varphi} _T((m-n)\omega_0)V[m ,n]\bigg|^2 \\=
\sum_{-\infty < k < \infty \atop max(-k,0)\leq n < \infty}\bigg|
H_g[n+k, n]- \widehat{\varphi} _T[k]V[n+k ,n]\bigg|^2.
\label{eq9}\end{eqnarray} The minimization is to be performed
subject to fixed energy constraint, given by\begin{eqnarray}2\sum
_{k = 1}^\infty \bigg| \widehat{\varphi}_T[k]\bigg|^2 + \bigg|
\widehat{\varphi}_T[0]\bigg|^2 = \mathcal E\end{eqnarray} Note
that $\mathcal E = \sum _{k = -\infty}^\infty \big|
\widehat{\varphi}_T[k]\big|^2$. The meaning of this energy
constraint $\mathcal E$ can be seen by applying the Parseval
theorem
\begin{eqnarray}\int_0^T \varphi ^2(t)dt = \frac{1}{2\pi}
\int_{-\infty} ^{\infty}|\widehat{\varphi}_T(\omega)|^2
d\omega\approx \frac{1}{2\pi} \sum_{k=-\infty}
^{\infty}|\widehat{\varphi}_T(k\omega_0)|^2\omega_0\end{eqnarray}
provided that $\omega_0$ is small compared to $\frac{1}{T}$.
Therefore, the constraint amounts to fixing the input energy
level.
\section{An Example of the Quantum Harmonic Oscillator Perturbed by an Anharmonic Potential}
In this section, the desired generator is a harmonic oscillator
plus an anharmonics perturbation proportional to $q^3$ and this is
realized by matching this generator to the interaction picture
generator of the harmonic oscillator plus the Hamiltonian of a
charge in a time varying electric field $E(t)$ [25, 31]. The
control input is the electric field $E(t)$. The Harmonic
oscillator is an extremely important and useful concept in the
quantum description of the physical world, and a good way to begin
to understand its properties is to determine the energy
eigenstates of its Hamiltonian. Its should be noted that, the
underlying Hilbert space is $\mathcal{H} = L^2(R)$ and position
and momentum operators $q$ and $p=-\iota\frac{\partial}{\partial
q}$ act in derive subspaces of this Hilbert space. The dynamics of
a single, one dimensional harmonic oscillator is governed by the
unperturbed Hamiltonian $H_0 = \frac{p^2}{2m} +
\frac{1}{2}m\omega_0^2q^2 $ and matching generator $H_g = H_0 +
\mu q^3$, where the perturbation parameter $\mu$ in the anharmonic
perturbation $\mu q^3$ is strictly speaking not a perturbation
parameter. It has been introduced just to illustrate how the
corresponding gate
\begin{eqnarray}U_g= e^{-\iota T(\frac {p^2+q^2}{2}+\mu q^3)}\end{eqnarray}
for small $\mu$, we can be approximated by the designed unitary
gate
\begin{eqnarray}W(T)= e^{-\iota\int_0^T
\varphi(t)\widetilde{V}(t)dt}\end{eqnarray} where
\begin{eqnarray}\widetilde{V}(t)= e^{\iota\frac{T}{2}(p^2+q^2)}q
e^{-\iota\frac{T}{2}(p^2+q^2)}\end{eqnarray} The time independent
perturbation theory is applied to the anharmonic Hamiltonian
$\frac{p^2+q^2}{2}+\mu q^3$ and calculate $e^{-\iota
T(\frac{p^2+q^2}{2}+\mu q^3)}$ up to some power of $\mu$ and then
approximate this gate by the time dependent oscillator gate
$e^{-\iota\int_0^T \varphi(t)\widetilde{V}(t)dt}$ but that would
require enormous computation and so we simply approximate the
truncated generator. This $\mu$ can be regarded as a perturbation
parameter for quantum time independent perturbation theory to
calculate the perturbed stationary states and hence the evolution
operator while $\epsilon$ can be regarded as a perturbation
parameter for quantum time dependent perturbation theory to
calculate the perturbed unitary evolution. These two evolution are
finally matched to determine the optimum time dependent
perturbation. The whole exercise aim at approximately $H_0$ plus a
nonlinear time independent potential with $H_0$ plus a linear time
dependent potential followed by evolution.
\par Assume $V = q$, that is $\epsilon\varphi(t)V = \epsilon
\varphi(t)q$. This corresponds to perturbing the oscillator by an
electric field $E(t) = \frac{-\epsilon \varphi(t)}{Q}$, where $Q$
is the electric charge on the oscillator. Defining the
annihilation and creation operators
\begin{eqnarray}a = \frac{p- \iota m\omega_0 q}{\sqrt{2m}}\end{eqnarray},
\begin{eqnarray}a^\dagger =\frac{p+ \iota m\omega_0
q}{\sqrt{2m}}\end{eqnarray} These operators yield $[a, a^\dagger]
= \omega_0$, $aa^\dagger = H_0 + \frac{1}{2}\omega_0$ and
$a^\dagger a = H_0 - \frac{1}{2}\omega_0$. Since, $a| 0 \rangle =
0$ in position space, it implies that \begin{eqnarray}(p- \iota
m\omega_0 q)| 0 \rangle = 0\end{eqnarray} The ground state wave
function $| 0 \rangle$ satisfies
\begin{eqnarray}
\bigg(\frac{d}{dq}+m\omega_0 q\bigg)| 0 \rangle = 0\end{eqnarray}
where solution is
\begin{eqnarray}| 0 \rangle = C e^{-m\omega_0 \frac{q^2}{2}}\end{eqnarray}
The normalized constant $C$ satisfies, that is,
\begin{eqnarray}|C|^2\int_{\Re}e^{-m\omega_0 q^2}dq = 1\end{eqnarray}
\begin{eqnarray} |C|^2 \sqrt{\frac{\pi}{m\omega_0}} = 1\end{eqnarray}
\begin{eqnarray}C = \bigg(\frac{m\omega_0}{\pi}\bigg)^\frac{1}{4}\end{eqnarray}
Hence, the $n^{th}$ order wave function is given by
\begin{eqnarray}|n\rangle= C_n a^{\dagger n}|0\rangle\end{eqnarray} where
\begin{eqnarray}|C|^2 \langle 0|a^n a^{\dagger n}|0 \rangle = \langle n|n \rangle = 1\end{eqnarray} and
\begin{eqnarray}a^n a^{\dagger n}= a^{n-1}\big([a, a^{\dagger n}]+ a^{\dagger n}
a\big) = \omega_0 n a^{n-1}a^{\dagger n-1} + a^{n-1}a^{\dagger
n}\end{eqnarray} Thus \begin{eqnarray}a^na^{\dagger n}|0\rangle =
\omega_0 n a^{n-1}a^{\dagger n-1}|0\rangle = ......= \omega _0^n
n!|0\rangle\end{eqnarray} and \begin{eqnarray}\langle
0|a^na^{\dagger n}|0\rangle = n!\omega_0^n\end{eqnarray} This
gives
\begin{eqnarray}|C_n|^2 = (n!)^{-1}w_0^{-n}\end{eqnarray}
\begin{eqnarray}C_n = \omega_0^{-\frac{1}{2}}(n!)^{-\frac{1}{2}}\end{eqnarray}
Thus,
\begin{eqnarray}  H_g[m, n] =\langle m| H_g|n\rangle = \langle
m| H_0+\mu q^3|n\rangle =
\bigg(n+\frac{1}{2}\bigg)\omega_0\delta[m-n]+\mu \langle m|
q^3|n\rangle\end{eqnarray}
$$\langle m|q^3|n\rangle = \langle m|\bigg(\frac{a^\dagger - a}{\iota \omega_0\sqrt{2m}}\bigg)^3|n\rangle
= \frac{\iota}{(2m)^\frac{3}{2}\omega_0^3}\langle m|(a^\dagger -
a)^3|n\rangle $$ where \begin{eqnarray} \langle m|(a^\dagger -
a)^3|n\rangle = \langle m|\big(a^{\dagger 3}-a^3-a^{\dagger 2}
a-a^\dagger a a^\dagger - aa^{\dagger 2}+a^2a^\dagger+aa^\dagger
a+a^\dagger a^2\big)|n\rangle\end{eqnarray} We wish to evaluate
$a|n\rangle$ and $a^\dagger|n\rangle$ in order to compute the
matrix elements of $q$, $q^2$, $q^3$ etc. Obviously
\begin{eqnarray}a|n\rangle = \lambda_n|n-1\rangle\end{eqnarray}
where the constant $\lambda_n$ is given by the normalization
condition
\begin{eqnarray}|\lambda_n|^2 = \langle n|a^\dagger a|n\rangle =
\langle n|H_0 - \frac{1}{2}\omega_0|n\rangle =
n\omega_0\end{eqnarray}Therefore $$ \lambda_n = \sqrt{n\omega_0}
$$ Likewise \begin{eqnarray}a^\dagger|n\rangle = \mu_n|n+1\rangle\end{eqnarray}
where \begin{eqnarray}|\mu_n|^2 = \langle n|a a^\dagger |n\rangle
= \langle n|H_0 + \frac{1}{2}\omega_0|n\rangle =
(n+1)\omega_0\end{eqnarray} Therefore
$$ \mu_n = \sqrt{(n+1)\omega_0} $$ Using these formulas for $a|n\rangle$ and
$a^\dagger|n\rangle$ applied successively, we can express
$a^ra^{\dagger s}|n\rangle$ and $a^{\dagger r}a^s|n\rangle$ as
linear combinations of $|n\pm k\rangle,\quad k=1, 2,\cdots$ and
hence from the orthonormality $\langle m|n\rangle=\delta[m-n]$, we
derive the following from which matrix elements of $q^3$ are
obtained.
\begin{eqnarray} \nonumber\langle m|a^{\dagger 3}|n\rangle &= \overline{\langle
n|a^{3}|m\rangle}=\omega_0^\frac{3}{2}(m(m-1)(m-2))^\frac{1}{2}\overline{\langle
n|m-3\rangle}
\\\nonumber&=\omega_0^\frac{3}{2}(m(m-1)(m-2))^\frac{1}{2}\delta[n-m+3]\end{eqnarray}
\begin{eqnarray} \nonumber\langle m|a^3|n\rangle &=
\omega_0^\frac{3}{2}(n(n-1)(n-2))^\frac{1}{2}\delta[m-n+3]\end{eqnarray}
\begin{eqnarray} \nonumber\langle m|a^{\dagger 2}a|n\rangle =
(m(m-1)n)^\frac{1}{2}\omega_0^\frac{3}{2}\langle m-2|n-1\rangle
=(m(m-1)n)^\frac{1}{2}\omega_0^\frac{3}{2}\delta[m-n-1]\end{eqnarray}
\begin{eqnarray} \nonumber\langle m|a^{\dagger}a a^{\dagger}|n\rangle =\omega_0^\frac{3}{2}
(m(n+1)^2)^\frac{1}{2}\langle m-1|n\rangle
=\omega_0^\frac{3}{2}\sqrt{m}(n-1)\delta[m-n-1]\end{eqnarray}
\begin{eqnarray} \nonumber\langle m|a a^{\dagger 2}|n\rangle =\omega_0^\frac{3}{2}
((m+1)(n+1)(n+2))^\frac{1}{2}\langle m+1|n+2\rangle
=\omega_0^\frac{3}{2}((m+1)(n+1)(n+2))^\frac{1}{2}\delta[m-n-1]\end{eqnarray}
\begin{eqnarray} \nonumber\langle
m|a^2 a^{\dagger}|n\rangle =\omega_0^\frac{3}{2}
((n+1)(m+1)(m+2))^\frac{1}{2}\langle m-2|n+1\rangle
=\omega_0^\frac{3}{2}((n+1)(m+1)(m+2))^\frac{1}{2}\delta[m-n-3]\end{eqnarray}
\begin{eqnarray} \nonumber\langle m|a a^{\dagger} a|n\rangle =\omega_0^\frac{3}{2}
((m+1)n^2)^\frac{1}{2}\langle m+1|n\rangle =\omega_0^\frac{3}{2}
n\sqrt{m+1}\delta[m-n+1]\end{eqnarray}
\begin{eqnarray} \nonumber\langle
m|a^{\dagger} a^2|n\rangle =\omega_0^\frac{3}{2}
(mn(n-1))^\frac{1}{2}\langle m-1|n-2\rangle
=\omega_0^\frac{3}{2}(mn(n-1))^\frac{1}{2}\delta[m-n+1]\end{eqnarray}
Thus, all the matrix element of $q^3$ and hence of $H_g$ can be
obtained. Our design method here is to achieve anharmonic gates
using harmonic gates perturbed by a time varying electric field
(such a system can be regarded as a harmonic oscillator with time
varying origin). Since \begin{eqnarray}\frac{p^2}{2m} +
\frac{1}{2}m\omega_0^2q^2 +\epsilon \varphi(t)q = \frac{p^2}{2m} +
\frac{1}{2}m\omega_0^2\bigg(q +\frac{\epsilon
\varphi(t)}{m\omega_0^2}\bigg)^2- \frac{\epsilon^2
\varphi^2(t)}{2m\omega_0^2}\end{eqnarray} we define $H_g[m, n] = 0
= V[m, n]$ if $m<0$ or $n<0$. Then the quantity to be minimized
after incorporating the energy constraint is (see eq. (4.29) and
(30)) $ E \big[\{\widehat{\varphi}_T[k],
\overline{\widehat{\varphi}_T[k]}, k\geq 1\},
\widehat{\varphi}_T[0], \lambda\big]$. It gives
\begin{eqnarray}\nonumber & \sum_{n,k\in Z}\big| H_g[n+k,
n]-\widehat{\varphi} _T[k]V[n+k ,n]\big|^2 -\lambda
\bigg\{\widehat{\varphi}_T[0]^2
+2\sum_{k=1}^\infty\big|\widehat{\varphi} _T[k]\big|^2-\mathcal
E\bigg\}\\\nonumber&=\sum_{n\geq 0} \big| H_g[n,
n]-\widehat{\varphi} _T[0]V[n, n]\big|^2+2\sum_{n\geq 0 k\geq 1}
\big| H_g[n+k, n]-\widehat{\varphi}_T[k]V[n+k ,n]\big|^2
\\ &-\lambda\bigg\{\widehat{\varphi}_T[0]^2+2\sum_{k=1}^\infty\big|\widehat{\varphi}
_T[k]\big|^2-\mathcal E\bigg\}\end{eqnarray} The independent
variables (real) with respect to which this minimization is to be
carried out are $\widehat{\varphi}_T[0]$,
$\Re\{\widehat{\varphi}_T[k]\}$, $\Im\{\widehat{\varphi}_T[k]\}$,
$k\geq 1$. Equivalently, treating $\{\widehat{\varphi}_T[k],
\overline{\widehat{\varphi}_T[k]}, k\geq 1\}$ and
$\widehat{\varphi}_T[0]$ as the independent variables to be
optimized in the error energy $E$ in eq. (4.57). The optimal
values of these variables are obtained from
\begin{eqnarray}\frac{\partial E}{\partial \overline{\widehat{\varphi}_T[k]}}=0,\quad \frac{\partial E}{\partial \widehat{\varphi}_T[k]}=0,\quad k\geq 1\end{eqnarray}
and \begin{eqnarray} \frac{\partial E}{\partial
\widehat{\varphi}_T[0]}=0\end{eqnarray} The two equations in eq.
(4.58) gives complex conjugate equations and hence it is
sufficient to retain just one of them, say
\begin{eqnarray}\nonumber\frac{\partial E}{\partial \overline{\widehat{\varphi}_T[k]}}=0, k\geq 1\end{eqnarray}
Using eq. (4.57), we get
\begin{eqnarray}-\sum_{n\geq 0}\bigg(H_g[n+k, n]-\widehat{\varphi} _T[k]V[n+k ,n]\bigg)\overline{V[n+k ,n]}
- 2\lambda\widehat{\varphi} _T[k]=0,\quad k\geq 1\end{eqnarray}
and
\begin{eqnarray}-\sum_{n\geq 0}\bigg( H_g[n, n]-\widehat{\varphi} _T[0]V[n, n]\bigg)V[n, n]
- 2\lambda\widehat{\varphi} _T[0]=0 \end{eqnarray}Using eq.
(4.61), we get
\begin{eqnarray}\nonumber\widehat{\varphi} _T[0]\bigg\{\bigg(\sum_{n\geq 0}V[n, n]^2\bigg)-\lambda\bigg\}-\sum_{n\geq 0} H_g[n, n]V[n, n]=0\end{eqnarray}
\begin{eqnarray}\widehat{\varphi} _T[0]= \frac{\sum\limits_{n\geq 0} H_g[n, n]V[n, n]}{\bigg\{\bigg(\sum\limits_{n\geq 0}V[n, n]^2\bigg)-\lambda\bigg\}} \end{eqnarray}
Eq. (4.60) gives
\begin{eqnarray}\nonumber\bigg\{\bigg(\sum_{n\geq 0}|V[n+k, n]|^2\bigg)-2\lambda\bigg\}\widehat{\varphi}_T[k]=\sum_{n\geq 0} H_g[n+k,n]\overline{V[n+k, n]},\quad k\geq 1\end{eqnarray}
\begin{eqnarray}\widehat{\varphi} _T[k]= \frac{\sum\limits_{n\geq 0} H_g[n+k, n]\overline{V[n+k, n]}}{\bigg\{\bigg(\sum_{n\geq 0}|V[n+k, n]|^2\bigg)-2\lambda\bigg\}}\end{eqnarray}
Define $A[k] =\sum_{n\geq 0} \bigg(H_g[n+k, n]\overline{V[n+k,
n]}\bigg)$, and $B[k] =\sum_{n\geq 0}\big|V[n+k, n]\big|^2,\quad
k\geq 0$, eqs. (4.62) and (4.63) can be expressed as
\begin{eqnarray}\widehat{\varphi} _T[0]
=\frac{A[0]}{B[k]-\lambda}\end{eqnarray}and
\begin{eqnarray}\widehat{\varphi} _T[k] =\frac{A[k]}{B[k]-2\lambda};\quad k\geq 1\end{eqnarray}
respectively. The Lagrange multiplier $\lambda$ (real) is
determined from the constraint equation $\frac{\partial
E}{\partial\lambda} = 0$, that is,
\begin{eqnarray}\widehat{\varphi} _T[0]^2 +2\sum_{k=1}^\infty|\widehat{\varphi} _T[k]|^2=\mathcal E \end{eqnarray}
From eqs. (4.64), (65) and eq. (4.66), the approximate numerical
solution for $0\leq k$, and $n\leq N$ is given by
\begin{eqnarray}\frac{|A[0]|^2}{|B[0]-\lambda|^2}+2\sum_{k=1}^\infty\frac{|A[k]|^2}{|B[k]-2\lambda|^2}=\mathcal E\end{eqnarray}
Equating the approximates of eq. (4.67) to zero, we get
\begin{eqnarray}\frac{|A[0]|^2}{|B[0]-\lambda|^2}+2\sum_{k=1}^N\frac{|A[k]|^2}{|B[k]-2\lambda|^2}= 0\end{eqnarray}
\begin{eqnarray}\nonumber &|A[0]|^2\prod_{k=1}^N|B[k]-2\lambda|^2
+2\sum_{k=1}^N\bigg\{|A[k]|^2\bigg(\prod_{m=1\atop m\neq
k}^N|B[m]-2\lambda|^2\bigg)\bigg\}\big(B[0]-\lambda\big)^2\\&=0\end{eqnarray}
Note that $|B[k]-2\lambda|^2$, or equivalently $|B[m]-2\lambda|^2$
appear in both eqs. (4.68) and (4.69). This factor arises by
differentiating the constraint component in eq. (4.57). This is a
polynomial of degree $2N$ in $\lambda$ which is solved using the
roots command in MATLAB $\{\widehat{\varphi} _T[k], k\geq 0\}$.
Therefore, the noise to signal energy ratio (NSER) is given by
\begin{eqnarray}\nonumber NSER =\frac{\bigg[2\sum\limits_{n\geq 0\atop k\geq 1, n+k\leq N}| H_g[n+k, n]-\widehat{\varphi}_T[k]
V[n+k,n]\bigg|^2+\sum\limits_{0\leq n\leq N}(H_g[n,
n]-\widehat{\varphi} _T[0]V[n, n])^2\bigg]} {2\sum\limits_{n\geq
0\atop k\geq 1, n+k\leq N}| H_g[n+k, n]|^2+\sum\limits_{0\leq
n\leq N}|H_g[n, n]|^2}\\\end{eqnarray} The above formula for the
NSER is given by the ratio of the minimum error energy between the
given generator $H_g$ and the designed generator to the energy of
the given generator. The minimum error energy is given by eq.
(4.29) with $\varphi_T(k)$ as the optimal modulating signal. It is
simply the Frobenius norm square of the difference between the
given generator matrix and a truncated version of the designed
generator matrix. The advantage of using the harmonic oscillator
as the unperturbed system manifests in requires the optimization
over only the discrete frequency samples of the control input
Fourier transform to be carried out in control to previous work,
where optimization over an entire continuous trajectory was
required. In other words rather than solving a linear integral
equation we required to solve only a matrix linear equation.
\section{Simulation Result Showing Gate Designed Error Energy}
In this section first, we realize controlled unitary gate and then
we plot noise to signal energy ratio (NSER), which is the
performance measure criterion of the proposed algorithm.
\subsection{Realization of Controlled Unitary Gate Using Generator Matching}
Controlled unitary gates act on two or more qubits where one or
more qubits act as a control for some operation [36, 41]. If the
control qubit is in the state $|0\rangle$ then the target qubit is
left unchange. The gate being implemented is the following
controlled unitary gate
\begin{eqnarray}|x_1x_2x_3\rangle\longrightarrow |x_1\rangle
U_1^{x_1}|x_2\rangle U_2^{x_1x_2}|x_3\rangle\end{eqnarray}
where $U_1 =\left(%
\begin{array}{cc}
  \alpha_1 & \beta_1 \\
  -\overline{\beta}_1  & \overline{\alpha} \\
\end{array}%
\right)$ and $U_2 =\left(%
\begin{array}{cc}
  \alpha_2 & \beta_2 \\
  -\overline{\beta}_2  & \overline{\alpha}_2 \\
\end{array}%
\right)$. In other words $U_1$ is applied to the second qubits iff
the first qubits is $1$ and $U_2$ is applied to the third qubits
iff both the first and second qubits are one. Another way to
express the gate action is via the following formulas (we choose
$x_3$ either $0$ or $1$) $$|00x_3\rangle \longrightarrow
|00x_3\rangle$$
$$|01x_3\rangle \longrightarrow
|01x_3\rangle$$
$$|10x_3\rangle \longrightarrow
|1\rangle U_1|0\rangle|x_3\rangle$$
$$|11x_3\rangle \longrightarrow
|1\rangle U_1|1\rangle U_2|x_3\rangle$$ A complete table of
three-qubits of controlled gate is given by
$$|000\rangle \longrightarrow |000\rangle$$
$$|001\rangle \longrightarrow |001\rangle$$
$$|010\rangle \longrightarrow |010\rangle$$
$$|011\rangle \longrightarrow |011\rangle$$
$$|100\rangle \longrightarrow \beta_1|110\rangle+\overline{\alpha}_1|100\rangle$$
$$|101\rangle \longrightarrow \beta_1|111\rangle+\overline{\alpha}_1|101\rangle$$
$$|110\rangle \longrightarrow \alpha_1\beta_2|111\rangle+\alpha_1\overline{\alpha}_2|110\rangle-
\overline{\beta}_1\beta_2|101\rangle-\overline{\beta}_1\overline{\alpha}_2|100\rangle$$
$$|111\rangle \longrightarrow \alpha_1\alpha_2|111\rangle-\alpha_1\overline{\beta}_2|110\rangle-
\overline{\beta}_1\alpha_2|101\rangle+\overline{\beta}_1\overline{\beta}_2|100\rangle$$
In matrix form the controlled gate $U_c$ is given by
\[
  U_c=
\left[{\begin{array}{cccccccc}
  1       & 0       & 0       & 0       & 0       & 0       & 0       & 0       \\
  0       & 1       & 0       & 0       & 0       & 0       & 0       & 0       \\
  0       & 0       & 1       & 0       & 0       & 0       & 0       & 0       \\
  0       & 0       & 0       & 1       & 0       & 0       & 0       & 0       \\
  0       & 0       & 0       & 0       & \overline{\alpha}_1   & 0   & -\overline{\beta}_1\overline{\alpha}_2  & \overline{\beta}_1\overline{\beta}_2   \\
  0       & 0       & 0       & 0       & 0  & \overline{\alpha}_1  & -\overline{\beta}_1\beta_2  & - \overline{\beta}_1\alpha_2   \\
  0       & 0       & 0       & 0       & \beta_1  & 0  & \alpha_1\overline{\alpha}_2  & -\alpha_1\overline{\beta}_2   \\
  0       & 0       & 0       & 0       & 0  & \beta_1  & \alpha_1\beta_2   & \alpha_1\alpha_2   \\
\end{array} }\right]
\]
Consider separable unitary gates, that is, $U_1$ is a unitary
operator in a Hilbert space $\mathcal{H}_1$, $U_2$ is a unitary
operator in a Hilbert space $\mathcal{H}_2$ and $U_1\otimes U_2$
is the separable unitary gate acting in the Hilbert space
$\mathcal{H}_1\otimes\mathcal{H}_2$. The unitary operator $U_1$
can be realized via a Hamiltonian $H_1$ in $\mathcal{H}_1$ and
$U_2$ via a Hamiltonian $H_2$ in $\mathcal{H}_2$. Then
\begin{eqnarray}U_1\otimes U_2 = e^{-\iota T H_1}\otimes e^{-\iota T
H_2}= e^{-\iota T(H_1\otimes I_2+I_1\otimes H_2)}\end{eqnarray}
For example, we can take $H_\alpha =
\frac{p_\alpha^2+q_\alpha^2}{2}$, $\alpha=1, 2$, that is, harmonic
oscillator Hamiltonians. Now suppose we perturb $U_1\otimes U_2$
to $U_g = (U_1\otimes U_2)(I+\iota\epsilon X)$, where $X$ is
Hermitian operator in $\mathcal{H}_1\otimes\mathcal{H}_2$. The
perturbed gate is the non-separable. To realize $U_g$, we may
perturbed the sum of independent oscillator Hamiltonians to
\begin{eqnarray}H(t)=H_1\otimes I_2+I_1\otimes H_2+\epsilon
\varphi(t)(q_1-q_2)^3\end{eqnarray} This $H(t)$ generates the
Schr\"odinger evolution operator solved by using eqs. (4.8) and
(4.9), we get
\begin{eqnarray}U_t=(U_1(t)\otimes U_2(t))\bigg(I-\iota \epsilon\int_0^T\varphi(t)\widetilde{V}_{12}(t)dt\bigg)\end{eqnarray}
where $$\widetilde{V}_{12}(t)= e^{\iota t H_0}V_{12}e^{\iota t
H_0}= (U_1^*(t)\otimes U_2^*(t))V_{12}(U_1(t)\otimes U_2(t))$$
where $H_0 = H_1\otimes I_2+I_1\otimes H_2$, $U_1(t)=e^{-\iota t
H_1}$, $U_2(t) = e^{-\iota t H_2}$ and $V_{12}= (q_1-q_2)^3$. We
can match the generator
$-\epsilon\int_0^T\varphi(t)\widetilde{V}_{12}(t)dt$ to $X$ and
realize the non-separable gate $U_g$, that is,
\begin{eqnarray}\min_{\atop\varphi(t), 0\leq t\leq T}\|X-\epsilon
\int_0^T\varphi(t)\widetilde{V}_{12}(t)dt\|^2\end{eqnarray} To
carry out the above minimization, we complete the matrix element
of $X$ and $\widetilde{V}_{12}(t)$ relative to the truncated basis
$|n_1, n_2\rangle = |n_1\rangle \otimes |n_2\rangle$, where $n_1,
n_2=0, 1, 2,\cdots, N$. The truncated basis state $|n_1\rangle$
and $|n_2\rangle$ are energy eigenstate of $H_1$ and $H_2$
respectively, given by
\begin{eqnarray}H_1|n_1\rangle=(n_1+\frac{1}{2})|n_1\rangle\end{eqnarray}
\begin{eqnarray}H_2|n_2\rangle=(n_1+\frac{1}{2})|n_2\rangle\end{eqnarray} A truncated version of the energy in eq. (4.75) is given
by
\begin{eqnarray}\sum _{0\leq n_1, n_2, m_1, m_2\leq N}|\langle n_1, n_2|X|m_1, m_2\rangle-\epsilon\int_0^T \varphi(t)\langle n_1, n_2|\widetilde{V}_{12}(t)|m_1, m_2\rangle dt|^2\end{eqnarray}
and we minimize this with respect to $\{\varphi(t)\}_{0\leq t\leq
T}$.\par Note that
\begin{eqnarray}\nonumber\langle n_1, n_2|\widetilde{V}_{12}(t)|m_1, m_2\rangle&=\langle n_1, n_2|e^{\iota t H_0}V_{12}(t)e^{-\iota t H_0}|m_1, m_2\rangle
\\&=e^{\iota(n_1-m_1+n_2-m_2)t}\langle n_1, n_2|V_{12}(t)|m_1,
m_2\rangle\end{eqnarray} and
$$\langle n_1, n_2|V_{12}(t)|m_1, m_2\rangle = \langle n_1, n_2|(q_1-q_2)^3|m_1,
m_2\rangle= \langle n_1, n_2|q_1^3-q_2^3-3q_1^2q_2+3q_1q_2^2|m_1,
m_2\rangle$$ So
\begin{eqnarray}\nonumber\langle n_1, n_2|V_{12}(t)|m_1, m_2\rangle &=\langle n_1|q_1^3|m_1\rangle\delta[n_2-m_2]
-\langle n_2|q_2^3|m_2\rangle\delta[n_1-m_1]\\&-3\langle
n_1|q_1^2|m_1\rangle\langle n_2|q_2|m_2\rangle+3\langle
n_1|q_1|m_1\rangle\langle n_2|q_2^2|m_2\rangle\end{eqnarray} Using
eqs. (4.79) and (4.80), we get
\begin{eqnarray}\nonumber\langle n_1, n_2|\widetilde{V}_{12}(t)|m_1, m_2\rangle&=e^{\iota(n_1-m_1+n_2-m_2)t}\langle n_1|q_1^3|m_1\rangle\delta[n_2-m_2]
-\langle n_2|q_2^3|m_2\rangle\delta[n_1-m_1]\\&-3\langle
n_1|q_1^2|m_1\rangle\langle n_2|q_2|m_2\rangle+3\langle
n_1|q_1|m_1\rangle\langle n_2|q_2^2|m_2\rangle\end{eqnarray} and
finally substituting the value of eq. (4.81) in eq. (4.74), we get
the realization of $(U_1\otimes U_2)(I+\iota X)$ as
\begin{eqnarray}\nonumber &\bigg\langle n_1, n_2\bigg|(U_1(t)\otimes U_2(t))\bigg(I-i\epsilon \int_0^T \varphi(t)\widetilde{V}_{12}(t)dt\bigg)\bigg|m_1,
m_2\bigg\rangle\\\nonumber
&=e^{-\iota(n_1+n_2+1)t}\bigg(\delta[n_1-m_1]\delta[n_2-m_2]-\iota\epsilon
\int_0^T\varphi(t)\langle n_1, n_2|\widetilde{V}_{12}(t)|m_1,
m_2\rangle\bigg)dt\\\end{eqnarray} The corresponding $N_1N_2\times
N_1N_2$ matrix $U$ can be formed by truncation so that eq. (4.81)
is the $(N_1n_1+n_2+1, N_1m_1+m_2+1)^{th}$ entry of $U$ where
$0\leq n_1, m_1\leq N_1-1$, and $0\leq n_2, m_2\leq N_2-1$. This
construction is equivalent to taking a matrix $Q \in
\mathcal{C}^{N_1N_2\times N_1N_2}$ and defining its matrix
elements $Q[n_1n_2, m_1m_2]=\langle e_{n_1}\otimes
f_{n_2}|Q|e_{n_1}\otimes f_{n_2}\rangle$ relative to the
lexicographically ordered basis of $\mathcal{C}^{N_1N_2}$, namely
\begin{eqnarray}\mathbb{B} = \{e_{n_1}\otimes f_{n_2}|0\leq n_1\leq N_1-1, 0\leq
n_2\leq N_2-1\}=\mathbb{B}_1\otimes\mathbb{B}_2\end{eqnarray}
where $\mathbb{B}_1 =\{e_{n_1}\}_{n_1=0}^{N_1-1}$ is an ordered
basis for $\mathcal{C}^{N_1}$ and $\mathbb{B}_2
=\{f_{n_1}\}_{n_2=0}^{N_2-1}$ is an ordered basis for
$\mathcal{C}^{N_2}$. By lexicographic ordering, we means that
$\{e_{n_1}\otimes f_{n_2}\}$ is the $(N_2n_1+n_2+1)^{th}$ element
of $\mathbb{B}$. Then, $Q[n_1n_2, m_1m_2]$ is the $(N_2n_1+n_2+1,
N_2m_1+m_2+1)^{th}$ element of the matrix of $Q$ relative to the
basis $\mathbb{B}$. In particular, if $Q=Q_1\otimes Q_2$, where
$Q_1\in \mathcal{C}^{N_1\otimes N_1}$ and $Q_2\in
\mathcal{C}^{N_2\otimes N_2}$ then $Q[n_1n_2, m_1m_2] =
Q[n_1m_1]Q[n_2m_2]$, that is, $[Q]_{\mathbb{B}}$ is the Kronecker
tensor product of $[Q_1]_{\mathbb{B}_1}$ and
$[Q_2]_{\mathbb{B}_2}$.
\subsection{Noise to Signal Energy Ratio (NSER)}
%
Figure 2 shows the Fourier transform of the optimal modulating
signal at Bohr frequency. As the Bohr frequency increases, the
Fourier transform of the optimal modulating signal also increases.
%
Figure 3 shows the noise to signal energy ratio with increasing
value of the truncation size $(N)$. Larger the value of truncation
size better is the approximation of the infinite dimensional
generator by the finite dimensional $(N+1)\times (N+1)$ matrix
obtained by truncation. The endpoint $(N = 500)$ of the graphs
therefore represent the best possible approximation of the
infinite dimensional gate generator $H_g$ by its truncated version
$[[\langle m|H_g|n\rangle]]_{0\leq m, n\leq N}$, which is an
approximation of the unitary matrix at that instant. It is given
by
\[
  H_g=
\left[{\begin{array}{cccccccc}
  1.0000       & 0       & 0       & 0       & 0       & 0       & 0       & 0       \\
  0       & 1.0000       & 0       & 0       & 0       & 0       & 0       & 0       \\
  0       & 0       & 1.0000       & 0       & 0       & 0       & 0       & 0       \\
  0       & 0       & 0       & 1.0000       & 0       & 0       & 0       & 0       \\
  0       & 0       & 0       & 0       & 1.0000       & 0       &-0.1437 & -0.0000 \\
  0       & 0       & 0       & 0       & 0        & 1.0000      & 0.0722  & -0.0000  \\
  0       & 0       & 0       & 0       & -0.1437  & 0.0722  & 1.0000  & -0.0383     \\
  0       & 0       & 0       & 0       & -0.0000  & -0.0000  & -0.0383  &  1.0000    \\
\end{array} }\right]
\]
We are therefore bound by constraint on $V$. The primary advantage
with using the matching generator philosophy is that even after
truncation to a finite dimensional subspace, Hermitianity of the
finite generator is not lost and hence the corresponding designed
gate remains unitary.

\section{An Example of Ion-trap Based Gate Design}
In this section, we apply the matching generator technique
developed in the previous section to an ion trap model consisting
of a spin $\frac{1}{2}$ particle interacting with a classical plus
a quantum magnetic field [45, 49]. The coupling of the ion spin to
the classical and quantum magnetic fields are modified by control
function of time which as in the previous sections, are optimized
with respect to their fourier transform samples. The quantum
magnetic field is assumed to be single model, that is, described
by a single creation and single annihilation operator. The overall
system plus quantum magnetic field is described by states
$|\alpha, n\rangle$ where $\alpha=\pm\frac{1}{2}$. are the spin
states of the ion and $n=0, 1, 2,\cdots$ label the quantum field
states (the quantum field is a single harmonic oscillator). The
presence of the quantum field enables us to design gates of a very
large size. Ion traps are used to simulate quantum gates.
Basically, an ion trap consists of sequence of ions each having
$+ve$ charge and with the ions coupled to an external
electromagnetic field. The net effect of this coupling is that
each ion gets harmonically coupled to the external world. If
$r_i=(x_i, y_i, z_i)$ is the position of the $i^{th}$ ion and
$p_i$ is its momentum, then the Hamiltonian of the sequence of
ions without the electromagnetic coupling is given by
\begin{eqnarray}H_0=\sum_{i=1}^N\frac{|p_i|^2}{2M}+\sum_{1\leq i<j\leq N}\frac{e^2}{|r_i-r_j|}\end{eqnarray}
The second term comes from Coulomb repulsion between the ions.
Electromagnetic coupling occurs via an additional potential
(harmonics) and is given by
\begin{eqnarray}V = \frac{1}{2}\sum_{i=1}^N(w_xx_i^2+w_yy_i^2+w_zz_i^2)\end{eqnarray}
A simplified model for the ion trap is to treat the unperturbed
ion and the electromagnetic field as having the Hamiltonian
$H_0=\frac{1}{2}\hbar \omega_0\sigma_z+\hbar\omega_0^\prime a\dag
a$, where ${a, a\dag}$ are the creation and annihilation operators
describe the electromagnetic photon field and $\frac{1}{2}\hbar
\omega_0\sigma_z$ describes the energy of a spin $\frac{1}{2}$
particle in a constant magnetic field along the $z$-axis. This
spin $\frac{1}{2}$ particle models the atom or ion. The
interaction Hamiltonian is given by
\begin{eqnarray}H_I=-\mu.B=\frac{\hbar}{2}(\Omega_1(t)\sigma_x+\Omega_2(t)\sigma_y)+\frac{\hbar}{2}\Omega_1(t)\sigma_x(a+a^\dag)\end{eqnarray}
where, we assume that the magnetic field consists time varying
components along the $x$ and $y$ axis. For simplicity of analysis,
we take $\Omega_2(t)=0$ so the magnetic field is only along the
$x$ axis. Then
\begin{eqnarray}H_I=\frac{\hbar}{2}\Omega(t)\sigma_x+\frac{\hbar}{2}\Omega(t)\sigma_x(a+a^\dag)\end{eqnarray}
Taking $\hbar=1$, the eigenstates of $H_0$ are $|\frac{1}{2},
n\rangle$, $|-\frac{1}{2}, n\rangle$, $n=0, 1, 2, \cdots$
\begin{eqnarray}H_0|\frac{1}{2},n\rangle=(\frac{\omega_0}{2}\sigma_z+\omega_0^{'}a^\dag
a)|\frac{1}{2}, n\rangle=(\frac{\omega_0}{2}+\omega_0^{'}
n)|\frac{1}{2}, n\rangle\end{eqnarray} and
\begin{eqnarray}H_0|-\frac{1}{2}, n\rangle=(-\frac{\omega_0}{2}+\omega_0^{'}
n)|-\frac{1}{2}, n\rangle\end{eqnarray} The infinitesimal
generator after introduction of the atom and field is the (in the
interaction picture) $W(t)=\int_0^Te^{\iota t H_0}H_I(t)e^{-\iota
t H_0}dt$ with matrix elements
$$\langle \frac{1}{2}, n|W|\frac{1}{2}, m\rangle, \langle \frac{1}{2}, n|W|-\frac{1}{2}, m\rangle,
\langle- \frac{1}{2}, n|W|\frac{1}{2}, m\rangle, \langle
-\frac{1}{2}, n|W|-\frac{1}{2}, m\rangle$$ Defining
$$E(\frac{1}{2}, n)=\frac{\omega_0}{2}+\omega_0^{'}n$$
$$E(-\frac{1}{2}, n)=-\frac{\omega_0}{2}+\omega_0^{'}n$$
All matrix elements are given by
$$\langle \frac{1}{2}, n|W|\frac{1}{2}, m\rangle=\int_0^T\langle \frac{1}{2}, n|H_I(t)|\frac{1}{2}, m\rangle
e^{\iota (E(\frac{1}{2}, n)-E(\frac{1}{2}, m))t}dt$$
$$\langle \frac{1}{2}, n|W|-\frac{1}{2}, m\rangle=\int_0^T\langle \frac{1}{2}, n|H_I(t)|-\frac{1}{2}, m\rangle
e^{\iota (E(\frac{1}{2}, n)-E(-\frac{1}{2}, m))t}dt$$
$$\langle -\frac{1}{2}, n|W|\frac{1}{2}, m\rangle=\int_0^T\langle -\frac{1}{2}, n|H_I(t)|\frac{1}{2}, m\rangle
e^{\iota (E(-\frac{1}{2}, n)-E(\frac{1}{2}, m))t}dt$$
$$\langle -\frac{1}{2}, n|W|-\frac{1}{2}, m\rangle=\int_0^T\langle -\frac{1}{2}, n|H_I(t)|-\frac{1}{2}, m\rangle
e^{\iota (E(-\frac{1}{2}, n)-E(-\frac{1}{2}, m))t}dt$$ where,
$$E(\frac{1}{2}, n)-E(\frac{1}{2}, m)=\omega_0^{'}(n-m)$$
$$E(\frac{1}{2}, n)-E(-\frac{1}{2}, m)=\omega_0+\omega_0^{'}(n-m)$$
$$E(-\frac{1}{2}, n)-E(\frac{1}{2}, m)=-\omega_0+\omega_0^{'}(n-m)$$
$$E(-\frac{1}{2}, n)-E(-\frac{1}{2}, m)=\omega_0^{'}(n-m)$$
\begin{eqnarray}\nonumber\langle \frac{1}{2}, n|H_I(t)|\frac{1}{2}, m\rangle &= \langle \frac{1}{2}, n|\frac{1}{2}\Omega(t)\sigma_x
+\frac{1}{2}\Omega(t)\sigma_x(a+a^\dag)|\frac{1}{2}, m\rangle\\&=
\Omega(t)\langle\frac{1}{2}|\sigma_x|\frac{1}{2}\rangle
(\frac{\delta[n-m]}{2}+\langle n|\frac{a+a^\dag}{2}|m
\rangle)\end{eqnarray}
Further, $$|\frac{1}{2}\rangle=\left(%
\begin{array}{c}
  1 \\
  0 \\
\end{array}%
\right),
\sigma_x=\left(%
\begin{array}{cc}
  0 & 1 \\
  1 & 0 \\
\end{array}%
\right)$$ so
$$\langle\frac{1}{2}|\sigma_x|\frac{1}{2}\rangle=\left(%
\begin{array}{cc}
  1 & 0 \\
\end{array}%
\right)
\left(%
\begin{array}{cc}
  0 & 1 \\
  1 & 0 \\
\end{array}%
\right)
\left(%
\begin{array}{c}
  1 \\
  0 \\
\end{array}%
\right)=
\left(%
\begin{array}{cc}
  1 & 0 \\
\end{array}%
\right)
\left(%
\begin{array}{c}
  0 \\
  1 \\
\end{array}%
\right)=0$$ We thus have
$$\langle \frac{1}{2}, n|H_I(t)|\frac{1}{2}, m\rangle=0$$
$$\langle \frac{1}{2}, n|H_I(t)|-\frac{1}{2}, m\rangle= \Omega(t)\langle \frac{1}{2}|\sigma_x|-\frac{1}{2}\rangle
[\frac{\delta[n-m]}{2}+\langle n|\frac{a+a^\dag}{2}|m \rangle]$$
Further more,
$$|-\frac{1}{2}\rangle=\left(%
\begin{array}{c}
  0 \\
  1 \\
\end{array}%
\right)$$
$$\langle\frac{1}{2}|\sigma_x|-\frac{1}{2}\rangle=\left(%
\begin{array}{cc}
  1 & 0 \\
\end{array}%
\right)
\left(%
\begin{array}{cc}
  0 & 1 \\
  1 & 0 \\
\end{array}%
\right)
\left(%
\begin{array}{c}
  1 \\
  0 \\
\end{array}%
\right)=
\left(%
\begin{array}{cc}
  1 & 0 \\
\end{array}%
\right)
\left(%
\begin{array}{c}
  0 \\
  1 \\
\end{array}%
\right)=1$$ and taking conjugates $\langle
-\frac{1}{2}|\sigma_x|\frac{1}{2}\rangle=1$. Thus
$$\langle \frac{1}{2}, n|H_I(t)|-\frac{1}{2}, m\rangle= \Omega(t)[\frac{\delta[n-m]}{2}+\frac{1}{2}\sqrt{m}\delta[n-m+1]+\sqrt{m+1}\delta[n-m-1]]$$
and
$$\langle -\frac{1}{2}, n|H_I(t)|-\frac{1}{2}, m\rangle=0$$ since $\langle -\frac{1}{2}|\sigma_x|-\frac{1}{2}\rangle=0$. Finally we get
\begin{eqnarray}\nonumber &\langle -\frac{1}{2}, n|H_I(t)|\frac{1}{2}, m\rangle = \Omega(t)\langle -\frac{1}{2}|\sigma_x|\frac{1}{2}\rangle
[\frac{\delta[n-m]}{2}+\langle n|\frac{a+a^\dag}{2}|m
\rangle]=\frac{1}{2}\Omega(t)[\delta[n-m]\\&+\sqrt{m}\delta[n-m+1]+\sqrt{m+1}\delta[n-m-1]]
\end{eqnarray}
The matrix of $H_I(t)$ relative to the truncated basis $\{\langle
\frac{1}{2}, n|, \langle -\frac{1}{2}, n|,
 \quad n=0, 1, 2,\cdots, N-1\}$ has then the block structure form
 $$\left[%
\begin{array}{cc}
  ((\langle \frac{1}{2}, n|H_I(t)|\frac{1}{2}, m\rangle))_{n, m} & ((\langle \frac{1}{2}, n|H_I(t)|-\frac{1}{2}, m\rangle))_{n, m} \\
  ((\langle -\frac{1}{2}, n|H_I(t)|\frac{1}{2}, m\rangle))_{n, m} & ((\langle -\frac{1}{2}, n|H_I(t)|-\frac{1}{2}, m\rangle))_{n, m} \\
\end{array}%
\right]$$
$$=\left[%
\begin{array}{cc}
  0 & ((\langle \frac{1}{2}, n|H_I(t)|-\frac{1}{2}, m\rangle))_{n, m} \\
  ((\langle -\frac{1}{2}, n|H_I(t)|\frac{1}{2}, m\rangle))_{n, m} & 0 \\
\end{array}%
\right]$$
$$=\frac{1}{2}\Omega(t)\left[%
\begin{array}{cc}
  0 & A \\
  A & 0 \\
\end{array}%
\right]$$ where,
$A=((\delta[n-m]+\sqrt{m}\delta[n-m+1]+\sqrt{m+1}\delta[n-m-1]))_{0\leq
n, m\leq N-1}$. What we need is the matrix of $e^{\iota t
H_0}H_I(t)e^{-\iota t H_0}$ relative to the truncated basis
$\{\langle \frac{1}{2}, n|, \langle -\frac{1}{2}, n|,
 \quad n=0, 1, 2,\cdots, N-1\}$. Based on the above argument, the block structured form of $H_I (t) $ is given by
$$=\left[%
\begin{array}{cc}
  0 & ((\langle \frac{1}{2}, n|H_I(t)|-\frac{1}{2}, m\rangle e^{(\iota(\omega_0+\omega_0^{'}(n-m))t})) \\
  ((\langle -\frac{1}{2}, n|H_I(t)|\frac{1}{2}, m\rangle e^{(\iota(-\omega_0+\omega_0^{'}(n-m))t})) & 0 \\
\end{array}%
\right]$$
$$=\left[%
\begin{array}{cc}
  0 & ((a[n, m]e^{(\iota(\omega_0+\omega_0^{'}(n-m))t})) \\
  ((a[n, m] e^{(\iota(-\omega_0+\omega_0^{'}(n-m))t})) & 0 \\
\end{array}%
\right]$$
Given a generator having the block structure $$\left[%
\begin{array}{cc}
  \textbf{0} & \textbf{C} \\
  \textbf{C} & \textbf{0} \\
\end{array}%
\right]$$ where $\textbf{C}^\star=\textbf{C}$, we can control the
magnetic field proportional to $\Omega(t), \quad 0\leq t\leq T$ so
that the generators have minimum distance from a given generator.
This is equivalent to requiring that
$\|\textbf{C}-\frac{1}{2}((a[n, m]\widehat{\Omega}_T(\omega_0
+\omega_0^{'}(n-m))))\|, \quad 0\leq n, m\leq N-1$ is a minimum,
where $\widehat{\Omega}_T=\int_0^T\Omega(t)e^{\iota\omega t}dt$.
Equivalently, we choose
$\widehat{\Omega}_T(\omega_0+\omega_0^{'}k),\quad |k|\leq N-1$, so
that
\begin{eqnarray}\sum_{n,m=0}^{N-1}|\textbf{C}[n, m]- \frac{1}{2}a[n, m]\widehat{\Omega}_T(\omega_0+\omega_0^{'}(n-m))|^2\end{eqnarray}
is a minimum, or equivalently so that
\begin{eqnarray}\sum_{|k|\leq N-1\atop max(0, -k)\leq m\leq min(N-1-k, N-1)}
|\textbf{C}[m+k, m]- \frac{1}{2}a[m+k,
m]\widehat{\Omega}_T(\omega_0+\omega_0^{'}k)|^2\end{eqnarray} is a
minimum. Eq. (4.93) is a trivial quadratic optimization problem,
which results in a linear equation for
$\widehat{\Omega}_T(\omega_0+\omega_0^{'}k), \quad |k|\leq N-1$ as
related to section 2 in the last problem of the eq. (4.29). Once
again we note that optimization needs to be carried out over only
the discrete frequency samples of $\widehat{\Omega}_T$ and not
over the entire time function trajectory $\Omega(t),\quad 0\leq
t\leq T$.
\section{Conclusions and Scope for Future Work}
A quantum gate is specified by a unitary matrix $U_g$ or
equivalently by its generator $H_g$ which is a Hermitian matrix
and satisfies $U_d = e^{-\imath H_g}$. Such a gate can be realized
by first realizing a generator $H_{g,N}=\frac{1}{N}H_g$ where $N$
is a large positive integer and then applying $U_{g,N} =
e^{-\imath H_{g,N}}$ $N$ times, that is, $U_{g,N}^N = U_g$. Based
on this general philosophy, we have perturbed the quantum system
with a Hamiltonian of the form $\epsilon\varphi(t) V$ where the
real function of modulating signal $\varphi(t)$, $t\in [0,T]$ is
in our command and $V$ is a suitably chosen Hermitian matrix.
After time $T$, the unitary evolution operator is $$U(T) =
U_0(T)\bigg(1-\iota\epsilon\int_0^T\varphi(t)\widetilde{V}(t)dt\bigg)+O(\epsilon^2)$$where
$U_0(T) = e^{-\iota T H_0}$ is the evolution operator of the
unperturbed system and $V(t) = U_0(t)V U_0(t)$. Taking $\epsilon =
\frac{1}{N}$, we get that the generator of the unitary matrix
$U_0(T)^{-1}U(T)$ is given upto $O(\frac{1}{N})$ by $$H_N =
\frac{1}{N}\int_0^T \varphi(t)\widetilde{V}(t)dt$$ that is
$U_0(T)^{-1}U(T)\approx 1-\iota H_N\approx e^{-\imath H_N}$. Now
$U_g^\frac{1}{N}=e^{-\iota\frac{H_g}{N}}$, realizing $U_g$ using
the quantum dynamics is equivalently to the matching
$$U_g^\frac{1}{N}\approx 1-\iota\frac{H_g}{N}\approx 1- \iota
H_N$$that is, $\frac{H_g}{N}\approx H_N$, or equivalently, $H_g
\approx\int_0^T \varphi(t)\widetilde{V}(t)dt$. The approximation
of $U_g$ is given by $$U_g\approx \bigg(I-\frac{i}{N}\int_0^T
\varphi(t)\widetilde{V}(t)dt\bigg)^N$$In the limit $N\rightarrow
\infty$, this becomes $$U_g\approx e^{-\iota\int_0^T
\varphi(t)\widetilde{V}(t)dt}$$ or
$$H_g\approx \int_0^T
\varphi(t)\widetilde{V}(t)dt$$ To get a unitary approximation, we
can use the Cayley transform which approximates $U_g$ by $$U_g
\approx\bigg( \frac{I-\frac{\iota}{2N}\int_0^T
\varphi(t)\widetilde{V}(t)dt}{I+\frac{\iota}{2N}\int_0^T
\varphi(t)\widetilde{V}(t)dt}\bigg)^N$$ The actual gate, realized
by the quantum evolution is $U(T) \approx U_0(T)(1-\iota H_N)$.
Since $H_N\approx \frac{H_g}{N}$, the approximation to $U_g$ using
quantum dynamics is given by $(U_0(T)^{-1}U(T))^N$. The function
$\varphi(t)$ is therefore chosen so that $\big\| H_g-\int_0^T
\varphi(t)\widetilde{V}(t)dt\big\|^2$ is a minimum subject to a
quadratic energy constraint on $\{\varphi(t)\}$. The solution for
$\varphi$ is easily expressed in the Fourier domain when $H_0$ is
the Harmonic oscillator Hamiltonian $\big(\frac{p^2+q^2}{2}\big)$.
The incorporation of the energy constraint leads to the associated
Lagrange multiplier being a root of a large degree polynomial and
this root is conveniently determined using MATLAB. Matching
unitary gates directly is problematic. It involves using
perturbation theory which may not result in a unitary gate
designed. On the other hand, matching generators will always give
a Hermitian approximation $H_g$ for the generator, even if we use
perturbation theory. Then the unitary gate designed $e^{-\iota
H_g}$ always will be unitary. Thus we have a marked advantage over
unitary matching. In the future, we shall display a better
approximation of the designed gate $U_g = e^{-\iota H_g}$ by using
multiple potentials, $ V_1,\ldots,V_p$ modulated by $p$ signal
$\varphi_1(t),\ldots, \varphi_p(t)$ resulting in the perturbing
potential being $\sum_{k=1}^p \varphi_k(t)V_k$ in Figure 1.
Finally we have introduced how by applying this technique to the
specific example of the ion trap model, we can practically realize
quantum gates in the laboratory.

\chapter{CONCLUSIONS}
\newpage
This thesis starts with introducing the basic dynamics of a
quantum system, which may be an atom, a finite set of atoms, a
finite state spin system, a harmonic oscillator, a finite set of
independent oscillators, or a quantum field. In all cases the
dynamics is described by an unperturbed Hamiltonian plus a small
perturbing Hamiltonian. The dynamics used to simulate the system
is the Schr\"odinger dynamics which involves analysis of the
unitary evolution operator  as a functional of control inputs
which modulate the perturbing Hamiltonian. The Dyson series
approximation enables us to get explicit formulas showing the
Volterra dependence of the evolution operator on the control
inputs. various cases of this general formalism have been
described in the thesis - one a finite state quantum system (like
a spin system) whose unperturbed Hamiltonian is a finite $N \times
N$ Hermitian matrix $H_0$ and the perturbing Hamiltonian is
$f(t)V$ or $\sum_{k=1}^d f_k(t)V_k$, where $f(t)$, $f_k(t)'s$ etc.
are control real scalar function of time. The resulting
approximate evolution operator $U(T|f)$ is matched to desired $N
\times N$ unitary gate $U_g$ by minimizing $$\|U(T|f)-U_g\|^2$$
with respect to f. Examples of how $H_0$ and $V$ may be
constructed from real physics are described. For example if $H_0$
is a harmonic oscillator $\frac{p^2+q^2}{2}$, then its matrix is
taken as a
truncated version $\left(\bigg(%
\begin{array}{c}
  \langle n|\frac{q^2+p^2}{2}|m\rangle \\
\end{array}%
\right)\bigg)_{0\leq n, m\leq N-1}$ (which is diagonal) where
$|n\rangle$, $n=0, 1, 2, \cdots$ are the normalized eigenstates of
$H_0$. If $V=\epsilon q^3$ (an anharmonic perturbing potential),
then the matrix of $V$ is $\epsilon\left(\bigg(%
\begin{array}{c}
   \langle n|q^3|m\rangle \\
\end{array}%
\right)\bigg)_{0\leq n, m\leq N-1}$ which is non-diagonal. Using
time independent perturbation theory, we calculate the matrix of
the gate $\langle n|e^{-\iota T (H_0+\epsilon V)}|m\rangle$. Then
we try to approximate, that is, realize this gate by perturbing
$H_0$ with a electric field potential $V(t)=\epsilon f(t)q$ by
using the Dyson series truncation. This formalism has also been
applied to the $3$-D harmonic oscillator by perturbing
$H_0=\sum_{k=1}^3\frac{q_k^2+p_k^2}{2}$ with an electromagnetic
field $E(t)$, $B(t)$ taking the approximate vector potential as
$\Phi(t, q)=-e(E(t), q)$ and $A(t, q)=\frac{1}{2}B(t)\times q$.
The resulting perturbed Hamiltonian is $$H(t)=\frac{(p+eA(t,
q))^2}{2}+\frac{q^2}{2}-e\Phi(t, q)$$ and using time dependent
perturbation theory, the truncated matrix elements
$$
\left(\bigg(%
\begin{array}{c}
  \langle n_1n_2n_3|U(t|E, B)|m_1m_2m_3\rangle \\
\end{array}%
\right)\bigg)_{0\leq n_1n_2n_3, m_1m_2m_3\leq N-1}
$$
are computed where $|n_1n_2n_3\rangle$ are the energy eigenstates
of $\frac{q^2+p^2}{2}$. A desired unitary gate $U_g$ is
approximated using $\langle n|U(t|E, B)|n$ by appropriate
selection of $E$, $B$. Here, we encounter for the first time, the
example of separable and non-separable gates. If $U_g=
U_{g1}\otimes U_{g2}\otimes U_{g3}$, that is, $U_g$ is a separable
gate, then we need to perturb the oscillator along each dimension
$H_{0k}=\frac{q_k^2+p_k^2}{2},\quad k=1, 2, 3$ individually by
perturbing each cannot Hamiltonian $H_{0k}$ individually. If
however $U_g=(U_{g1}\otimes U_{g2}\otimes U_{g3})(I+\iota\epsilon
X_I)\approx (U_{g1}\otimes U_{g2}\otimes U_{g3})e^{\iota\epsilon
X_I}$ is a weakly non-separable gate then we may first realize
$U_{g1}\otimes U_{g2}\otimes U_{g3}$ using independent oscillators
and then realize $U_g$ by using the non-independent
electromagnetic perturbation $A(t, q)$ and $\Phi(t, q)$. Simple
optimization routines have been written for calculating the
optimum perturbation subject to energy constraints. The final
application of the perturbed Hamiltonian idea is to gate design in
the ion trap model. Here by $H_0$, the unperturbed Hamiltonian
consists of the sum of a spin $\frac{1}{2}$ particle Hamiltonian
interacting with a constant magnetic field $H_{s0}=k\sigma_z$ and
the quantum electromagnetic field Hamiltonian
$H_0=\sum_{k=1}^Na_k^\dag a_k$ consists of $N$ field oscillators.
The interaction between the quantum magnetic field and the spin
$\frac{1}{2}$ particle has the form
$$
V=\sum_{k=1}^N(\overrightarrow{\sigma},v(k,
t))a_k+(\overrightarrow{\sigma}, \overline{v(k, t)})a_k^\dag
$$
where $v(k, t)$ are complex valued functions having the form $v(k,
t)=v_0(k)e^{\iota \omega_k t}$ and $v_0(k)$ represent the transfer
function frequency samples through which the quantum magnetic
field passes before interacting with the spin $\frac{1}{2}$
particle, $\overrightarrow{\sigma}=(\sigma_x, \sigma_y, \sigma_z)$
acts in the system Hilbert space $H_s=\mathbb{C}^2$, while the
$a_k^{'s}$ act in the both field space $L^2(\mathbb{R}_+)^{\otimes
N}$. The idea is to apply the Dyson series approximation to arrive
at the formula for the total unitary evolution operator $U(T)$ on
the system plus bath space $\mathbb{C}^2\bigotimes
L^2(\mathbb{R}_+)^{\otimes N}$ and design the filter frequency
samples $\{v_0(k)\}$ to get the best possible approximation of
$U(t)$ to a given $U_g$. By using the large number of degrees of
freedom of quantum field theory, the dimension of the designed
gate can be made vary large. Further, given an input system plus
bath field density $\rho_s(0)\bigotimes \rho_F(0)$ We can
calculate, using the Dyson series expansion, the system density
after time $T$ namely
$$\rho_s(T)=\Tr_F(U(T)(\rho_S(0)\bigotimes \rho_F(0)U^*(T))$$
that is, by tracing out over the bath field variables. Then design
the $v_0(k)^{'s}$ to get good match of $\rho_s(T)$ with a given
$\rho_{sg}$. This latter problem enables us to realize a given
mixed state rather than a unitary gate.

\renewcommand{\bibname}{REFERENCES}

\newpage
\addcontentsline{toc}{chapter}{LIST OF PUBLICATIONS}
\newpage
\begin{center}
\section*{\Huge{LIST OF PUBLICATIONS}}
\end{center}
\begin{enumerate}
\subsection*{Papers in International Journals:}

   \item Kumar Gautam, Tarun Kumar Rawat, Harish Parthasarathy, Navneet Sharma and Varun Upadhyaya, "Realization of the Three-Qubit Quantum Controlled Gate Based on Matching Hermitian Generators", Quantum Information Processing (Springer), 16, 113 2017. 10.1007/s11128-017-1564-4

    \item Kumar Gautam, Tarun Kumar Rawat, Harish Parthasarathy and
Navneet Sharma, "Realization of commonly used quantum gates using
perturbed harmonic oscillator", in Quantum Inf. Process, 14
(2015), 3257-3277.
    \item Kumar Gautam, Garv Chauhan, Tarun Kumar Rawat, Harish
Parthasarathy and Navneet Sharma, "Realization of quantum gates
based on three dimensional harmonic oscillator in a time varying
electromagnetic field", in Quantum Inf. Process, 14 (2015),
3279-3302.
    \item Ashwani Kumar Sharma, Harish Parthasarathy , Dharmendra
Upadhyay and Kumar Gautam, "Design and Realization of Quantum
based Digital Integrator", International Journal of Electronic and
Electrical Engineering. ISSN 0974-2174, Volume 7, Number 5 (2014),
pp. 449- 454.
\end{enumerate}

\newpage
\addcontentsline{toc}{chapter}{BIO-DATA}
\newpage
\begin{center}
\section*{\Huge{BIO-DATA}}
\end{center}

Kumar Gautam was born on October 06, 1985 at Begusarai (Bihar). He
received the degree of AMIETE in Electronics and Telecommunication
in 2008 from Grad IETE, New Delhi. He received the degree of M.
Tech. in Electronics system and Communication in 2010 from NIT
Rourkela, Odissa. He worked as Assistant Professor in Electronics
and Communication Engineering Department at M.R.C.E. faridabad
from 2010-2011. In January 2012, he joined as T.R.F. in
Electronics and Communication Engineering Department at Netaji
Subhas Institute of Technology, New Delhi for pursuing the degree
of Ph.D. He enrolled for Ph.D. in Electronics and Communication
Engineering Department, University of Delhi under the supervision
of Prof. Harish Parthasarathy and Dr. Tarun Kumar Rawat.
Presently, he is working as Assistant Professor at IILM-CET,
Greater Noida UP.

\end{document}